\documentclass{pasa} %
\usepackage{graphicx}	% Including figure files
\usepackage{amsmath}	% Advanced maths commands
\usepackage{amssymb}	% Extra maths symbols
\usepackage{etoolbox}
\usepackage{threeparttable, tablefootnote}

\title[ASKAP Observations: the GAMA 23 Field]{ASKAP Commissioning Observations of the GAMA 23 Field}
\author[D. A. Leahy et al.]{
Denis A. Leahy$^{1}$\thanks{E-mail: leahy@ucalgary.ca},
A.M. Hopkins$^{2}$,
R. P. Norris$^{3,4}$,
J. Marvil$^{4,5}$,
J.D. Collier$^{3,4,26}$,
E.N. Taylor$^{6}$,
J.R. Allison$^{7}$,
C. Anderson$^{4}$,
M. Bell$^{8}$,
M. Bilicki$^{9,10}$,
J. Bland-Hawthorn$^{11}$,
S. Brough$^{12}$,
M.J.I. Brown$^{13}$,
S. Driver$^{14}$,
G. Gurkan$^{15}$,
L. Harvey-Smith$^{4,3}$,
I. Heywood$^{16,17}$,
B.W. Holwerda$^{18}$,
J. Liske$^{19}$,
A. R. Lopez-Sanchez$^{2}$,
D. McConnell$^{4}$,
A. Moffett$^{20}$,
M.S. Owers$^{21}$,
K.A. Pimbblet$^{22}$,
W. Raja$^{4}$,
N. Seymour$^{23}$,
M.A. Voronkov$^{4}$,
and L. Wang$^{24,25}$
\affil{$^{1}$Department of Physics and Astronomy, University of Calgary,  Calgary, Alberta, T2N 1N4, Canada}
\affil{$^{2}$Australian Astronomical Optics, Macquarie University,105 Delhi Rd, North Ryde, NSW 2113, Australia}
\affil{$^{3}$Western Sydney University, Locked Bag 1797, Penrith, NSW, 2751, Australia}
\affil{$^{4}$CSIRO Astronomy and Space Science, Australia Telescope National Facility, PO Box 76, Epping, NSW 1710, Australia}
\affil{$^{5}$National Radio Astronomical Observatory, P.O. Box O, 1003 Lopezville Road, Socorro, NM 87801-0387, USA}
\affil{$^{6}$Centre for Astrophysics and Supercomputing, Swinburne University of Technology, Hawthorn, Vic 3122, Australia}
\affil{$^{7}$Sub-Dept. of Astrophysics, Department of Physics, University of Oxford, Denys Wilkinson Building, Keble Rd., Oxford, OX1 3RH, UK}
\affil{$^{8}$University of Technology Sydney, 15 Broadway, Ultimo NSW 2007, Australia}
\affil{$^{9}$Leiden Observatory, Leiden University, P.O. Box 9513, NL-2300 RA 
Leiden, The Netherlands}
\affil{$^{10}$National Centre for Nuclear Research, Astrophysics Division, P.O. Box 447, PL-90-950 {\L}\'{o}d\'{z}, Poland}
\affil{$^{11}$Sydney Institute for Astronomy, School of Physics A28, University of Sydney, NSW 2006, Australia}
\affil{$^{12}$School of Physics, University of New South Wales, NSW 2052, Australia}
\affil{$^{13}$School of Physics and Astronomy, Monash University, Clayton, VIC 3800, Australia}
\affil{$^{14}$International Centre for Radio Astronomy Research (ICRAR), University of Western Australia, Stirling Highway, Perth, Western Australia}
\affil{$^{15}$CSIRO Astronomy and Space Science, PO Box 1130, Bentley WA 6102, Australia}
\affil{$^{16}$Astrophysics, Department of Physics, University of Oxford, Keble Road, Oxford OX1 3RH, UK}
\affil{$^{17}$Department of Physics and Electronics, Rhodes University, PO Box 94, Grahamstown, 6140, South Africa}
\affil{$^{18}$Department of Physics and Astronomy, 102 Natural Science Building, University of Louisville, Louisville KY 40292, USA}
\affil{$^{19}$Hamburger Sternwarte, Universitat Hamburg, Gojenbergsweg 112, 21029 Hamburg, Germany}
\affil{$^{20}$Department of Physics and Astronomy, Vanderbilt University, USA}
\affil{$^{21}$Department of Physics and Astronomy, Macquarie University, NSW 2109, Australia}
\affil{$^{22}$E.A.Milne Centre for Astrophysics, University of Hull, Cottingham Road, Kingston-upon-Hull, HU6 7RX, UK}
\affil{$^{23}$International Centre for Radio Astronomy Research, Curtin University, Bentley WA 6102, Australia }
\affil{$^{24}$SRON Netherlands Institute for Space Research, Landleven 12, 9747 AD, Groningen, The Netherlands}
\affil{$^{25}$Kapteyn Astronomical Institute, University of Groningen, Postbus 800, 9700 AV, Groningen, The Netherlands}
\affil{$^{26}$
The Inter-University Institute for Data Intensive Astronomy (IDIA), Department of Astronomy, University of Cape Town, Rondebosch, 7701, South Africa}
}

\jid{PASA}
\doi{10.1017/pas.\the\year.xxx}
\jyear{\the\year}

\usepackage{aas_macros}
\usepackage{hyperref} 
\hypersetup{colorlinks,citecolor=blue,linkcolor=blue,urlcolor=blue}
\hypersetup{draft}

\begin{document}
\begin{frontmatter}
\maketitle

\begin{abstract}
We have observed the G23 field of the Galaxy And Mass Assembly (GAMA) survey using the
Australian Square Kilometre Array Pathfinder (ASKAP) in its commissioning phase, to validate
the performance of the telescope and to characterize the detected galaxy  
populations.
This observation covers $\sim$48 deg$^2$ with synthesized beam of 32.7$^{\prime\prime}$ by 17.8$^{\prime\prime}$ at 936\,MHz,
and $\sim$39 deg$^2$ with synthesized beam of 15.8$^{\prime\prime}$ by 12.0$^{\prime\prime}$  at 1320\,MHz.
At both frequencies, the r.m.s. (root-mean-square) noise is $\sim$0.1 mJy/beam. 
We combine these radio observations with the GAMA galaxy data, which includes spectroscopy of galaxies that are i-band selected with a magnitude limit of 19.2.
Wide-field Infrared Survey Explorer (WISE) infrared (IR) photometry is used to determine which galaxies host an active galactic nucleus (AGN).
In properties including source counts, mass distributions, and IR vs. radio luminosity relation, 
the ASKAP detected radio sources behave as expected.
Radio galaxies have higher stellar mass and luminosity in IR, optical and UV than other galaxies. 
We apply optical and IR AGN diagnostics and find that they disagree for $\sim$30\% of the galaxies in our sample.
We suggest possible causes for the disagreement. Some cases can be explained by optical extinction of the AGN, but
for more than half of the cases we do not find a clear explanation. 
Radio sources are more likely ($\sim$6\%) to have an AGN than radio quiet galaxies ($\sim$1\%), 
but the majority of AGN are not detected in radio at this sensitivity. 
\end{abstract}

\end{frontmatter}

\section{Introduction}
The moderate redshift Universe (0.1$\lesssim z\lesssim1$) has in the past decade become 
accessible for large surveys of galaxies at optical and  IR wavelengths.
However, most radio surveys have sampled different populations from these, mainly because 
most bright radio sources represent the synchrotron emission from active galactic nuclei (AGN) (e.g. \citealt{2014ARA&A..52..589H}), which are relatively uncommon in optical and IR surveys.
This is now changing, as the largest area radio surveys are now reaching the sensitivity where they are dominated by low-redshift star forming galaxies (SFG) (e.g. \citealt{2017Norris}), in which the radio emission is 
powered by star formation processes. 
Thus, it is an opportune time to compare radio and galaxy surveys to find the same objects emitting in radio and optical/IR bands.

The Evolutionary Map of the Universe (EMU) \citep{2011Norris} is a wide-field radio continuum survey planned for the new Australian Square Kilometre Array Pathfinder (ASKAP) telescope \citep{2008SJohnson}. 
The primary goal of EMU is to make a deep (r.m.s. $\sim$10$\mu$Jy/beam) radio continuum survey of the entire Southern sky at 1.3 GHz, extending as far North as +30 declination, with a resolution of 10 arcsec. 
EMU is expected to detect and catalogue about 70 million galaxies, including typical SFG up to z$\sim$1, powerful starbursts to even greater redshifts, and AGN to the edge of the visible Universe. 
Here we report early EMU observations, taken during the commissioning phase of ASKAP, 
with the goal of characterizing the galaxy and radio galaxy samples.

Radio galaxies can be divided into two classes: SFG and AGN. 
The radio emission from SFG is synchrotron emission of cosmic-ray electrons accelerated by shocks from supernovae, and so the radio emission is confined to the inner few kpc of a galaxy, and is seen in the radio image as a single unresolved component. 
The radio emission is relatively weak, is strongly correlated with the far-IR flux, and is typically detected from galaxies in the nearby Universe, with a median redshift $z<0.5$.
In AGN, the radio emission is synchrotron emission emitted by the jets and lobes of relativistic electrons accelerated from the environs of a supermassive black hole, and may appear as a single, double, or triple component, or sometimes as a complex structure with visible jets. 
The radio emission can be very strong, and easily detectable up to very high redshifts. 

Although these two types of radio source are powered by very different physical mechanisms, it is notoriously hard to distinguish between them from the radio data alone \citep{2013PASA...30...20N}. 
While multiple-component sources and strong high-redshift sources are probably AGN, a single component source at low redshift, or with no measured redshift, may be either AGN or SFG. 

Optical line ratios have been used as a diagnostic to classify galaxies as SFG or AGN since it was proposed
by  \citet{1981BPT}.  
We use this Baldwin-Phillips-Terlevich (BPT) diagnostic in the form developed by \citet{2001KewleySBmodel},
which uses the line ratios [OIII]$\lambda$5007/$H{\beta}$ and [SII]$\lambda\lambda$6717,6731/$H{\alpha}$
\citep[e.g.,][]{2013HopkinsGAMAspectra}.
The line ratios are sensitive to the spectral hardness of the radiation which ionizes the gas, and can be used to determine when the
radiation is too spectrally hard to originate from star formation, i.e., the ionizing radiation can only be produced by an AGN.
We refer to the ([OIII]$\lambda$5007/$H{\beta}$) vs ([SII]$\lambda\lambda$6717,6731/$H{\alpha}$) diagnostic as the BPT diagram and refer
to this set of lines as the BPT lines.

AGN generate thermal radiation in the accretion disk surrounding a supermassive black hole. 
Much of this  is reradiated in IR by dust with a wide range of temperatures, resulting in characteristic red mid-IR colours which
can be distinguished from the mid-IR colours resulting from stars.
The WISE mid-IR colour criterion W1-W2>0.8 was introduced by \citet{2011Jarrett}, and is used here to aid in
distinguishing AGN from other galaxies.

In this paper we present radio data obtained with ASKAP during its commissioning phase.
One main emphasis is to verify that the radio observations are detecting real radio sources rather than imaging artifacts.
The second is to use the radio data, in conjunction with optical and IR data, to study galaxy populations.

The optical data used here is from the GAMA survey \citep{2011DriverGAMA}, and the IR data from the
WISE all-sky survey  \citep{2010Wright}. GAMA is a spectroscopic survey of galaxies at low to moderate redshift ($z\lesssim1$).  
The GAMA regions of the sky are accessible to EMU, thus GAMA is ideal for
 providing optical data on the galaxies observed by EMU.
With the combined radio, optical and IR data of the galaxy surveys, we investigate their masses, redshifts, and states of activity (SFG vs AGN). 
Our goal is to measure properties which distinguish different galactic populations, using the unique combination of depth and sky coverage of ASKAP.

This paper is organized as follows. 
The ASKAP radio data analysis, GAMA optical data and WISE IR data are described in Section~\ref{sec:obs}.
In Section~\ref{sec:multiwave},  we identify optical and IR counterparts to the radio sources
and describe the statistical properties of the various sets, including redshifts and stellar masses derived from the optical data. 
In Section~\ref{sec:AGNSFG}, we apply optical and IR AGN diagnostics to classify galaxies and radio sources and
compare redshift, mass and luminosity distributions of the different subsets.
The properties of the radio sources and of the differences in AGN classification using optical and IR methods are 
discussed in Section~\ref{sec:disc}.
A summary of the main results is given in  Section~\ref{sec:summary}. 

We calculated luminosity distance from redshift,
using standard cosmological parameters: flat universe with $H_0$=70 km/s/Mpc, $\Omega_m$=0.27, $\Omega_{\Lambda}$=0.73.
Stellar masses were determined using a \citet{2003PASP..115..763C} stellar initial mass function.

\begin{figure}
     \includegraphics[width=\columnwidth]{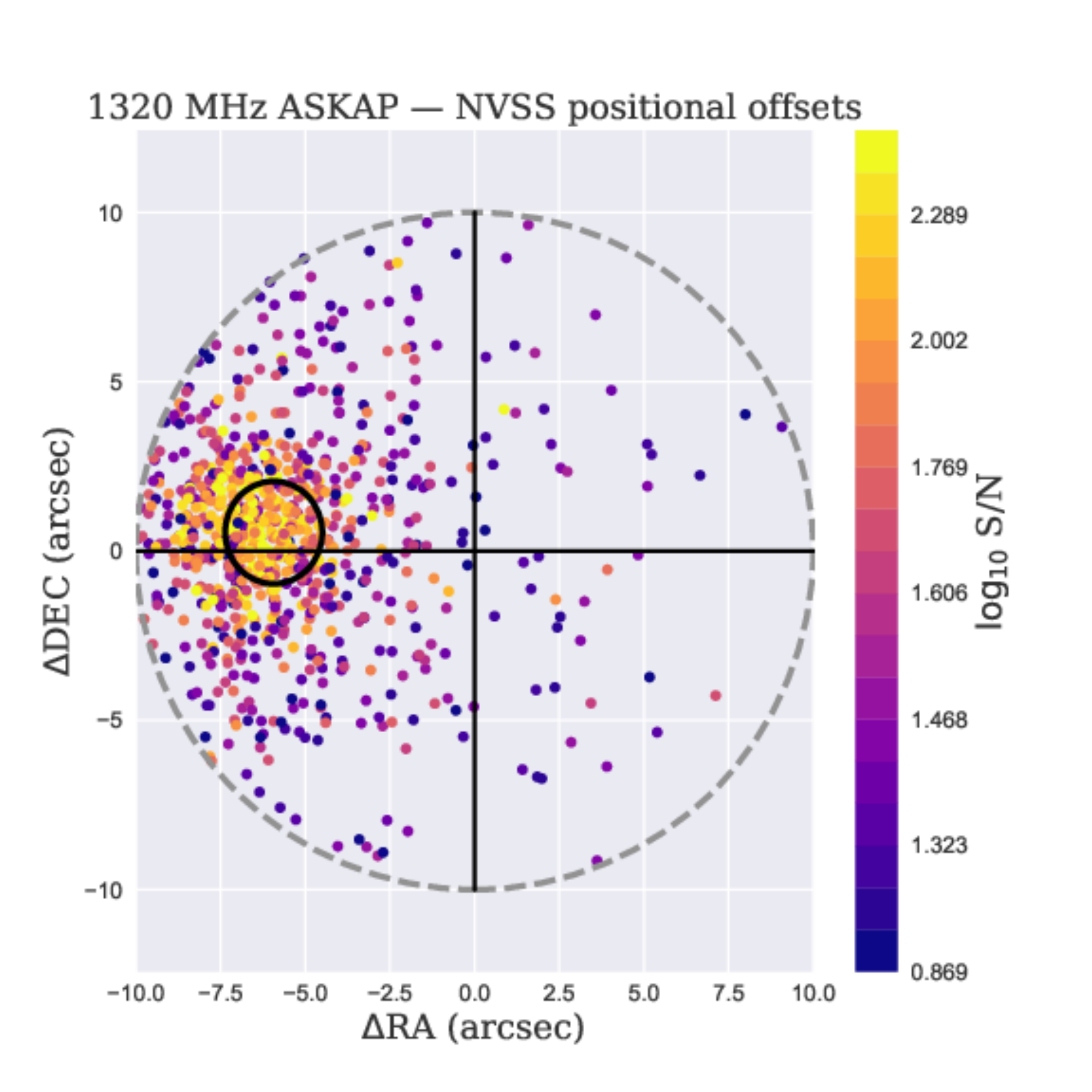} 
          \includegraphics[width=\columnwidth]{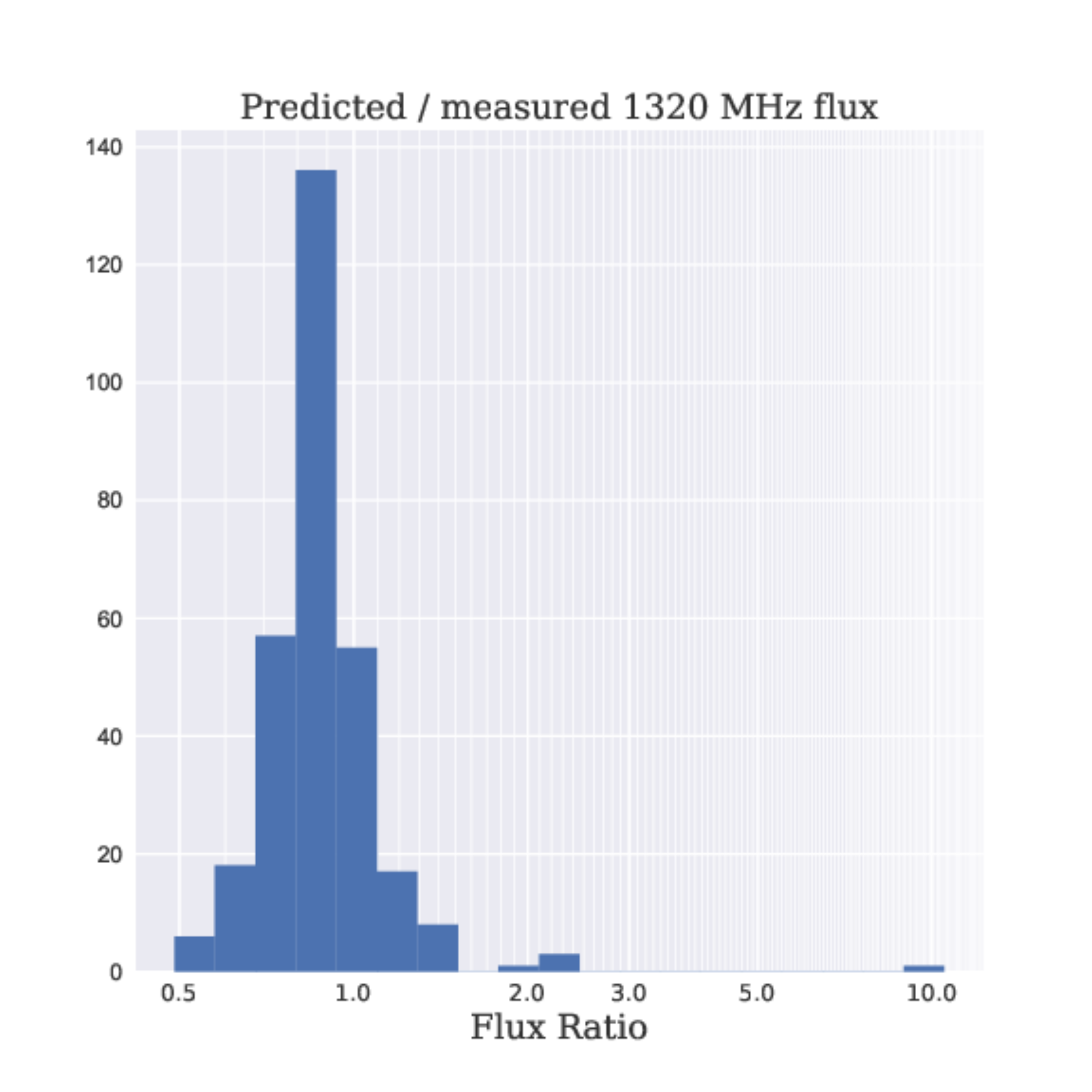} 
    \caption{Top panel: offsets of the  raw 1320\,MHz ASKAP positions from the NVSS positions for sources detected in both surveys.
    The color bar gives the scale for signal-to-noise (SN) of individual ASKAP sources.
The offsets are small compared to the synthesized beam size of 15.8$^{\prime\prime}$ by 12.0$^{\prime\prime}$  at 1320\,MHz.
The median 1320\,MHz-NVSS offset of -5.69$^{\prime\prime}$ in R.A. and 0.26$^{\prime\prime}$ in Dec. was used to correct the raw ASKAP 1320\,MHz positions.
 The median 936\,MHz-NVSS offset of -6.71$^{\prime\prime}$ in R.A. and 1.04$^{\prime\prime}$ in Dec. was used to correct the raw ASKAP 936\,MHz positions.
Bottom panel: ratios of NVSS-SUMSS predicted 1320\,MHz flux density to ASKAP raw measured 1320\,MHz flux density for sources measured  in all 3 surveys. 
The median ratio for 1320 MHz sources of 0.8830 was used to correct the ASKAP 1320\,MHz flux densities.
The median ratio   for 936 MHz sources of 0.8810 was used to correct the ASKAP 936\,MHz flux densities. 
      }
    \label{fig:figoffset}
\end{figure}

\begin{figure*}
	\includegraphics[scale=1.06]{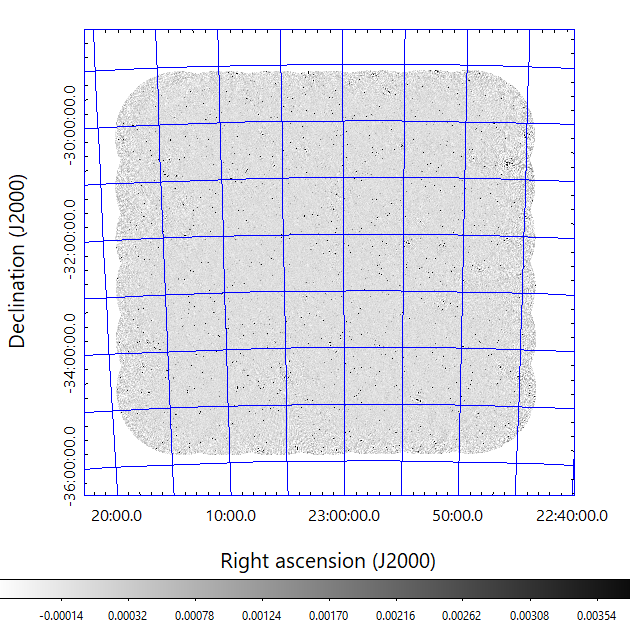}
    \caption{The 936\,MHz final processed radio image for the GAMA 23 region.
    This covers a sky area of $\sim7.3^\circ$ in R.A. by $\sim6.8^\circ$ in Dec.  The greyscale bar at the bottom is in units of Jy/beam.}
    \label{fig:fig936image}
\end{figure*}

\begin{figure*}
	\includegraphics[scale=1.07]{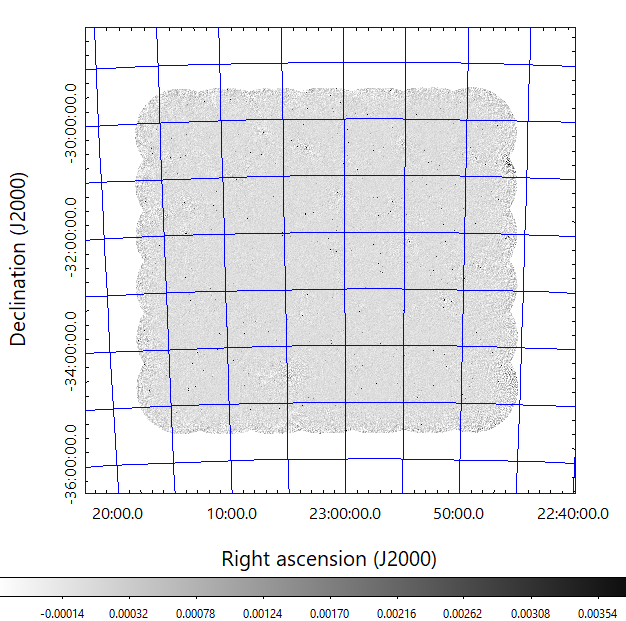}
    \caption{The 1320\,MHz final processed radio image for the GAMA 23 region.
    This covers a sky area of $\sim6.6^\circ$ in R.A. by $\sim6.1^\circ$ in Dec.  The greyscale bar at the bottom is in units of Jy/beam.}
    \label{fig:fig1320image}
\end{figure*}

\section{Observations and Data Analysis}\label{sec:obs}

\subsection{Radio data}

The Australian Square Kilometre Array Pathfinder \citep[ASKAP;][]{2008SJohnson}
 is a new radio telescope approaching completion in Western Australia.  
 It has a maximum baseline of 6 km, and operates with a bandwidth of $\simeq$300 MHz over the frequency range 700 to 1800 MHz. 
 Each of the 36 antennas is equipped with a Phased Array Feed  \citep[PAF;][]{2012Schinckel}
giving it a field of view of $\sim$35 deg$^{2}$, resulting in a high survey speed. 
ASKAP's continuum survey is the Evolutionary Map of the Universe \cite[EMU][]{2011Norris},
which is expected to generate a catalogue of about 70 million galaxies at 1100 MHz.

At the time of the observations described here, the telescope was in a commissioning phase \citep{2016McConnell}.
Only 12 of the antennas were used, with a maximum baseline of 2 km. 
An observation of PKS $B1934-638$ was performed immediately adjacent in time to each observation of the target field 
for purposes of instrumental calibration.  Each calibration observation contained one calibrator scan of duration 
$\sim$5 minutes at the center of each of the 36 PAF beams.  
Using these data, the ASKAPsoft task ``cbpcalibrator'' was run to solve for the instrumental bandpass of each beam 
and to set the initial flux density scale based on the model of \citet{1994reynolds}.   
As is customary with ASKAP, a phase reference calibrator was not included in the schedule due to the large overhead 
that would be incurred by observing it at the center of each beam. 
Instead, initial phases were transferred from the observation of the bandpass calibrator and further refined using 
self-calibration of the target field, and the astrometry was adjusted to match an external reference catalogue.

The observing parameters are listed in Table \ref{tab:askap}.
 The PAF beam configuration was 36 beams in a 6$\times$6 square, with two observations A and B offset by half a beam width, at a position angle of 45$^\circ$.
The 936\,MHz map covers a larger region  ($\sim$7.3$^\circ$ by  $\sim$6.8$^\circ$) than the 1320\,MHz image ($\sim$6.6$^\circ$ by  $\sim$6.1$^\circ$)
because of the larger primary beam size at 936\,MHz. 
The data were reduced in ASKAPsoft using W-snapshots \citep{2012SPIE.8500E..0LC}, which is a hybrid wide-field algorithm combining W-projection and warped snapshot imaging.
The image was cleaned (deconvolved) using the Cotton-Schwab algorithm and a single (delta function) spatial scale. 

The resulting synthesized beam was 32.7$^{\prime\prime}$ by 17.8$^{\prime\prime}$  at 936\,MHz and 15.8$^{\prime\prime}$ by 12.0$^{\prime\prime}$  at 1320\,MHz.
The synthesized beam of the higher frequency observations is more circular than the beam at lower frequency 
because of differences in the U-V coverage, which is due primarily to differences in hour angle coverage and 
different amounts of antenna-based flagging.
In both cases, the output images consisted of 7784 $\times$ 7387 pixel images, where each pixel is a 4-arcsec square.

\begin{table*}
%\begin{threeparttable}
\caption{ASKAP Observation Details}
	\label{tab:askap}
\begin{tabular}{lllll}
\hline
Scheduling  & Frequency & Bandwidth &  Date & Observing  \\ 
Block ID & (MHz) & (MHz) & & time (hrs)   \\ 
\hline
2827 & 1320 & 144 & 03-Dec-2016 &  5 \\  
2831 & 1320 & 144 & 04-Dec-2016 & 5   \\      
2949 & 936 & 192 & 16-Dec-2016 & 4  \\   
2955 & 936 & 192 & 17-Dec-2016 & 4 \\  
2961 & 936 & 192 & 18-Dec-2016 & 4  \\  
\hline
\end{tabular}  
\end{table*}

Because these data were taken at an early stage of ASKAP commissioning, our flux density and astrometry calibration were still in a preliminary state, 
and so these had to be corrected after processing. This was done by comparing the ASKAP images with the  NVSS \citep{1998Condon} and SUMSS \citep{1999Bock} catalogues. 
We used only non-blended, unresolved sources detected at signal-to-noise (hereafter SN) above 5$\sigma$ in ASKAP, NVSS and SUMSS,
and which were not flagged as problematic.

The offsets of the  raw  1320 MHz ASKAP positions from the NVSS positions are shown in the top panel of Figure~\ref{fig:figoffset}.
The scatter plot for 936 MHz is similar. 
The medians of the 1320-NVSS and 936-NVSS position offsets were used as the position
corrections for the 1320 MHz and 936 MHz images, respectively. 
The  1$\sigma$ astrometric accuracy was estimated by using the median absolute deviation, converted to 1$\sigma$.
At 936 MHz the resulting 1$\sigma$ accuracy is 3.00$^{\prime\prime}$ in R.A. and 3.40$^{\prime\prime}$ in Dec.
and at 1320 MHz the resulting 1$\sigma$ accuracy is 2.37$^{\prime\prime}$ in R.A. and 2.64$^{\prime\prime}$ in Dec.

The flux density scale was calibrated by using sources that were in both NVSS and SUMSS. 
We interpolated between SUMSS (843 MHz) and NVSS (1420 MHz) flux densities for each source assuming a power-law spectrum. 
We tabulated the ratios of the calculated 936 MHz and 1320 MHz NVSS-SUMSS flux densities
to the raw 936 MHz and 1320 MHz ASKAP flux densities, respectively.  
The histogram of these ratios for 1320 MHz is shown in the bottom panel of Figure~\ref{fig:figoffset}.
The median calculated to raw 936 MHz flux density ratio was 0.8810 and the median for 1320 MHz was 0.8830.
These median ratios were used as flux density correction factors for the ASKAP 936 MHz and 1320 MHz images, respectively.

\begin{figure}
     \includegraphics[width=\columnwidth]{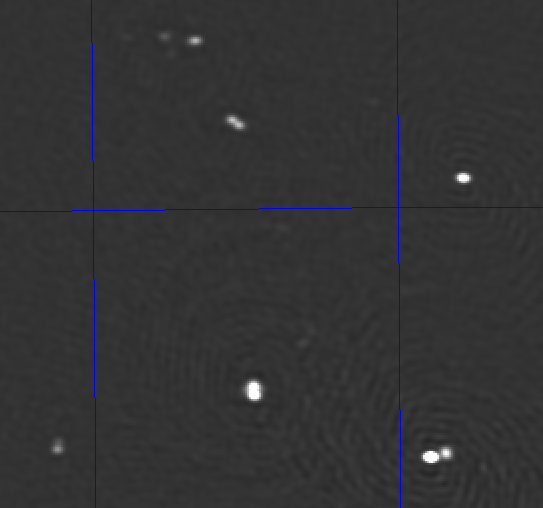}
    \caption{ Enlargement of the 1320-MHz image in an area centered on R.A. 23H 01M 10S, Dec. -32$^\circ$13$^{\prime}$30$^{\prime}$
in a region containing some double radio sources.
The grid spacing is 0.2$^\circ$. 
   The intensity greyscale is linear, from -5 to +20 m Jy/beam.
      }
    \label{fig:fig1320Dbl}
\end{figure}

\begin{figure}
     \includegraphics[width=\columnwidth]{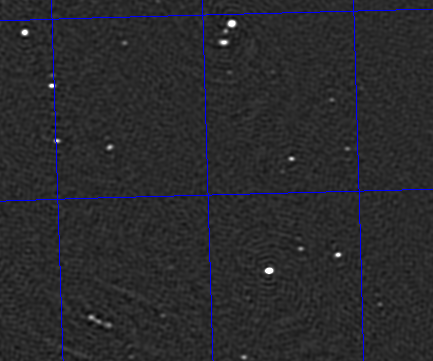}
    \caption{ Enlargement of the 1320-MHz image in an area centered on R.A. 23H 10M 24S, Dec. -33$^\circ$10$^{\prime}$30$^{\prime}$,
 in a region containing two radio triples (central source plus two lobes).
The grid spacing is 0.2$^\circ$ in R.A and in Dec. 
    The intensity greyscale is linear, from -1 to +5 mJy/beam.
       }
    \label{fig:fig1320Tpl}
\end{figure}

The resulting images are shown in Figures~\ref{fig:fig936image} and ~\ref{fig:fig1320image}.  
The root-mean-square (r.m.s.) noise of source free regions in both 936\,MHz and 1320\,MHz images is $\sim$0.1 mJy/beam.
To illustrate the image quality obtained with ASKAP in its commissioning phase, we show images on spatial scales of tens of arcminutes.
Figure~\ref{fig:fig1320Dbl} and Figure~\ref{fig:fig1320Tpl} show regions with typical r.m.s. noise level in the maps (0.1 mJy).
The first illustrates some double radio sources. The latter shows one
triple source (central source plus two lobes) at the top centre of the image and a fainter triple source at the lower left.

\subsubsection{Source extraction}\label{sec:srcex}

\begin{figure}
     \includegraphics[width=\columnwidth]{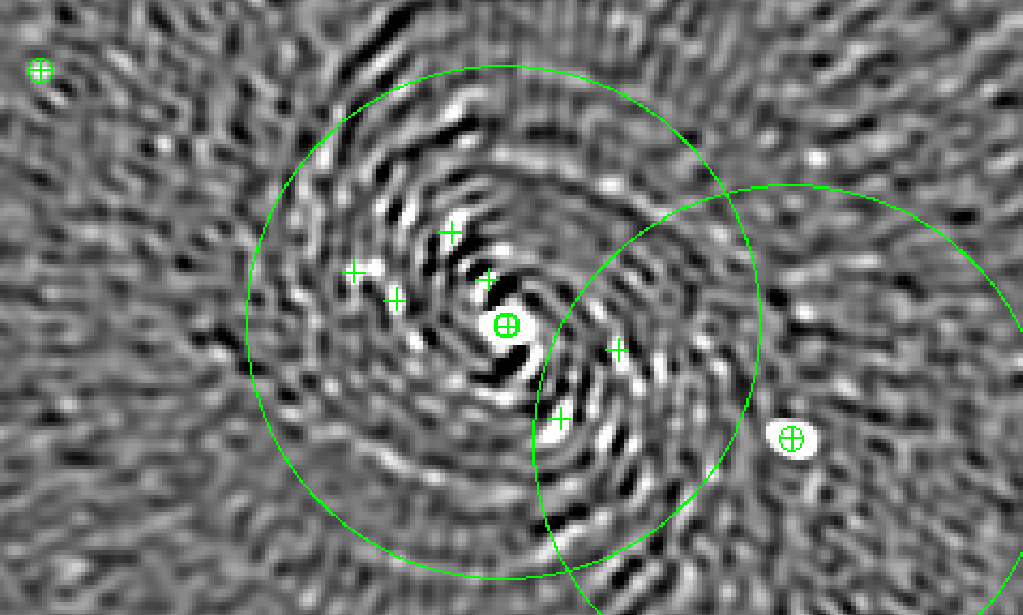}
    \caption{ Enlargement of the 936\,MHz image centred on a bright component (integrated flux density 0.78 Jy) which exhibits a spiral shaped pattern of residuals. 
    The large circles
    are 240$^{\prime\prime}$ in radius. The crosses show the detected components (4 sigma above background) before the correction. The small circles show the
   components after the correction. The intensity greyscale is expanded to show faint features, and is linear from -1 mJy/beam (black) to +1.5 mJy/beam (white).
   Most sources are saturated at this scale. 
       }
    \label{fig:figresiduals}
\end{figure}

We make the distinction between components, which are individual point-like detections in a radio image, and sources,
which can be either a single component or composed of multiple components, such as a double-lobed radio galaxy.
The Aegean software \citep{2012HancockASCL} was used for component  extraction.
We used a $4\,\sigma$ threshold, resulting in a list of 6053 components at 936\,MHz, and 3827 at 1320\,MHz.

Small but significant residuals, of order 1\% of the peak flux density, are found around bright components. 
The residuals occur in a spiral pattern 
which extends out to a radius of about 6$^{\prime}$ ($\sim$10 times the synthesized beam).
Figure~\ref{fig:figresiduals} shows a bright component exhibiting such residuals, the resulting false components, and our correction for them.
The flux densities of the false components were all less than 1.25\% of the integrated flux density of the bright central component at 936\,MHz, 
and less than 1.6\% at 1320\,MHz.
We started with the brightest component in the map and labelled it as the ``source''.  
All components within 6$^{\prime}$ radius of the ``source''  which were fainter than 1.25\% (at 936\,MHz) 
or 1.6\% (at 1320\,MHz) of the ``source" were added to a list of possible false components.
This was repeated for the second brightest component in the map, 
and so on, until a complete list of possible false components was created. 
The false components were removed from the original component list except for those few not consistent with the spiral beam residual pattern.
After artifact removal, the catalogue contained 5968 components at 936\,MHz, and 3757  at 1320\,MHz.

\subsubsection{Source counts}

\begin{figure}
	\includegraphics[width=\columnwidth]{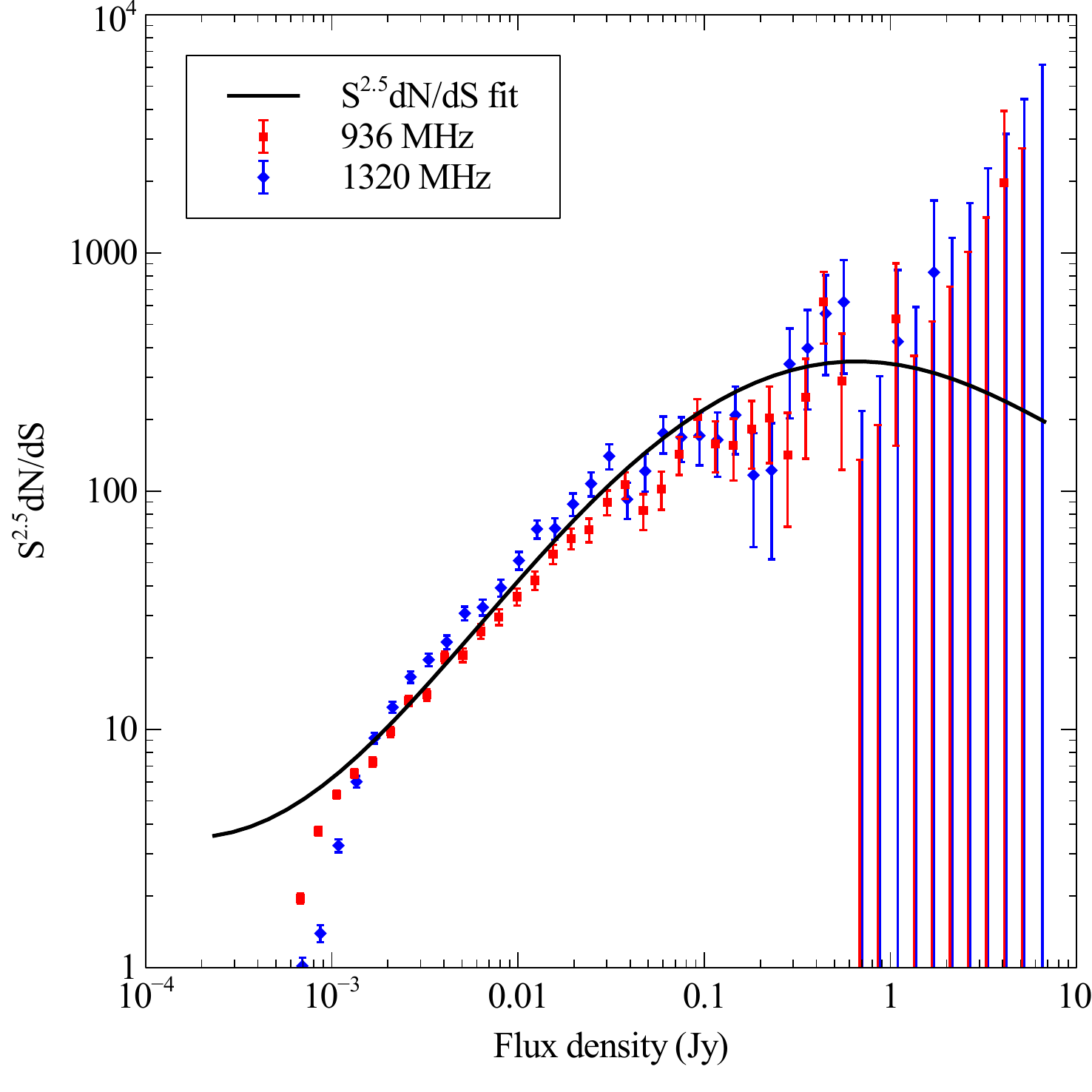}
    \caption{
    Source counts normalized by the Euclidean slope, $S^{2.5}dN/dS$, from the 936\,MHz and 1320\,MHz observations, with Poisson error bars. 
    The smooth line is a polynomial fit to the compilation of published source count data \protect\citep{2003HopkinsPhoenix}.
    }
    \label{fig:figsrccounts}
\end{figure}

The source counts normalized by the Euclidean slope ($S^{2.5}dN/dS$ vs. flux density), are shown in Figure~\ref{fig:figsrccounts} for 936\,MHz and 1320\,MHz.
The smooth line is a fit to the compilation of published source count data from \citet{2003HopkinsPhoenix}.
The flux densities for 936\,MHz sources and 1320\,MHz sources were adjusted to 1.4 GHz flux densities using a spectral index of
$\alpha=-0.7$. 

The data for both frequencies match well the fit line between $\sim1.2$ mJy and $\sim 0.2$ Jy. 
Above 0.2 Jy, the counts are dominated by small number statistics, reflected in the large error bars. 
Below $\sim 1.2$ mJy, the observations fall below the expected source counts, indicating that they are incomplete at this level, particularly at 1320\,MHz.
The map r.m.s. is 0.1 mJy excluding the edges.  
The r.m.s. rises nearly linearly from to $\sim$ 0.1 mJy at $\sim$ 0.5$^\circ$ from the edges to $\sim$ 0.5 mJy at the edges.
It also rises from $\sim$ 0.1 mJy at $\sim$ 0.25$^\circ$ away from $\sim$ 0.5 Jy sources to $\sim$ 1 mJy at $\sim$ 0.05$^\circ$ away.
The fraction of the map area with r.m.s above   $\sim$ 0.1 mJy is $\sim$0.2.
 Thus a significant fraction of sources below $\sim$ 1 mJy are not detected at the $4\,\sigma$ level.

\begin{table}
	\caption{Number of sources, by morphology, in the radio source catalogue}
	\label{tab:radiostats}
	\begin{tabular}{lccccccc} 
		\hline
		Frequency &  total             & single  &  double   & triple       \\
		       &       &  & (2 lobes) & (core+2 lobes)     \\		
		\hline
		936\,MHz  & 5791 & 5710 & 56 & 25\\ 
		1320\,MHz & 3589 & 3451 & 109 & 29 \\
		\hline
	\end{tabular}
\end{table}

\subsubsection{Identifying complex radio sources}

\begin{figure}
     \includegraphics[width=\columnwidth]{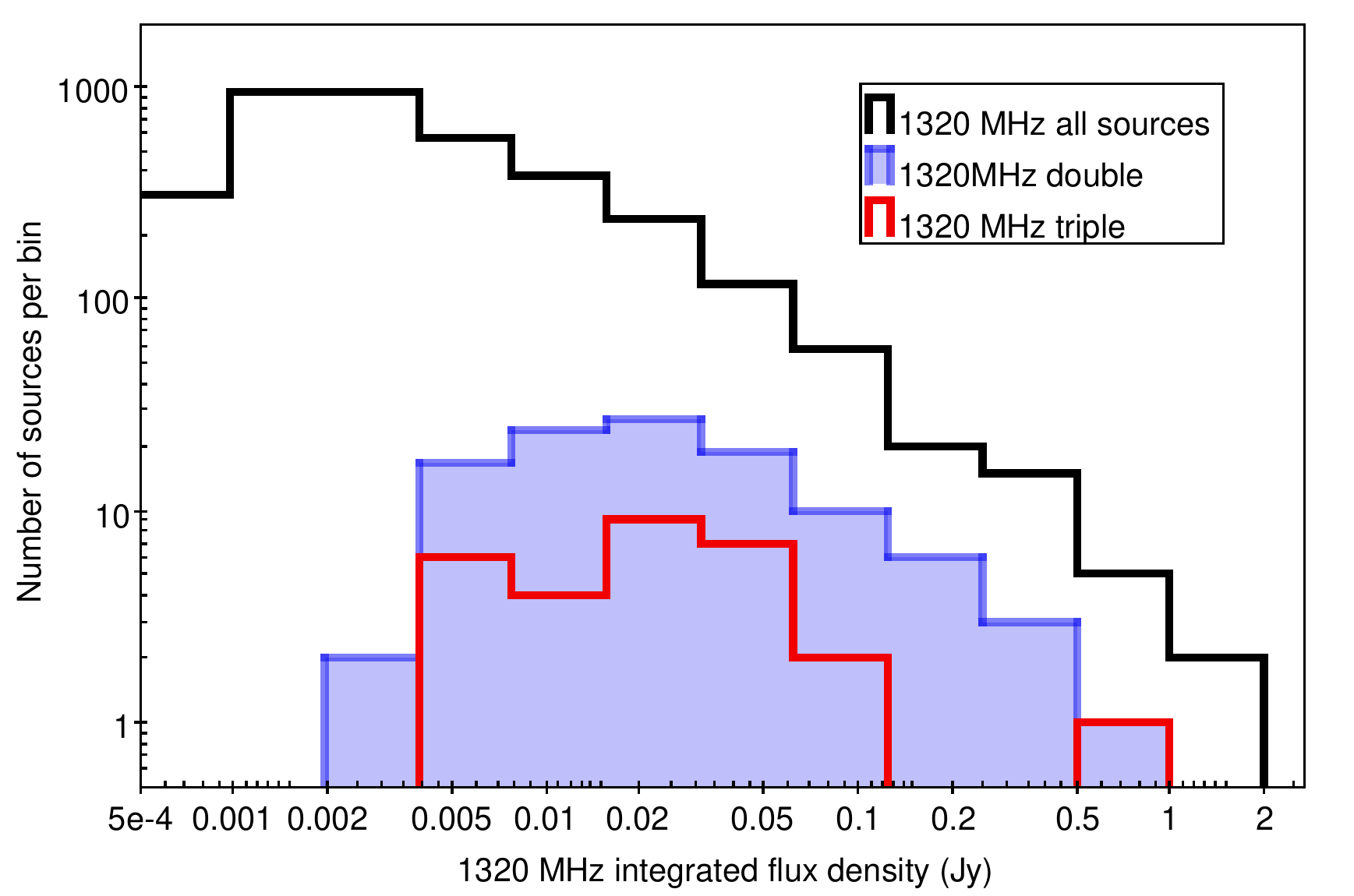} 
    \caption{ Distribution of integrated 1320 MHz flux densities of all sources (black line), double sources (blue shaded histogram) and triple sources (red line).
   The all source category is dominated by single sources. The doubles are brighter on average than the triples and both are much brighter than single sources.  
       }
    \label{fig:figSglDblTpl}
\end{figure}

About 95\% of radio sources at the mJy level, which we term ``simple'', 
consist of a single component which is coincident with the optical host galaxy \cite[e.g.,][]{2006Norris}.
The remaining $\sim$5\% are complex sources, such as double-lobed radio galaxies which appear on the radio image as two isolated components typically separated by 
tens of arc seconds, with the optical host galaxy located roughly halfway between them.
Cross-matching simple sources with optical/IR data is straightforward, 
and can be achieved using any of several algorithms and tools such as
those available within TopCat \citep{2011TaylorM}. 

 To identify the complex sources, we compiled one list of candidate complex sources from
the list of ``islands'' generated by Aegean, and a second list of candidates by using 
TopCat to identify pairs and groups of more than two sources within a radius of 6 arcmin.
There was a significant overlap ($\sim50-70\%$) between these two lists of candidates.
The difference between the two lists illustrates the difficulty of using a single algorithm to identify complex sources.
The visual inspection of Aegean islands and TopCat groups resulted in a revised list of singles, doubles and triples.
The singles include those components that were put in islands or groups by the software, but had no visual evidence of association. 
All islands and groups of 4 or more were found to consist of a double or triple plus unrelated singles. 

We tested the doubles against the Magliocchetti criterion \citep{1998Magliocchetti}, 
for which the flux density ratio of the two lobes is required to be between 1/4 and 4,
and the separation is required to be $<$~100$^{\prime\prime}$ $\times\sqrt{S_{tot}/100~{\rm mJy}}$. 
We used a flux density ratio between 1/3 and 3, to err on the side of rejecting sources,
but only found one source with flux density ratio between 1/4 and 1/3 (or 3 and 4) that also passed
the separation test.
Sources that failed this test were split into components and added back to the singles list. 

We scanned the full ASKAP radio images for radio doubles or triples not identified by either Aegean or TopCat algorithms,
and found several more radio doubles and triples, typically those with large angular separation.
The resulting numbers of single, double and triple radio sources are shown in Table 2. 

To estimate host galaxy positions and integrated flux densities for the double sources, we calculated the flux density-weighted centroid and the total integrated flux density. 
For the triples, we assumed the host position to be that of the component closest to the centre.
The distribution of 1320 MHz integrated flux densities for the double sources, triple source and full set of singles, 
doubles and triples is show in Fig.~\ref{fig:figSglDblTpl}.

\subsection{Comparison with NVSS and SUMSS surveys}

\begin{figure*}
     \includegraphics[scale=0.67]{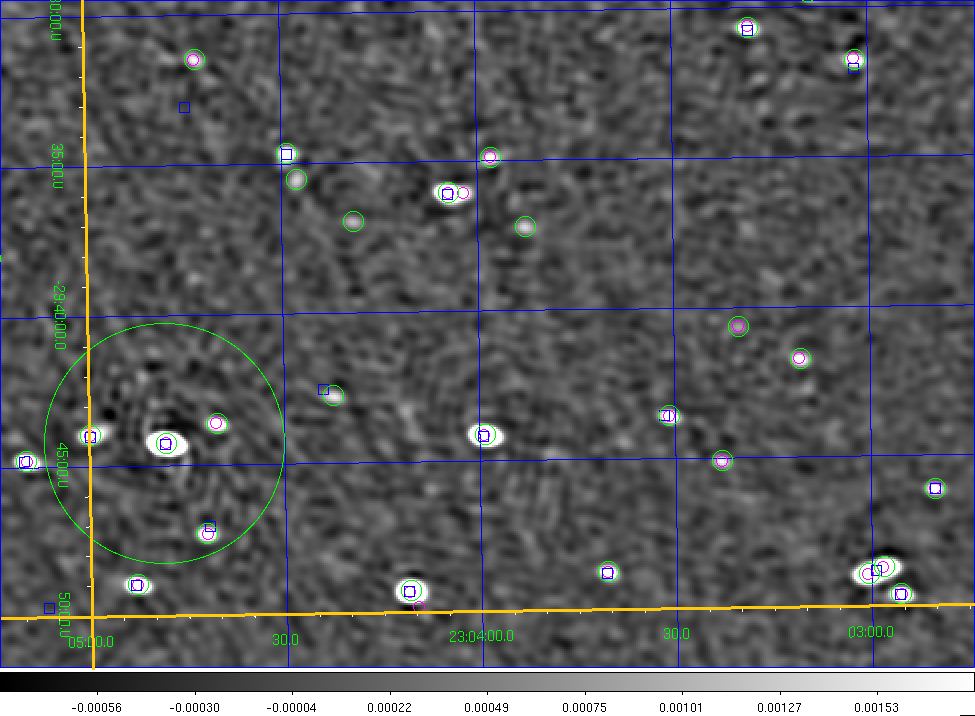}
    \caption{ Enlargement of the 936-MHz image in an area centered on R.A. 23m 04m, Dec. -29$^\circ$40$^{\prime}$.
     The small green circles show the 936-MHz detected sources, the large green circle is the area around a bright source within which false components were removed
     (see text, Section~\ref{sec:srcex}).  
     Overlaid are the NVSS 1420 MHz sources (small magenta circles) and SUMSS 843 MHz souces (blue squares).
    The intensity greyscale is linear, from -1 to +2 mJy/beam, which is expanded to show the background noise level.
The grid spacing in Dec. is 5$^{\prime}$ and the grid spacing in R.A. is 30s.
       }
    \label{fig:fig937NVSS_SUMSS}
\end{figure*}

We compared our 936-MHz and 1320-MHz images and components with the 1420 MHz NVSS and SUMSS 843 MHz catalogues
to verify the ASKAP imaging and source detection algorithms. 

Figure~\ref{fig:fig937NVSS_SUMSS} shows a region from the 936-MHz map comparing the 936-MHz detected components with components from
the NVSS and SUMSS catalogues. 
In the full area covered by the 936-MHz image, there are  5897 936-MHz components, 2718 NVSS components and 1152 SUMSS components.
There are 2209 936-MHz - NVSS cross-matches within 20$^{\prime\prime}$,  879 936-MHz - SUMSS matches and 943 NVSS - SUMSS matches. 
In the full area covered by the 1320-MHz image, there are 3756 1320-MHz components, 1724 NVSS components and 735  SUMSS components.
There are 1532 1320-MHz - NVSS cross-matches within 20$^{\prime\prime}$,  702 1320-MHz - SUMSS matches and 701 NVSS - SUMSS matches. 

All bright components ($\gtrsim 20$ mJy) are detected in all 3 data sets,  ASKAP, SUMSS and NVSS. 
The number of faint matches is consistent with expectations for detections of real components with noise, 
given the different sensitivities of the data.

\subsection{Optical Data} \label{sec:GAMAdata}

The GAMA G23 imaging and  i-band photometric data is from the VST Kilo-Degree Survey (KiDS, \citealt{2013ExA....35...25D}), 
and ZYJHKs-bands data is from the VISTA Kilo-degree Infrared Galaxy survey (VIKING, \citealt{2013Msngr.154...32E}). 
The GAMA photometry is described in \citet{2016MNRAS.455.3911D} 
and was obtained from  the GAMA Data Management Unit (DMU) 
``G23InputCatv6''. All magnitudes are in the AB system. 
The G23 spectroscopic sample was VST KiDS i-band selected with a magnitude limit of 19.2. 
The GAMA spectral catalogue is described in  \citet{2013HopkinsGAMAspectra}. 
The data reduction, spectral analysis and redshift completeness 
are described in \citet{2015LiskeGAMA}.  The spectral line information is from the GAMA DMU ``SpecLineSFRv5''.
There are 37359 galaxies with spectra and redshifts for the area overlapping the 936\,MHz field.

Because full SED-derived stellar mass estimates were not yet available for G23, we obtained stellar mass estimates for G23 galaxies by
 matching each G23 galaxy to  galaxies with similar 
 redshift  and iZYJHKs 
 photometry in the GAMA equatorial catalogues, and assuming they had a similar mass.
We use a Gaussian weighting kernel, with size of 0.005 in the redshift dimension and sizes of the photometric 1$\sigma$ errors 
in the photometric dimensions, to find the average mass-to-light ratio of the `similar' reference galaxies, and assigned that
mass to the G23 galaxy.
This idea is similar in spirit to kernal-density estimation (KDE) 
 photometric redshift estimation of \citet{2017MNRAS.466.1582W}.
We have validated this approach by testing our ability to recover the SED-fit mass estimates for half of the equatorial
catalogue, using the other half of the catalogue as a reference sample.
We find that the global random error in the mass estimates derived in this way is of
the order of 0.15 dex, compared to a median formal uncertainty of 0.12 dex for the SED-fit mass estimates described in \citet{2011Taylor}.
This yields 35353 galaxies with mass determinations. 

\subsection{IR data}

The IR data were taken from the ALLWISE catalogue of the WISE satellite \citep{2010Wright}.
 A detailed description of the WISE mission and catalogue is on-line at 
the Infrared Science Archive (IRSA) at NASA/IPAC (http://irsa.ipac.caltech.edu).
The catalogue (hereafter referred to as WISE) contains Vega magnitudes and errors for wavebands at 3.4, 4.6, 12, and 22 micron 
wavelengths (bands W1, W2, W3, W4). 
The angular resolution is 6.1$^{\prime\prime}$, 6.4$^{\prime\prime}$, 6.5$^{\prime\prime}$, and 12.0$^{\prime\prime}$ in the four bands, respectively.
The w1mpro, w2mpro, w3mpro, w4mpro magnitudes and errors were used in the current analysis.
The appendix of \citet{2014Cluver} discusses 
the effect for resolved (nearby) sources on the WISE mpro magnitudes. The w1mpro and w2mpro magnitudes are fainter than isophotal
magnitudes by $\sim0.4$. Here we are using W1-W2 colour which is much less affected for resolved sources.

\begin{table*}
	\centering
	\caption{Population measures for 936\,MHz and 1320\,MHz sources,  G23 galaxies$^{a}$,
	and cross-matches with WISE photometry.}
	\label{tab:TBLbasicstats}
	\begin{tabular}{lccccc} % four columns, alignment for each
		\hline
		Category &  No. & Redshift &  $log(M/M_{\odot})$ & W1 mag & 936\,MHz f.d.(Jy) \\
		                &   &  mean(SD) & mean(SD) & mean(SD) & mean(SD) \\		
		\hline
		936\,MHz  & 5791 & n/a & n/a & n/a & 0.0095(0.0413) \\ 
		1320\,MHz  & 3589 & n/a & n/a & n/a & n/a \\  
		936/1320\,MHz & 3289 & n/a & n/a & n/a & 0.0124(0.0506) \\ 
		G23 spectra/mass  & 35353 & 0.220(0.118)  & 10.37(0.62) & n/a & n/a \\ 
		G23 mass/lineflux  & 28179 & 0.181(0.073)  & 10.24(0.60) & n/a & n/a \\ 
		G23 mass/4lineflux$>0^{b}$  & 15165 & 0.171(0.075)  & 9.99(0.60) & n/a & n/a \\	
		G23 mass/lineflux$<0^{c}$  & 13014 & 0.194(0.069)  & 10.52(0.46) & n/a & n/a \\	
		G23 mass/4lineflux$<0^{d}$  & 1785 & 0.186(0.072)  & 10.61(0.44) & n/a & n/a \\		
		936/G23 spectra/mass & 978 & 0.221(0.169)  & 10.70(0.62) & n/a & 0.0064(0.0322) \\ 
		936/G23 mass/lineflux & 684 & 0.148(0.083)  & 10.51(0.61) & n/a & 0.0043(0.0192) \\ 
		936/G23 mass/4lineflux$>0$ & 410 & 0.127(0.075)  & 10.25(0.59) & n/a & 0.0037(0.0233) \\
 		936/G23 mass/lineflux$<0$ & 274 & 0.181(0.084)  & 10.90(0.41) & n/a & 0.0051(0.0102) \\
		936/G23 mass/4lineflux$<0$ & 45 & 0.184(0.068)  & 10.98(0.39) & n/a & 0.0096(0.0160) \\
		G23/WISE spectra/mass & 31470 & 0.223(0.118) & 10.41(0.59) & 15.39(0.81) & n/a \\
		G23/WISE mass/lineflux & 25234 & 0.184(0.072) & 10.28(0.57) & 15.42(0.84) & n/a \\
		G23/WISE mass/4lineflux$>0$ & 13149 & 0.174(0.074) & 10.05(0.57) & 15.69(0.80) & n/a \\
		G23/WISE mass/lineflux$<0$ & 11985 & 0.194(0.068) & 10.54(0.45) & 15.13(0.79) & n/a \\
		G23/WISE mass/4lineflux$<0$ & 837 & 0.186(0.075) & 10.64(0.43) & 14.85(0.93) & n/a \\
		936/G23/WISE spectra/mass & 889 & 0.228(0.170) & 10.74(0.59) & 14.24(0.88) & 0.0067(0.0336) \\	 
		936/G23/WISE mass/lineflux & 610 & 0.152(0.082) & 10.56(0.59) & 14.01(0.92) & 0.0043(0.0199) \\
		936/G23/WISE mass/4lineflux$>0$ & 352 & 0.133(0.075) & 10.30(0.57) & 14.04(0.88) & 0.0036(0.0246) \\
		936/G23/WISE mass/lineflux$<0$ & 258 & 0.179(0.084) & 10.92(0.40) & 13.97(0.98) & 0.0053(0.0104) \\	
		936/G23/WISE mass/4lineflux$<0$ & 23 & 0.177(0.078) & 11.03(0.37) & 13.89(0.94) & 0.0130(0.0181) \\	
		\hline
	\end{tabular}
	\begin{tablenotes}\footnotesize
	\item[*] a. The GAMA23 stellar masses were based on stellar masses from the GAMA equatorial fields as described in Section~\ref{sec:GAMAdata}.
\item[*] b. 4lineflux$>0$ means H$\alpha$,  H$\beta$, [OIII]$\lambda$5007 and [SII]$\lambda\lambda$6717,6731 all have lineflux$>0$.
\item[*] c. lineflux$<0$ means any one of H$\alpha$,  H$\beta$, [OIII]$\lambda$5007 and [SII]$\lambda\lambda$6717,6731 has lineflux$<0$.
\item[*] d.  4lineflux$<0$ means H$\alpha$,  H$\beta$, [OIII]$\lambda$5007 and [SII]$\lambda\lambda$6717,6731 all have lineflux$<0$.
\end{tablenotes}
\end{table*}

\section{Multi-wavelength analysis of the ASKAP radio sources and G23 galaxies}\label{sec:multiwave}

\begin{figure}
     \includegraphics[width=\columnwidth]{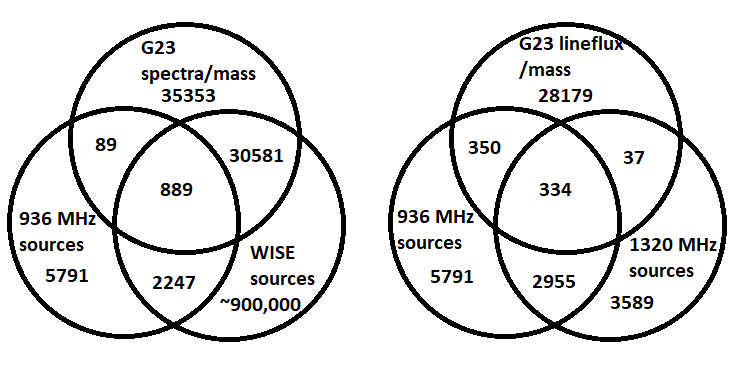}
    \caption{ Number of matches between 936\,MHz sources, 1320\,MHz sources, G23 galaxies and WISE sources. 
    The central set of 889 in the left diagram was obtained by first matching the 
35353 G23 galaxies with spectra and masses to the WISE
    sources, then matching the resulting 30581+889 galaxies to the  5791 936-MHz sources. The central set of 334 in the right diagram was obtained by first matching 5791 936-MHz
    and 3589 1320-MHz sources, then matching those 2955+334 to 28179 G23 galaxies with linefluxes and masses.
       }
    \label{fig:Venn}
\end{figure}

\subsection{Source cross-matching}

We cross-matched the 936\,MHz and 1320\,MHz catalogues of radio sources to the G23 galaxies  
by matching positions within a specified radius using TopCat.
For each cross-match, we first plotted the separation distribution for the real catalogues and for the case where one of
the catalogues had the positions randomized. Then we compared the separation distributions to determine an optimum
radius for separating real from random matches.
The 936 MHz and 1320 MHz catalogues were matched using a radius of $7^{\prime\prime}$. 
The 1320 MHz  catalogue was matched with G23 using radius $5^{\prime\prime}$.
The 936 MHz  catalogue was matched with G23 using radius $7^{\prime\prime}$.   
Because the 1320\,MHz sources were essentially a subset of the 936\,MHz sources,
we mainly report results for the larger set of 936\,MHz cross-matches.

\begin{figure}
	\includegraphics[width=\columnwidth]{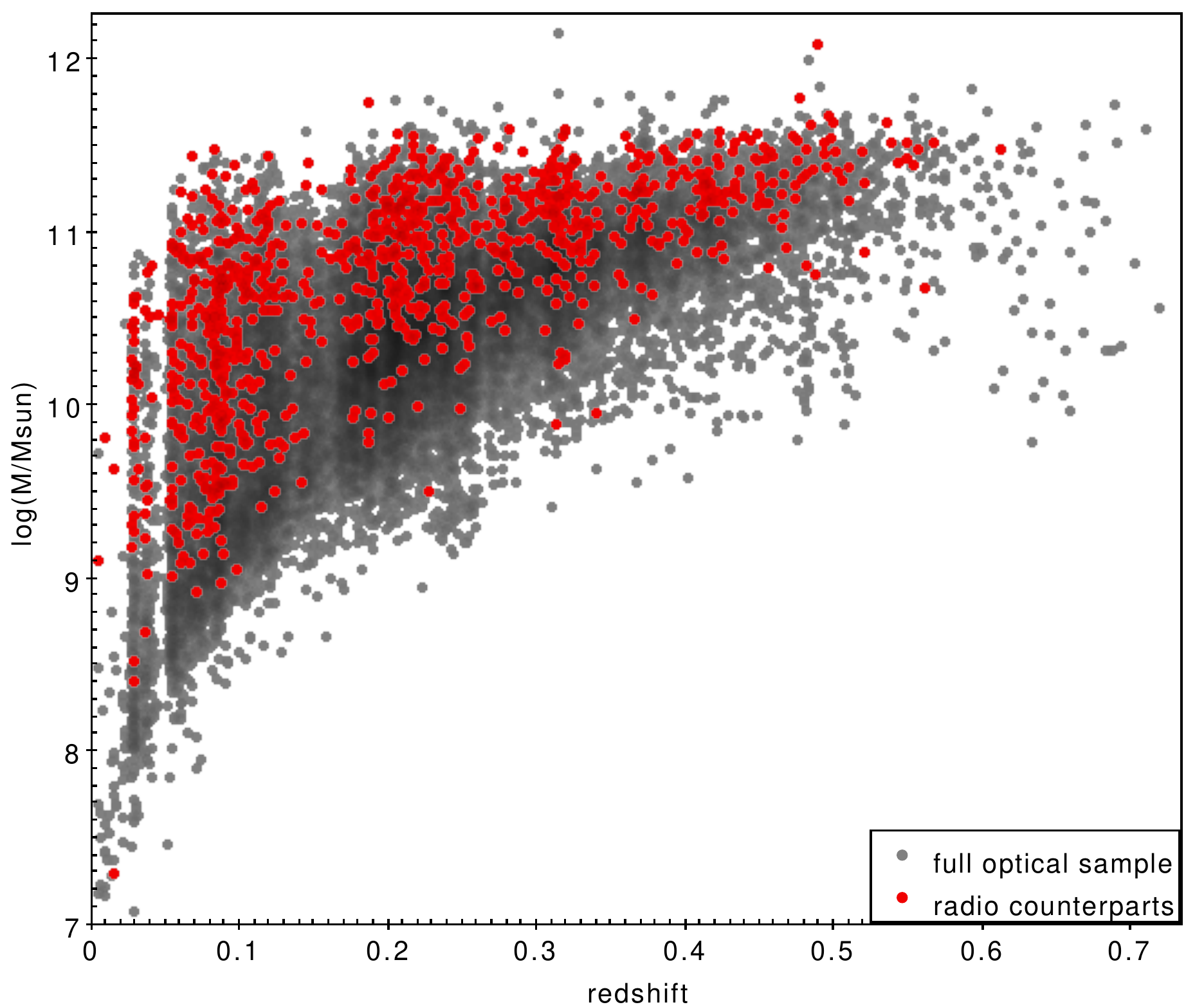} 
	\includegraphics[width=\columnwidth]{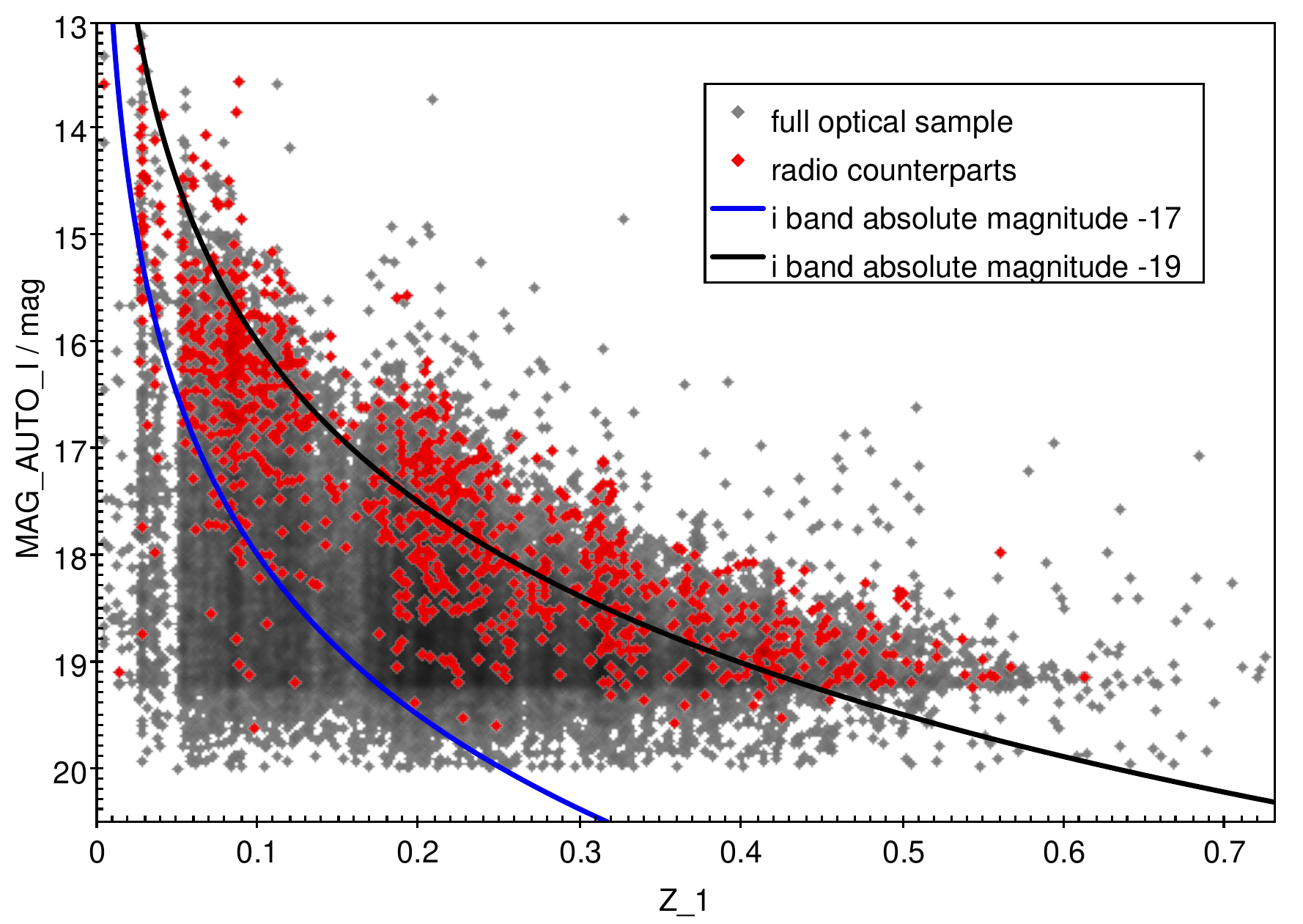} 
	\includegraphics[width=\columnwidth]{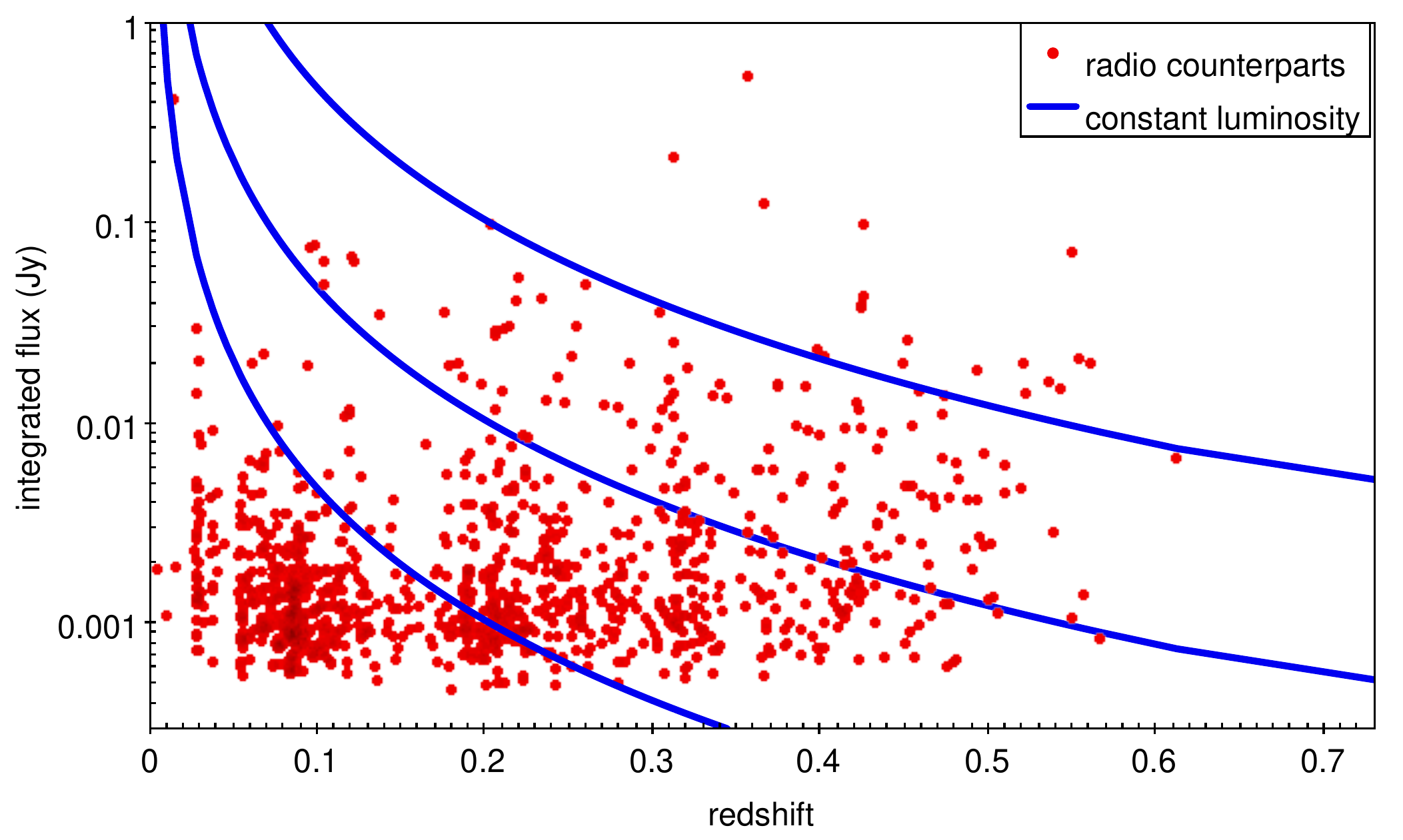} 
   \caption{
    Top panel: Stellar mass in units of log($M/M_{\odot}$) vs redshift for the 936\,MHz/G23 sources (red), 
    and for all G23 galaxies (grey). 
    This shows that galaxies with detected radio emission are mainly galaxies with larger masses,
    and shows the effect of redshift-dependent completeness on the lower limit of detected G23 galaxies and radio counterparts.
    Middle panel: i-band magnitude vs redshift for the galaxy sample and for the galaxies with detected 936\,MHz radio sources. 
    The lines are for a galaxy of absolute i band magnitudes $-17$ and $-19$.   
    Bottom panel: Integrated 936\,MHz flux density vs redshift for the G23 galaxies with radio counterparts (red points). 
    The blue lines are for sources of constant radio luminosity (differing by one order of magnitude).  Radio counterparts which would
    have flux densities below the limits of the current ASKAP data fall below a horizontal line at $\sim5\times10^{-4}\,$Jy. Radio counterparts which
    would be beyond the current redshift limit of the G23 galaxies fall to the right of a line at $z\sim0.6$.
}
    \label{fig:figmassVSz}
\end{figure}

We cross-matched the 936 MHz and 1320 MHz  catalogues and the matched 936 MHz/1320 MHz catalogue 
with WISE using a radius of  $4^{\prime\prime}$. This is smaller than
for the cross-match of radio catalogues with G23 because of the much higher source density on the sky for WISE than G23.

G23 was cross-matched with WISE using radius of $1.5^{\prime\prime}$. 
This gives low contamination (0.84\%), and yields a large fraction of cross-matches (87.6\%). 
For the GAMA G12 and G15 fields, \citet{2014Cluver} found similar results: 77\% (for G12) and 88\% (for G15) of the galaxies 
had WISE W1 and W2 detections at $>5\sigma$.

We compared the cross-match of the 936 MHz/WISE catalogue to G23 with the  cross-match of the 936 MHz catalogue with the  G23/WISE catalogue. 
The latter yields better results:  
the number cross-matches is somewhat reduced, but there is much lower contamination with random cross-matches.
The reason for this is the superior spatial resolution of both G23 positions (sub-arcsec) 
and WISE positions  ( $\simeq0.2-0.3^{\prime\prime}$) compared to the uncertainties of the radio positions ( $\sim5^{\prime\prime}$). 
\footnote{
The WISE catalogue contains sources detected at $\ge$5$\sigma$, thus does not contain all galaxies with emission in the WISE bands. 
Forced photometry at the positions of the G23 galaxies would likely have found many $<5\sigma$ WISE sources associated with
the galaxies, although that is beyond the scope of the current work.
}

From these different catalogues, we then extracted the following subsets:
\begin{enumerate}
\item all galaxies with spectra and masses;
\item all galaxies with spectra, masses and BPT line flux measurements;
\item all galaxies with  spectra, masses and all BPT lines in emission;
\item all galaxies with  spectra, masses and at least one BPT line in absorption;
\item all  galaxies with spectra, masses and all BPT lines in absorption.
\end{enumerate} 
Summaries of the numbers of sources in the different subsets are given in Table~\ref{tab:TBLbasicstats} 
and, in part, in Figure~\ref{fig:Venn}.
All galaxies with spectra have redshifts.

\subsection{Properties of cross-matched sources}

A mass-redshift diagram for G23 galaxies and their 936\,MHz radio counterparts is given in the top panel of Figure~\ref{fig:figmassVSz}.
At all redshifts, the radio sources occur in the high-mass end of the galaxy mass distribution. 
The radio source mean masses 
 (`936/G23 spectra/mass' and `936/G23/WISE spectra/mass')
are significantly higher than for the full galaxy population (`G23 spectra/mass'), by  factors of 2.14 and 2.34, respectively.
The 936\,MHz and 1320\,MHz mass distributions are not significantly different. 

The middle panel of Figure~\ref{fig:figmassVSz} shows i-band magnitude vs redshift for the G23 galaxies and for radio counterparts compared to lines of
constant i-band absolute magnitude $-17$ and $-19$. 
The radio sources are associated with the optically most luminous galaxies, consistent with them having higher than average stellar masses. 
The fact that the radio counterparts extend down to the detection limit of the G23 region (a horizontal line at i band magnitude of 19.2) implies that
the radio observations are fairly complete for redshifts $z<0.6$.

The bottom panel of Figure~\ref{fig:figmassVSz} shows integrated radio flux density vs redshift for the 936\,MHz radio counterparts. 
Lines of constant radio luminosity are shown. Cosmological volume increases rapidly with redshift 
so the number of luminous radio sources (e.g. within a band between a pair of blue lines) should increase rapidly.
This illustrates increasing incompleteness with redshift of radio counterparts and of galaxies.

There are two reasons why we do not observe more radio counterparts:
\begin{itemize} 
\item
Radio counterparts are not detected below the lower edge of the observed distribution because of the limited sensitivity of the radio observations.
\item
Radio counterparts are not observed at high redshift ($z>0.6$) because of the lack of such galaxies in G23.
The gradual decrease in radio counterparts between redshift 0.1 and 0.6 is consistent with the decrease in the number (i.e. completeness) of optically measured galaxies with redshift, as seen in the top and middle panels.
\end{itemize}
Both factors contribute to lack of observed radio counterparts. 
The number of observed radio sources (5791 at 936\,MHz) is much larger than the number with optical counterparts (978). 
Those without optical galaxy counterparts may lie either at $z<0.6$ with their optical counterpart fainter than the G23 magnitude limit, or at $z>0.6$. 
The relatively high radio flux density limit of these observations suggests that most are likely to be high radio luminosity AGNs at high redshifts \citep[e.g.,][]{2008MNRAS.386.1695S} 
or, possibly, dust obscured galaxies at lower redshifts ($z\lesssim 0.6$).

Radio sources, such as broad-line AGN and star forming galaxies, typically have bluer near-ultraviolet (NUV)$-r$
colours than the overall galaxy population \citep{2014ARA&A..52..589H}. 
We used  NUV magnitudes from the G23 input catalogue derived from GALEX observations \citep{2015LiskeGAMA}. 
The mean NUV magnitude is 21.11 for galaxies detected at 936\,MHz compared to 21.94 for all galaxies.
For NUV magnitudes brighter than 19, the probability of a galaxy to have radio emission is higher by a factor of $\sim5$.
For SFG, brighter UV magnitudes imply higher star formation rates \citep{2017ApJ...847..136B}. The UV could be
from accretion disks of AGN, however recent measurements of the AGN UV luminosity function \citep{2017Ricci}\footnote{
\citet{2017Ricci} also noted that the SF in AGN contributed a significant fraction of the total UV emitted by AGN.} 
imply that  the UV input to the universe from SFG is significantly larger than from AGN. 
Thus the brighter UV from radio-detected galaxies in most cases is dominated by star formation.

The brightest emission line galaxies in either H$\alpha$ or H$\beta$ are mostly detected as 936\,MHz sources:
8 out of the 10 brightest H$\alpha$ emitters and 7 out of the 10 brightest H$\beta$ emitters are 936\,MHz sources 
whereas only 1 in 25 galaxies with H$\alpha$ and H$\beta$ emission lines is a 936\,MHz source.
Thus, radio sources are highly over-represented in the population of brightest H$\alpha$ and H$\beta$ galaxies. 

\begin{figure}
	\includegraphics[width=\columnwidth]{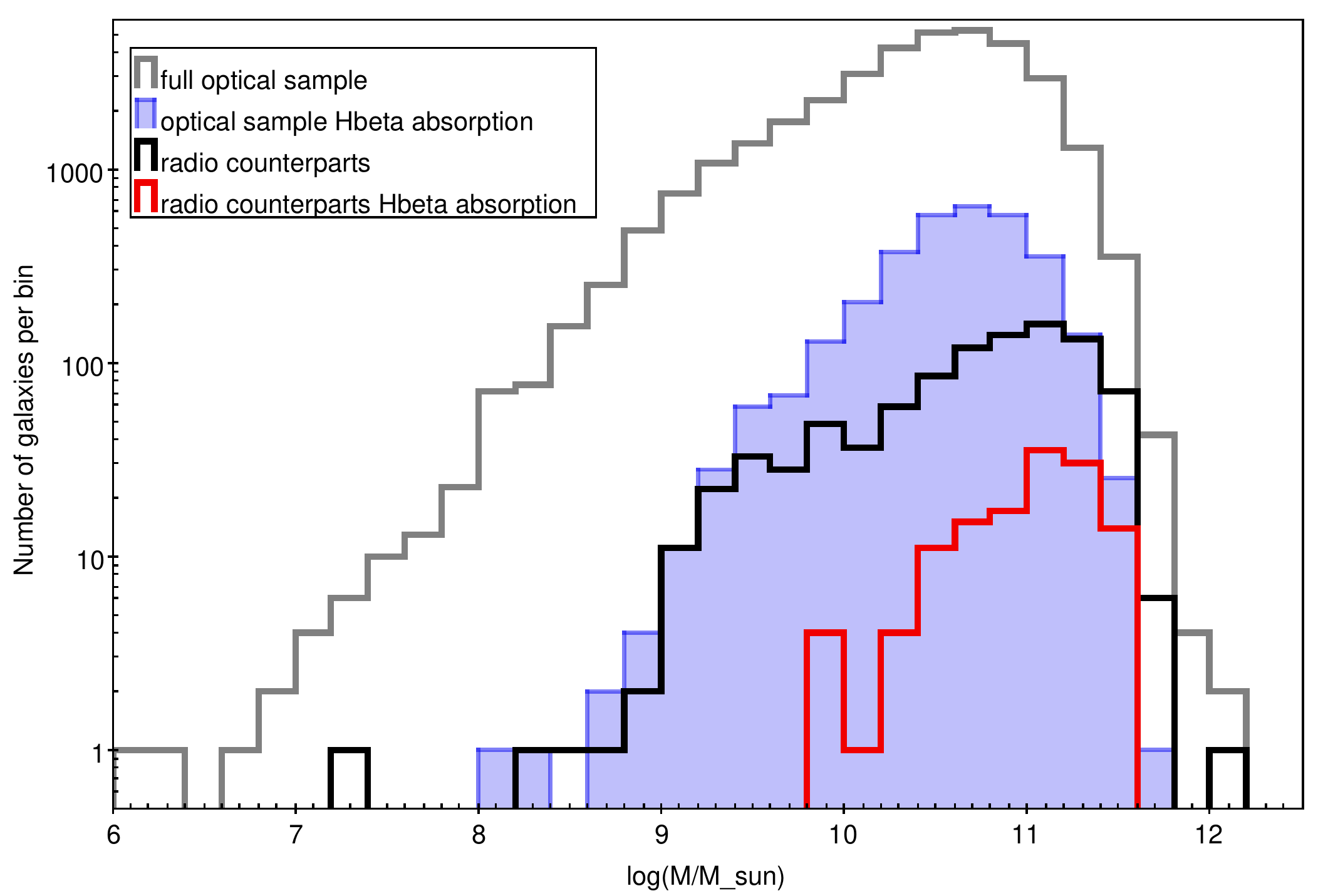} 
		\includegraphics[width=\columnwidth]{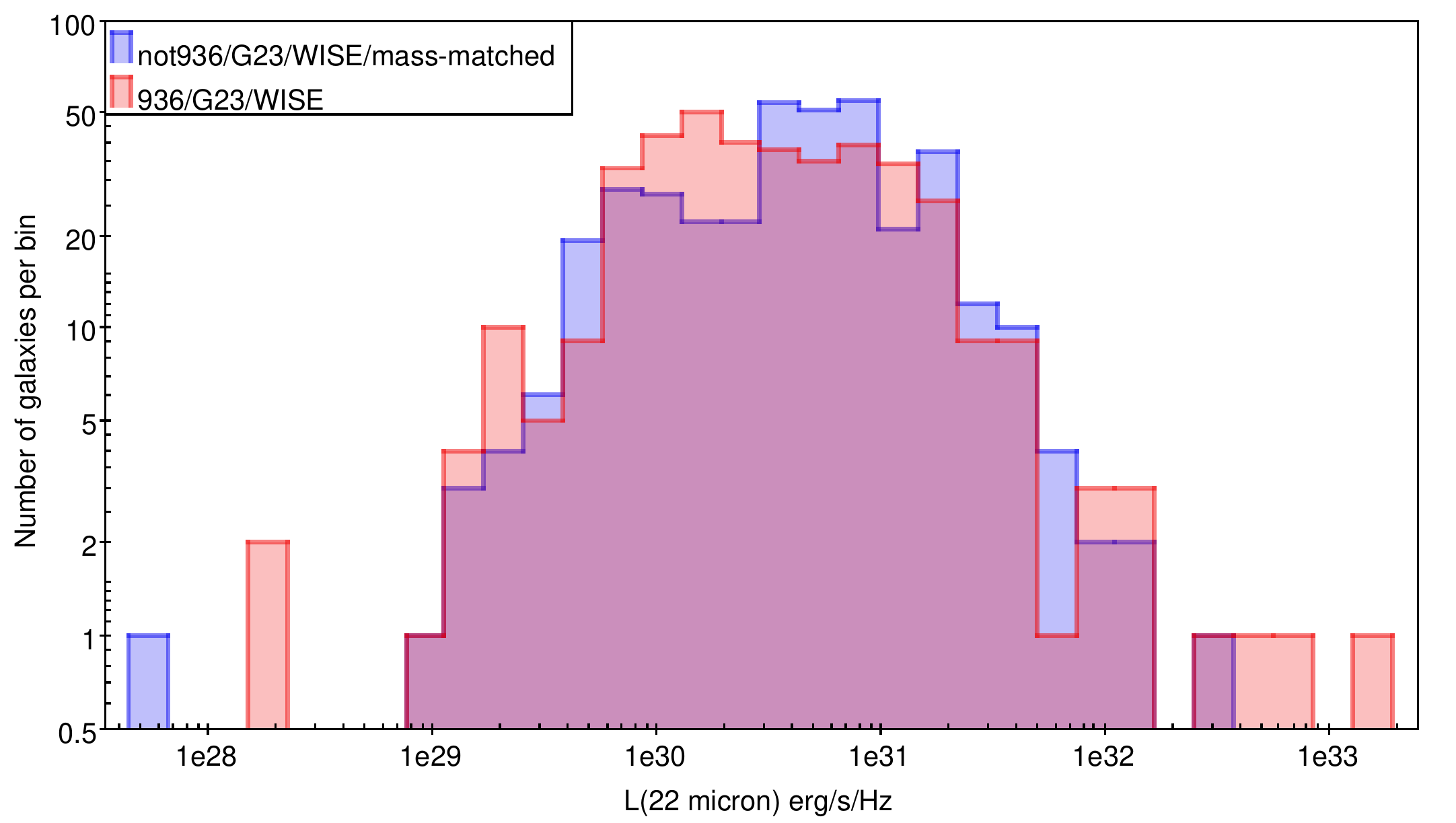}
    \caption{Top panel:
    The distribution of stellar masses  (number of galaxies per mass bin, units of log($M/M_{\odot}$)) for radio counterparts 
     (936/G23 set) with 
    $H\beta$ in absorption (red), and for all radio counterparts (black).
    Overlaid on these are the distributions for G23 galaxies with $H\beta$ in absorption (blue), and 
    for all G23 galaxies (grey). 
    This illustrates (i)  the significantly larger masses for galaxies with $H\beta$ in absorption compared to
    all galaxies; and (ii) the shift to even larger masses for both sets of galaxies when they have 936\,MHz radio emission. 
    Bottom panel: comparison of the 22$\mu m$ luminosity distribution for radio-emitting galaxies (red histogram) with that 
    for a mass-matched set of non-radio galaxies (blue histogram).
    }
    \label{fig:figmassHBabsdist}
\end{figure}

Compared to the full optical sample, the fraction of strong H$\beta$ absorption line systems is larger by a factor of $\simeq2$ for the radio counterparts.
The mass distributions are shown in the top panel of Figure~\ref{fig:figmassHBabsdist}.
The mean log(M/$M_{\odot}$) is 10.37 for the whole galaxy sample, 10.68 for the optical galaxy H$\beta$-absorption-line sample,
and 11.00 for the 936\,MHz detected H$\beta$-absorption-line sample.
The stronger H$\beta$ absorption line strengths for 936\,MHz galaxies are related to the higher masses for 936\,MHz H$\beta$ absorption line systems. 
This is similar to the result from \citet{2005MNRAS.362...25B} that the fraction of galaxies that host a radio AGN rises strongly with host stellar mass.
While the radio flux density limit of the current data is significantly lower than in that work, and the current set of radio sources includes a larger fraction of SFG, 
it is clear that the highest mass systems include those with old stellar populations, and their radio emission is likely driven by an AGN.
 
\begin{figure}
		\includegraphics[width=\columnwidth]{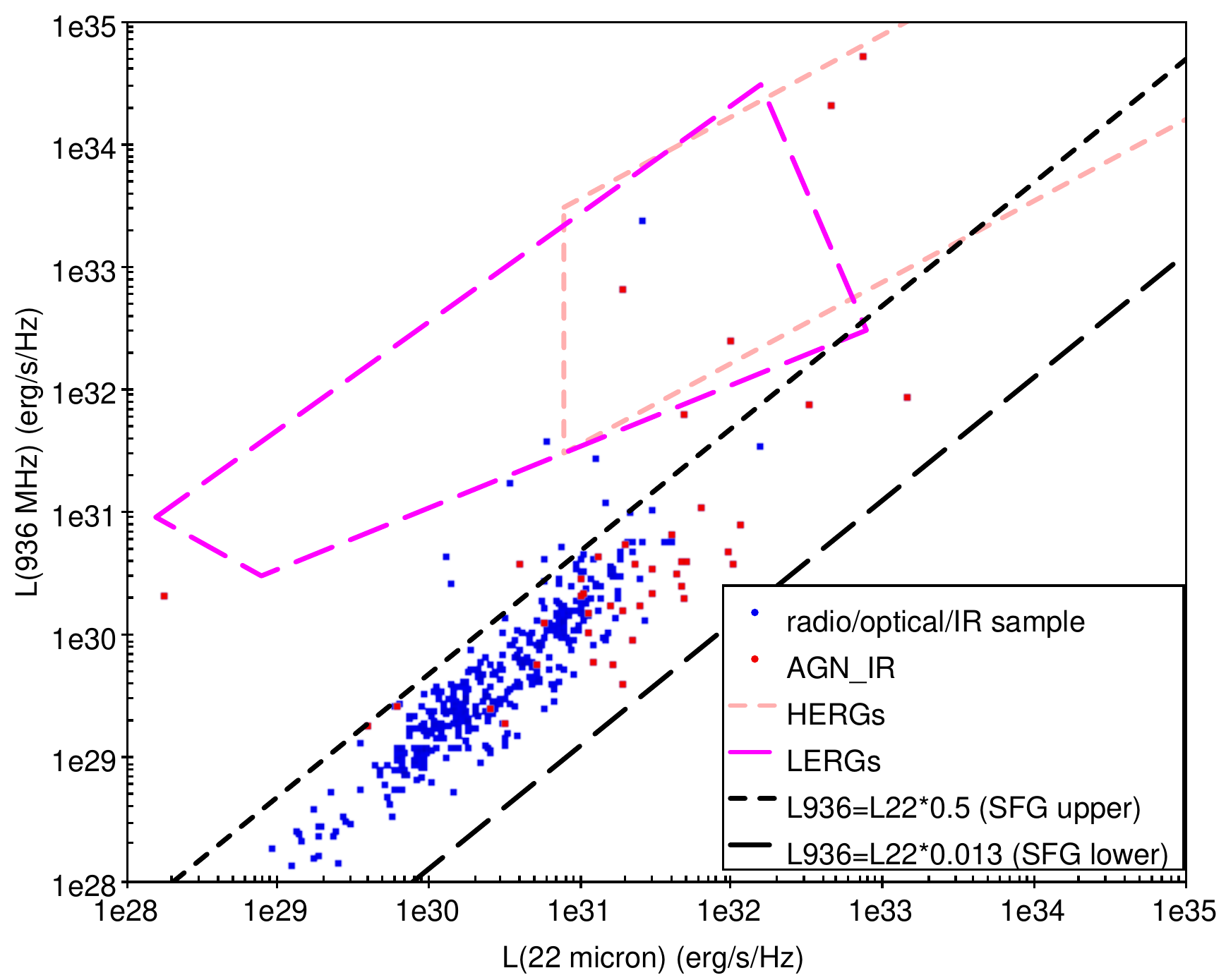} 
		\includegraphics[width=\columnwidth]{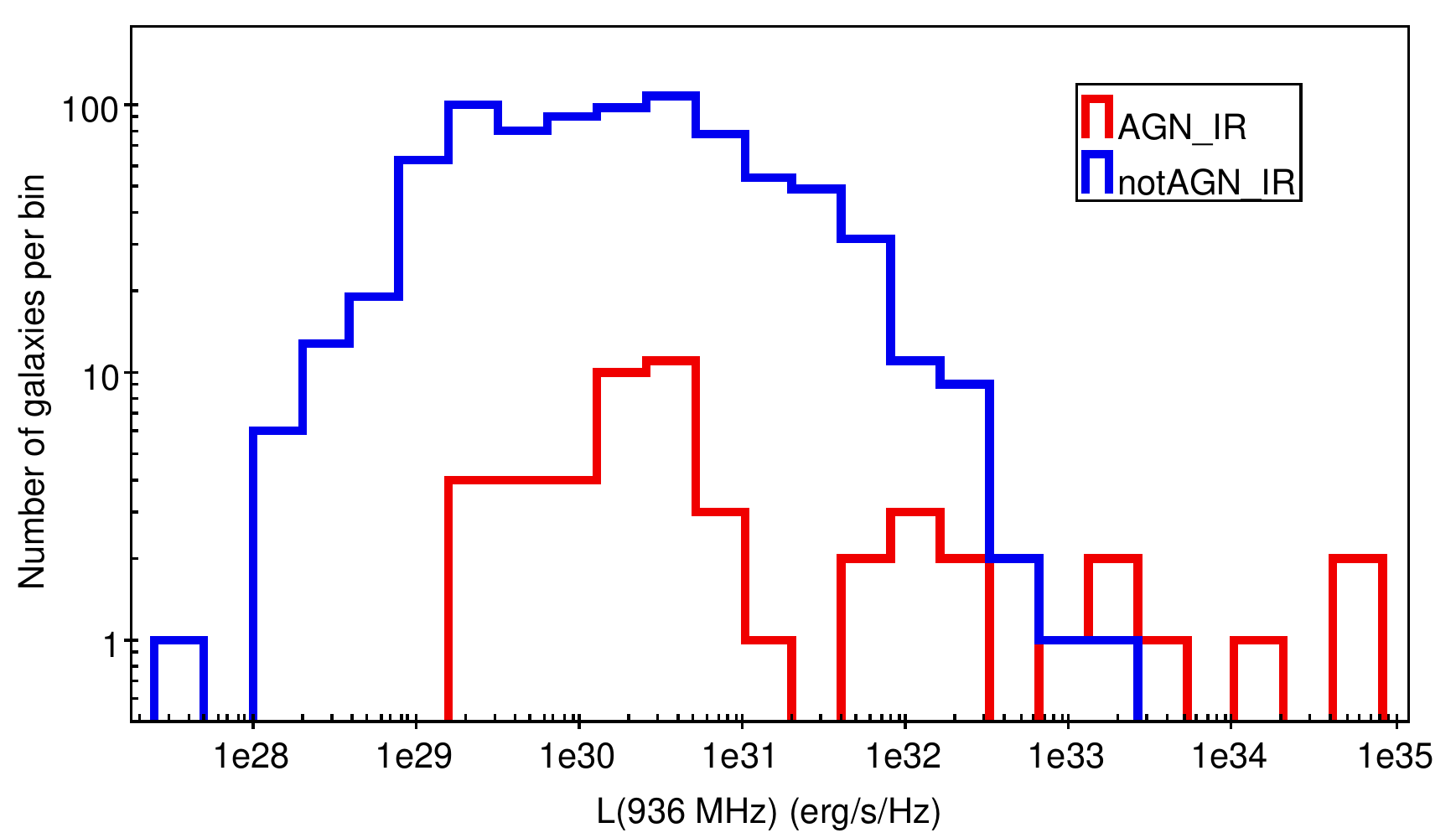} 
    \caption{
Top panel:   936\,MHz luminosity vs W4 band ($22~\mu m$) luminosity, both in units of erg s$^{-1}$ Hz$^{-1}$, 
    for the 936/G23/WISE sources with WISE W4  
    signal-to-noise (SN) $\ge$3 (blue points) 
    and for the AGN$_{IR}$ subset (red points). 
     Most sources ($96\%$, 436 of 452) fall in the band for star-forming galaxies, delimited by the two lines shown 
  (derived from \protect\citealt{2009Rieke}).
    The regions outlined by the pink and magenta dashed lines indicate the regions occupied by HERGs and LERGs
\protect\citep{2014MNRAS.438.1149G}. The sources detected here mainly have lower $L_{936~MHz}$ than HERGs and LERGs.
 Bottom panel: 936 MHz luminosity distribution for AGN$_{IR}$ and notAGN$_{IR}$  from the 936/G23/WISE sample.}
\label{fig:figL22vL936}
\end{figure}

 Radio sources have much stronger mid-IR emission than non-radio galaxies in general (see Table~\ref{tab:TBLbasicstats}).  
The  936/G23/WISE  sources have a mass about a factor of $\sim4$ higher than the G23/WISE galaxy population.
We created a sample of not936/G23/WISE galaxies by excluding the radio-emitters.
Then we selected a mass-matched set of non-radio galaxies by choosing the best match by mass from the not936/G23/WISE
set for each radio galaxy in the 936/G23/WISE set. 
We verified that the mass distribution of the not936/G23/WISE/mass-matched set is essentially identical to that of the 936/G23/WISE set.
The resulting $22~\mu m$ luminosity distributions of the radio and non-radio emitters is shown in the bottom panel of
 Figure~\ref{fig:figmassHBabsdist}.
There is no difference, except for the 3 radio sources with highest $22~\mu m$ luminosity, so we can conclude that the higher 
mid-IR emission of radio sources is consistent with being associated with increased mass of the galaxy.

Both $22~\mu m$ and $\sim$1.4\,GHz luminosities are star formation rate indicators for SFG (e.g. \citealt{2009ApJ...703.1672K},
\citealt{2009Rieke}, \citealt{2017ApJ...847..136B}, \citealt{2017ApJ...850...68C}). 
For AGN, they are indicators of AGN activity \citep{2014MNRAS.438.1149G}.
We show the tight correlation between 22$~\mu m$ and 936\,MHz luminosities ($L_\nu$, units erg/s/Hz) for the 936/G23/WISE galaxies in the top panel of 
Figure~\ref{fig:figL22vL936}.\footnote{The conversion from W4 magnitude to flux density in Jy is given in the WISE explanatory supplement 
on the NASA/IPAC/IRSA website. The effective wavelength of the W4 filter has been recalibrated to $23~\mu m$ \citep{2014PASA...31...49B}.
No K-corrections were applied to obtain $22~\mu m$ luminosities in this study.}
We also show the range of ratio 936\,MHz to $22~\mu m$ luminosity found by \cite{2009Rieke} for SFG by
the two black dashed lines.
 68\% (95\%) of the data falls within $\pm$0.3 dex ($\pm$0.6 dex) of a 936\,MHz to $22~\mu m$ luminosity ratio of 0.13.
 Consistent with our results, \cite{2017ApJ...847..136B} finds that 150MHz, 1.4GHz and $22~\mu m$ luminosities as a function of H$\alpha$ luminosity have scatter of ~0.2 dex.
Thus it is seen that the majority of our radio/optical/IR galaxies are consistent with their radio and $22~\mu m$ emission 
powered by star formation.
    
\subsection{AGN and SFG diagnostics}\label{sec:AGNSFG}

\subsubsection{IR AGN diagnostic}

The WISE mid-IR colour criterion W1-W2>0.8 was introduced by \citet{2011Jarrett} to distinguish AGN from
other galaxies.
\citet{2012ApJ...753...30S} found that AGN can be identified, with 95$\%$ reliability and 78$\%$ completeness, using this criterion.
Sources with W1-W2$>$0.8 ($>1\sigma$) are labelled AGN$_{IR}$ and
those with W1-W2$<$0.8 ($>1\sigma$) are  labelled notAGN$_{IR}$.
The W1-W2 test gives an AGN$_{IR}$ classification for 374 sources and a notAGN$_{IR}$ classification for 29983 sources. 

 The AGN$_{IR}$ and notAGN$_{IR}$ galaxies have strikingly different properties.
 Figure~\ref{fig:figL22vL936} (top panel) shows the AGN$_{IR}$ constitute $\sim$1/2 of the outliers from the radio vs. $22~\mu m$ relation, 
  whereas they are only a small fraction (0.08) of the radio and $22~\mu m$ emitting galaxies. 
 The bottom panel of  Figure~\ref{fig:figL22vL936} shows the radio luminosity distributions of 
 AGN$_{IR}$ and notAGN$_{IR}$.
The AGN$_{IR}$ are more luminous on average by a factor of 156 at 936\,MHz.

\begin{figure}
	\includegraphics[width=\columnwidth]{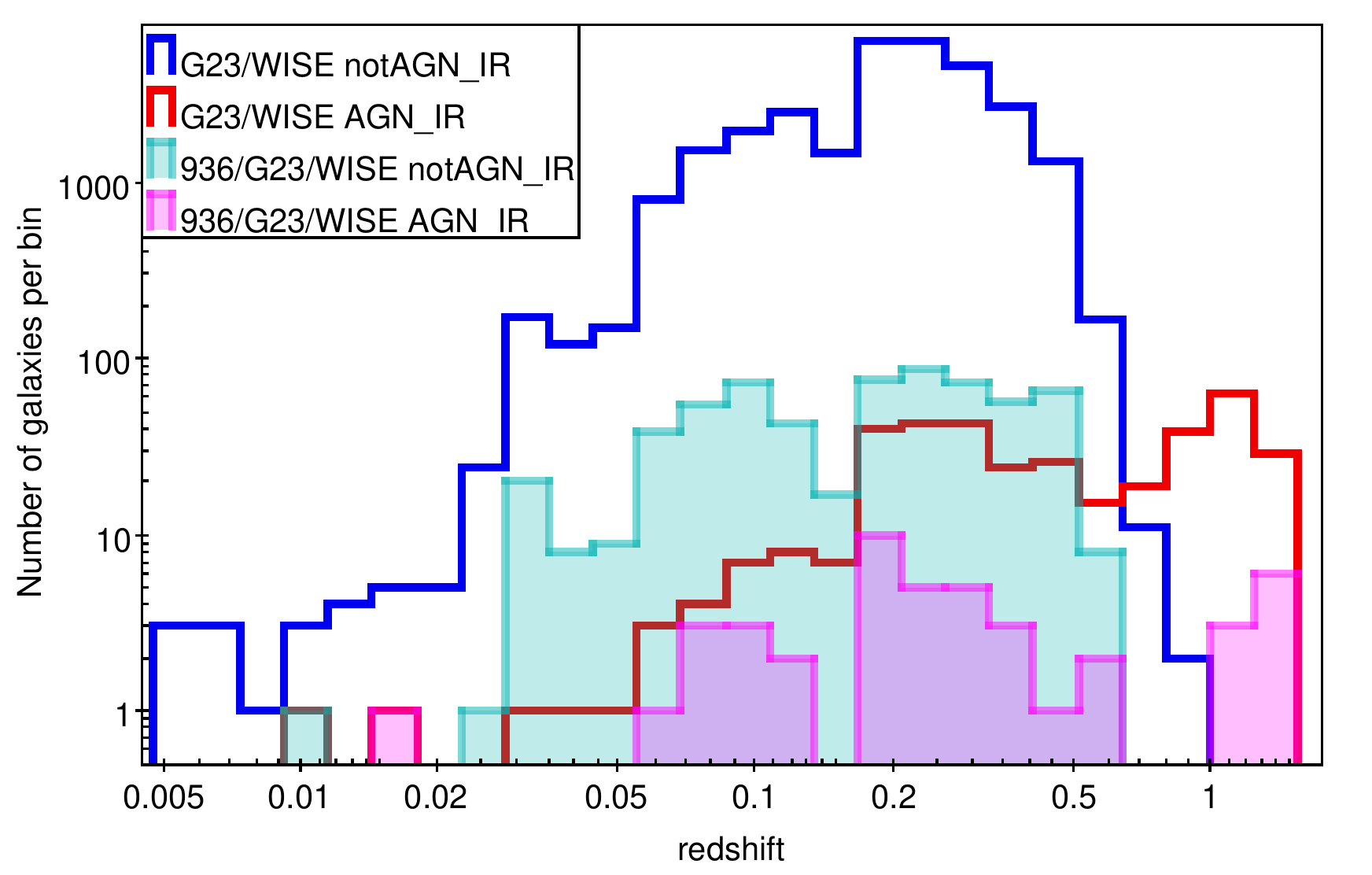} 
	\includegraphics[width=\columnwidth]{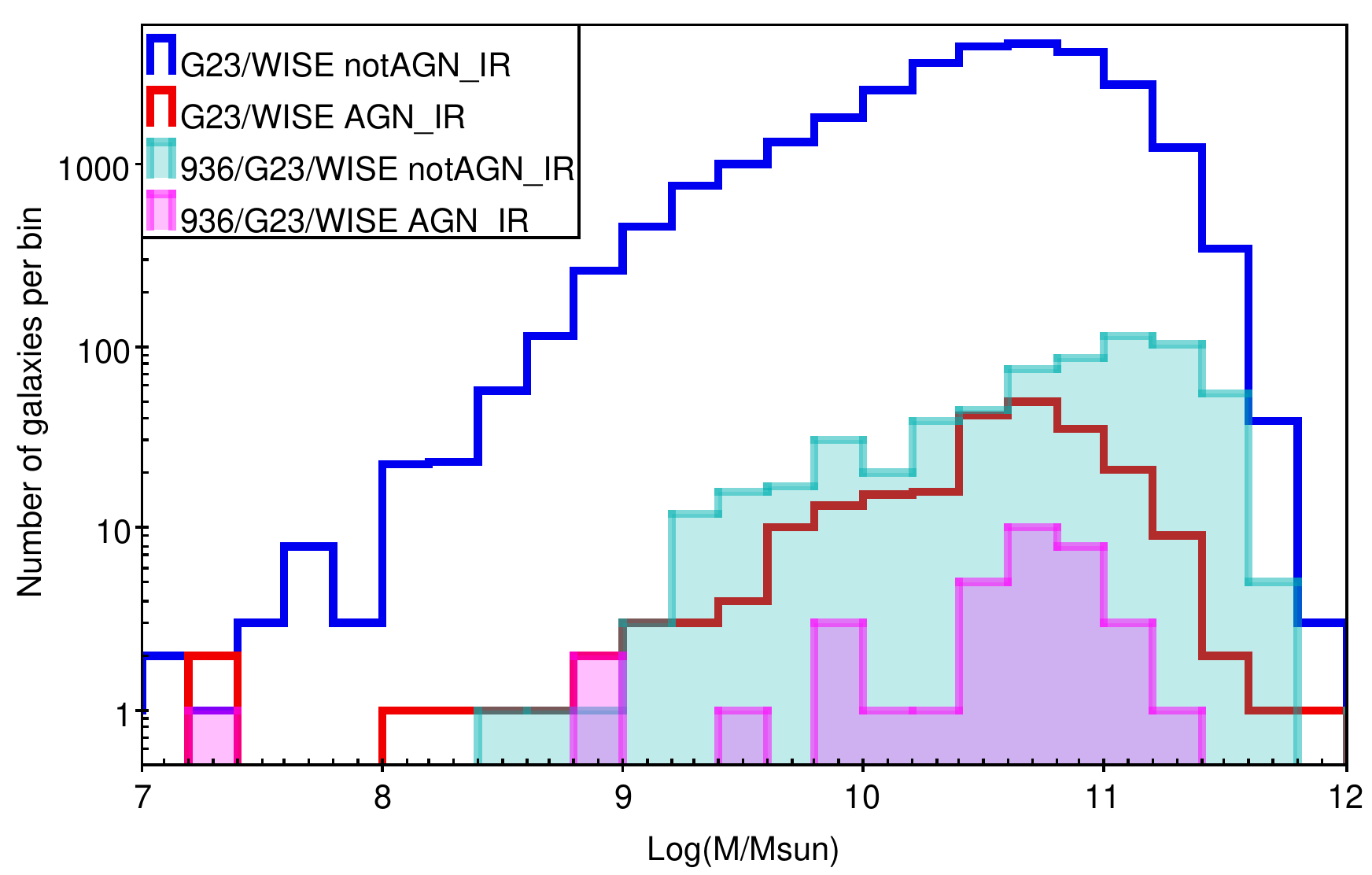} 
	\includegraphics[width=\columnwidth]{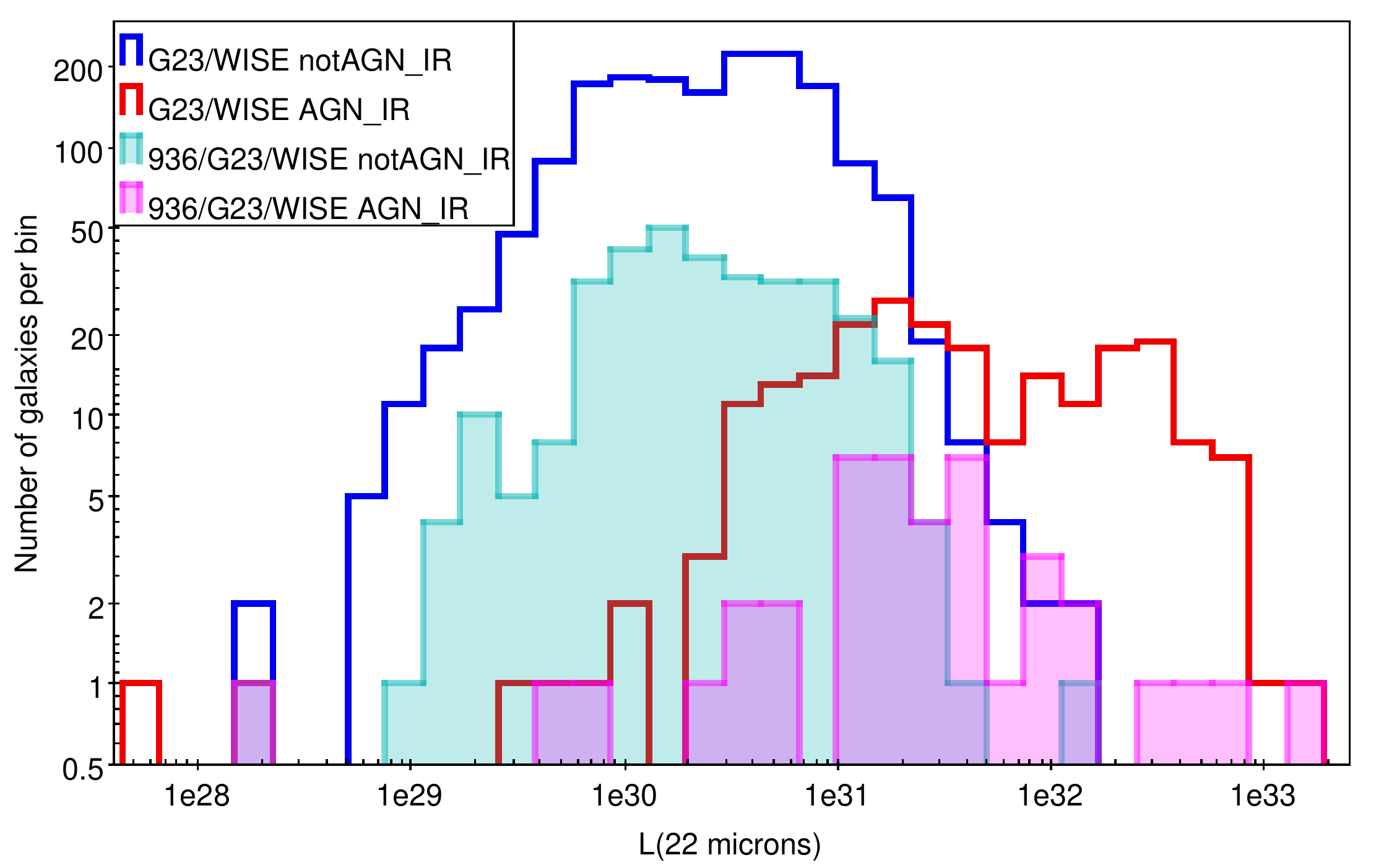}
    \caption{
    Comparison of the optical/IR galaxy sample with 936\,MHz emitting subsets. Shown are distributions of: 
    redshift (top panel); mass (second panel); and $22~\mu m$ luminosity, L($22~\mu m$), in erg/s/Hz (third panel, for subset with w4snr$>$3).
        Sets shown are G23/WISE notAGN$_{IR}$ (blue), G23/WISE AGN$_{IR}$ (red), 
    936/G23/WISE notAGN$_{IR}$ (cyan) and 936/G23/WISE AGN$_{IR}$ (magenta). 
    }
    \label{fig:figzw4}
\end{figure}

 Figure~\ref{fig:figzw4} shows the redshift, mass and $22~\mu m$ luminosity distributions for G23/WISE and 936/G23/WISE samples.
The mean redshift of AGN$_{IR}$ is much higher than for notAGN$_{IR}$ (see Table~\ref{tab:TBLw1w2BPTstats}).
The mean mass of G23/WISE AGN$_{IR}$ is nearly the same as for notAGN$_{IR}$, and the mass distributions are nearly the same
 (Figure~\ref{fig:figzw4} middle panel).
 The mean $22~\mu m$ luminosity of AGN$_{IR}$ is higher by a factor $\sim25$ than for notAGN$_{IR}$ (for the G23/WISE w4snr$>$3 sample).

For the 936\,MHz detected G23/WISE galaxies (whole and w4snr$>$3 subsample, see Table~\ref{tab:TBLw1w2BPTstats}), 
the AGN$_{IR}$ redshifts and 936\,MHz radio flux densities are higher (by factor $\sim3$ for redshift and $\sim8$ for flux density) than for notAGN$_{IR}$.
There is an even more extreme difference considering luminosities: 
the $22~\mu m$ luminosities are higher by a factor of 20 than for notAGN$_{IR}$  (for the w4snr$>$3 subsample).
The AGN$_{IR}$ mean masses are similar for the  whole 936/G23/WISE sample and the subsample with w4snr$>$3.
However the notAGN$_{IR}$ mean mass is higher (by a factor of 2.0)  for the whole 936/G23/WISE sample compared to the subsample with w4snr$>$3.

The fact that such strong differences in redshift and luminosity are seen between the AGN$_{IR}$ and notAGN$_{IR}$ categories
indicates that the contamination of AGN$_{IR}$ by notAGN galaxies and of  notAGN$_{IR}$ by AGN is not a large
fraction of either category.
Contamination of notAGN$_{IR}$ by low luminosity AGN would not strongly
affect the redshift and radio luminosity difference we see between the AGN$_{IR}$ and notAGN$_{IR}$ sets.

\subsubsection{Optical AGN/SFG diagnostic}
\label{AGNSFG}

\begin{figure}
	\includegraphics[width=\columnwidth]{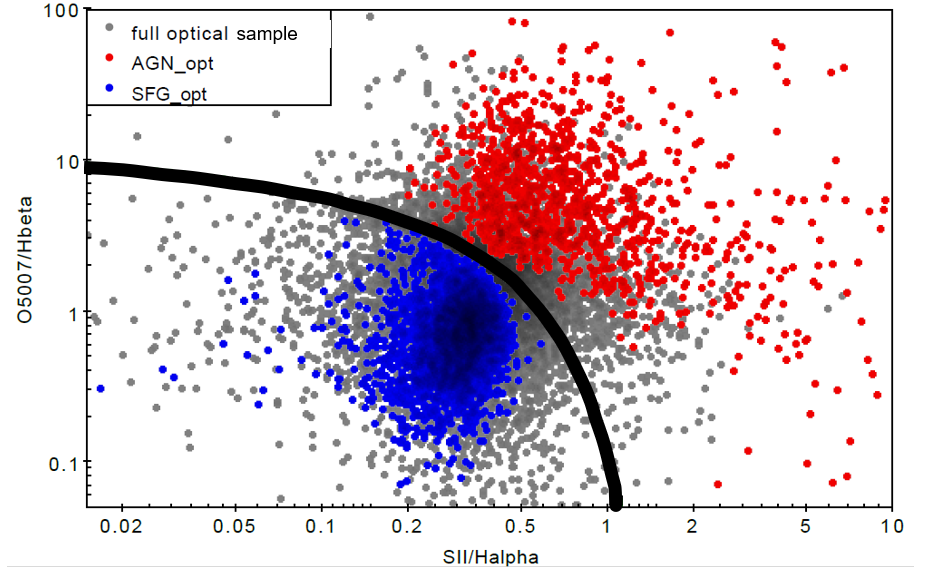}
    \caption{
    The BPT diagram, [OIII]$\lambda$5007/$H{\beta}$ vs 
    [SII]$\lambda\lambda$6717,6731/$H{\alpha}$, for the set of G23 galaxies with all of the required lines 
    in emission (grey). 
    The dividing line between AGN and SFG  \protect\citep{2001KewleySBmodel}  is shown in black.
    The galaxies for which the error box lies entirely on the AGN side or on the SFG side of the dividing line 
   are labelled AGN$_{opt}$ (red) or SFG$_{opt}$ (blue), respectively.
    }
    \label{fig:figBPTerr}
\end{figure}

As a complement to the W1-W2 colour criterion, we use the BPT diagnostic to distinguish AGN from SFG.
Requiring observed BPT line fluxes rejects all galaxies which have $z>0.315$ because the [SII]$\lambda\lambda$6717,6731 lines are redshifted out of the observed spectra.
This has the side effect of rejecting high mass galaxies ($\sim$1/3 to 2/3 of those above $5\times10^{10} M_{\odot}$).

To classify a galaxy as AGN$_{opt}$ or SFG$_{opt}$, we required the entire error box in the BPT diagram to lie on the 
AGN$_{opt}$ or SFG$_{opt}$ side of the diagnostic BPT line. 
Figure~\ref{fig:figBPTerr} shows the BPT diagram.
Galaxies which have an ambiguous classification because of line flux uncertainties are shown by the grey points.

For the G23 sample, about half (Table 3) of the galaxies with measured line fluxes have one or more of the BPT lines in absorption, 
and of the $\sim$15,000 with emission lines, $\sim$4400 had high enough SN to be classified by the BPT diagnostic.
Table~\ref{tab:TBLw1w2BPTstats} gives the numbers of AGN and SFG for different subsets.

\begin{table*}
	\centering
	\caption{Properties$^{a}$ of SFG and AGN subsets using BPT and W1-W2 criteria.}
	\label{tab:TBLw1w2BPTstats}
	\begin{tabular}{lcccccc}
		\hline
		Category &  No. & Redshift &  $log(M/M_{\odot})$ & W1 mag & 936\,MHz f.d.(Jy) & $L_{\nu}(22~\mu m)$ (erg/s/Hz) \\
		           &    &  mean(SD) & mean(SD) & mean(SD) & mean(SD)  & mean(SD)  \\	
		               \hline
		G23 & & & & & &  \\
		\hline
		SFG$_{opt}$ & 3310  &  0.176(0.070) & 10.12(0.44) & n/a & n/a & n/a \\ 
	        AGN$_{opt}$ & 1112 &  0.158(0.080) & 10.04(0.72) & n/a & n/a & n/a  \\	
%	           & & & & \\
                \hline
	        936/G23 & & & & & & \\
	        \hline
		SFG$_{opt}$  & 187 &  0.119(0.067) & 10.24(0.53) & n/a & 0.0024(0.0059) & n/a \\
	        AGN$_{opt}$ & 66 &  0.139(0.076) & 10.44(0.55) & n/a & 0.0022(0.0028) & n/a \\	
		\hline
		G23/WISE & & & & & & \\
		\hline
		notAGN$_{IR}$ & 29983 &  0.220(0.102) & 10.44(0.57) & 15.35(0.76) & n/a & n/a  \\
	        AGN$_{IR}$ & 374 &  0.591(0.411) & 10.47(0.65) & 14.87(0.99)  & n/a & n/a \\
		SFG$_{opt}$ & 2202 &  0.173(0.070) & 10.12(0.42) & 15.43(0.84) & n/a & n/a \\
	        AGN$_{opt}$ & 765 &  0.165(0.078) & 10.16(0.67) & 15.28(0.89) & n/a & n/a  \\
%	           & & & & \\
                \hline
	        936/G23/WISE & & & & & & \\
	        \hline
		notAGN$_{IR}$ & 813 &  0.216(0.132) & 10.77(0.58) & 14.22(0.86) & 0.0049(0.0212) & n/a \\
	        AGN$_{IR}$ & 51 &  0.432(0.426) & 10.46(0.71) & 14.34(1.06)  & 0.037(0.108) & n/a  \\
		SFG$_{opt}$  & 140 &  0.115(0.065) & 10.25(0.52) & 13.86(0.76) & 0.0018(0.0031) & n/a \\
	        AGN$_{opt}$ & 56 &  0.151(0.076) & 10.51(0.47) & 13.95(0.86) & 0.0022(0.0030) & n/a \\		
		\hline
	        G23/WISE & & & & & & \\
	        	        	         w4snr$\ge$3 & & & & & &\\
	        \hline
		notAGN$_{IR}$ & 1718 &  0.158(0.095) & 10.37(0.48) & 14.50(0.83) & n/a & 4.75e30(8.29e30)\\
		AGN$_{IR}$ & 224 &  0.492(0.370) & 10.60(0.60) & 14.43(0.91) & n/a & 1.17e32(1.92e32)\\
		AGN$_{IR}$/AGN$_{opt}$ & 27 &  0.202(0.061) & 10.61(0.31) & 14.27(0.77) & n/a & 1.47e31(1.14e31)\\
		AGN$_{IR}$/SFG$_{opt}$ & 17 &  0.224(0.052) & 10.48(0.45) & 14.45(1.17) & n/a & 2.10e31(3.03e31)\\
		AGN$_{IR}$/lineflux$<$0 & 19 &  0.173(0.085) & 10.26(0.93) & 13.87(1.38) & n/a & 1.31e31(1.77e31)\\
	        \hline
	        936/G23/WISE  & & & & & &\\
	        	         w4snr$\ge$3 & & & & & &\\
	        \hline
		notAGN$_{IR}$ & 333 &  0.128(0.088) & 10.36(0.54) & 13.78(0.84) & 0.0034(0.0295) & 4.90e30(9.81e30)\\	
		AGN$_{IR}$ & 43 &  0.318(0.313) & 10.47(0.71) & 14.18(1.06) & 0.0318(0.0099) & 9.83e31(2.48e32)\\   
		AGN$_{IR}$/AGN$_{opt}$ & 13 &  0.178(0.064) & 10.59(0.36) & 14.07(0.65) & 0.0018(0.0007) & 1.64e31(1.39e31)\\
		AGN$_{IR}$/SFG$_{opt}$ & 1 &  0.236(n/a) & 10.69(n/a) & 14.64(n/a) & 0.0023(n/a) & 1.33e31(n/a)\\
	AGN$_{IR}$/lineflux$<$0 & 3 &  0.129(0.061) & 10.68(0.51) & 11.66(0.66) & 0.0026(0.0010) & 1.49e31(1.76e30)\\ 
			\hline
	\end{tabular}
		\begin{tablenotes}\footnotesize
	\item[*] a. The GAMA23 stellar masses were based on stellar masses from the GAMA equatorial fields as described in Section~\ref{sec:GAMAdata}.
\end{tablenotes}	   
\end{table*}

\begin{figure}
	\includegraphics[width=\columnwidth]{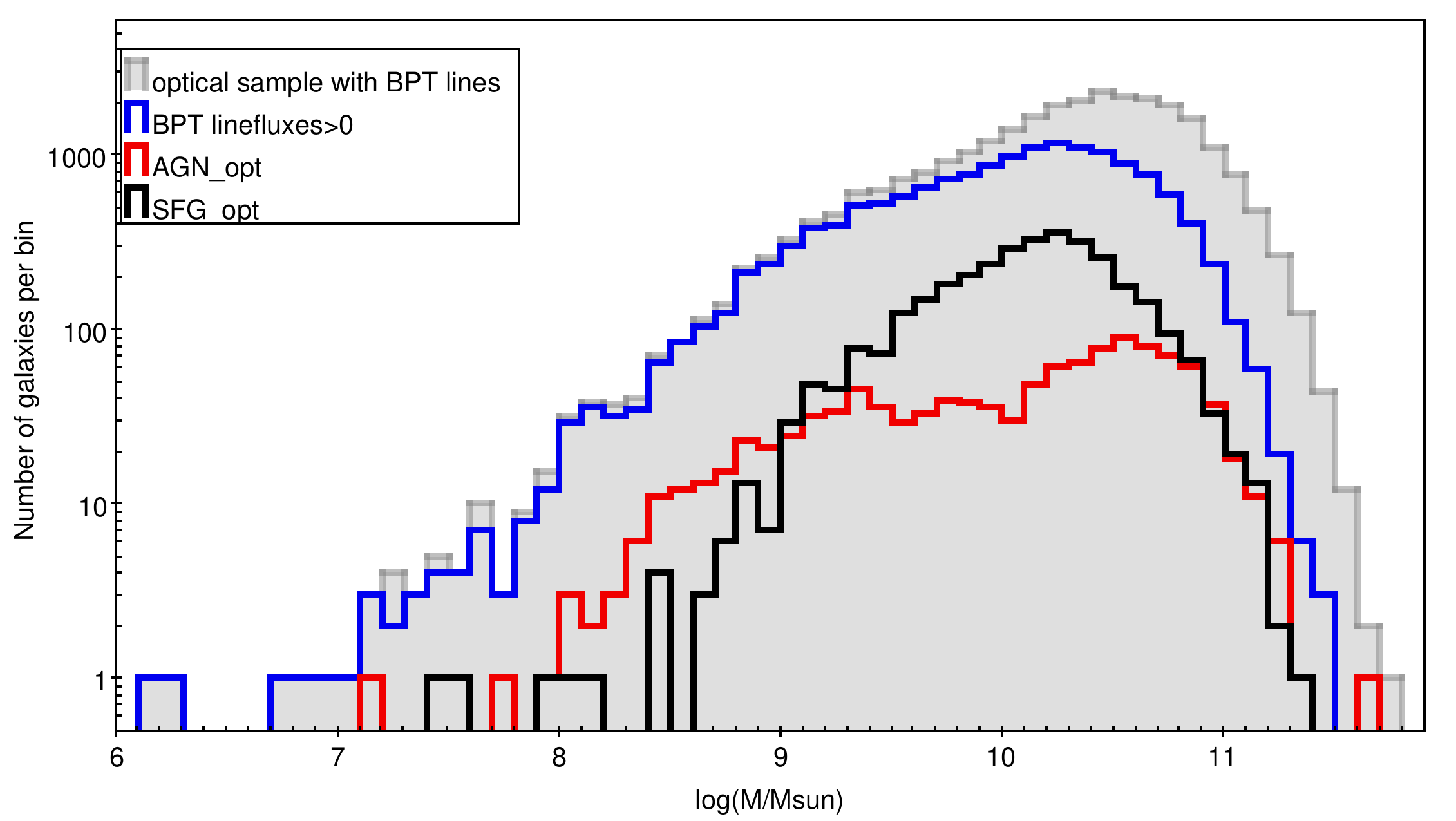} 
	\includegraphics[width=\columnwidth]{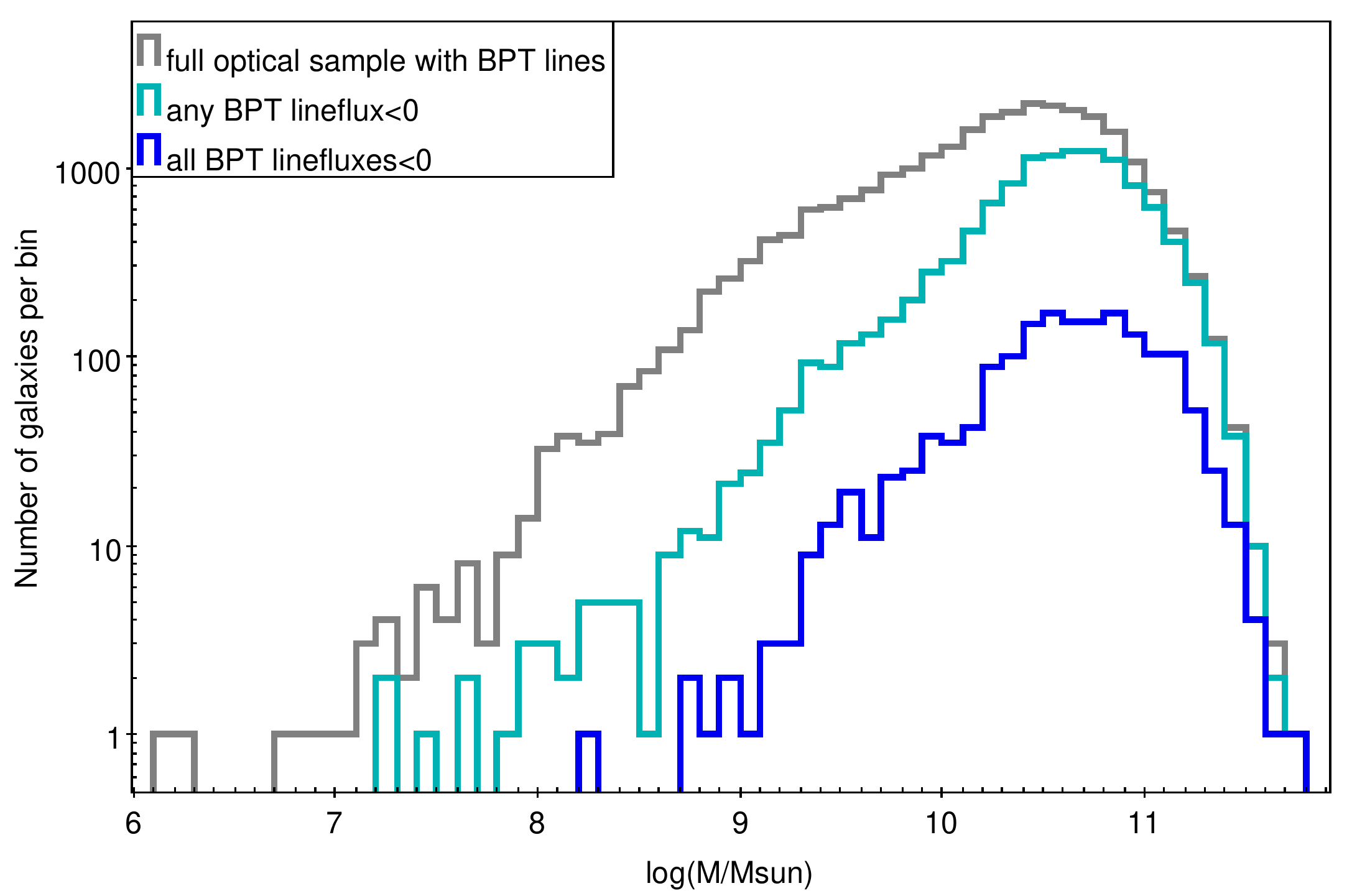} 
    \caption{
    The distribution of masses  (number of galaxies per mass bin, in units of log($M/M_{\odot}$))
    for the different categories of G23 galaxies, all with measured BPT line fluxes. 
     Top panel: the full sample (grey), galaxies with all 4 BPT lines in emission (blue), 
     SFG$_{IR}$ with BPT lines in emission (black), and AGN$_{opt}$ with BPT lines in emission (red).
     Bottom panel: the full sample (grey), galaxies with at least one BPT line in absorption (cyan), 
     and galaxies with all 4 BPT lines in absorption (blue).
The top panel illustrates that galaxies with all BPT lines in emission have lowest mean mass, and
      AGN$_{opt}$ have a broader, likely bimodal mass distribution. 
     The bottom panel illustrates that galaxies with all BPT lines in absorption have highest mean mass.
    }
    \label{fig:figmassDistBPT}
\end{figure}

The mass distributions for the different categories are shown in the top and bottom panels of Figure~\ref{fig:figmassDistBPT}.
In order, from highest mean mass to lowest mean mass, the categories are:
galaxies with all BPT lines in absorption;  galaxies with at least one BPT line in absorption; 
SFG$_{opt}$; AGN$_{opt}$; and galaxies with all BPT lines in emission.
The AGN$_{opt}$-classified galaxies have the broadest mass distribution, inconsistent with single-peaked distribution
and characteristic of a bimodal mass distribution. 
A $\chi^2$ fit to the distribution shows it requires two components: a high mass component peaking at 
$\sim3\times10^{10}M_{\odot}$, and a low mass component peaking at $\sim2\times10^{9}M_{\odot}$.

To compare the BPT criterion with the W1-W2 criterion, we use the G23/WISE sample.
The requirement of a WISE detection in effect rejects low mass galaxies (see Table~\ref{tab:TBLbasicstats}): 
it results in rejection of $\sim$1/6 or more of the galaxies below $3\times10^9 M_{\odot}$.
 Table~\ref{tab:TBLw1w2BPTstats} shows that the increase in mean mass for G23/WISE compared to G23 depends on the AGN classification: 
G23/WISE AGN$_{opt}$ has mean mass higher than G23 AGN$_{opt}$ by factor 2.24; and
G23/WISE  SFG$_{opt}$ has  mean mass higher than G23  SFG$_{opt}$ by factor 1.35.

\begin{figure}
	\includegraphics[width=\columnwidth]{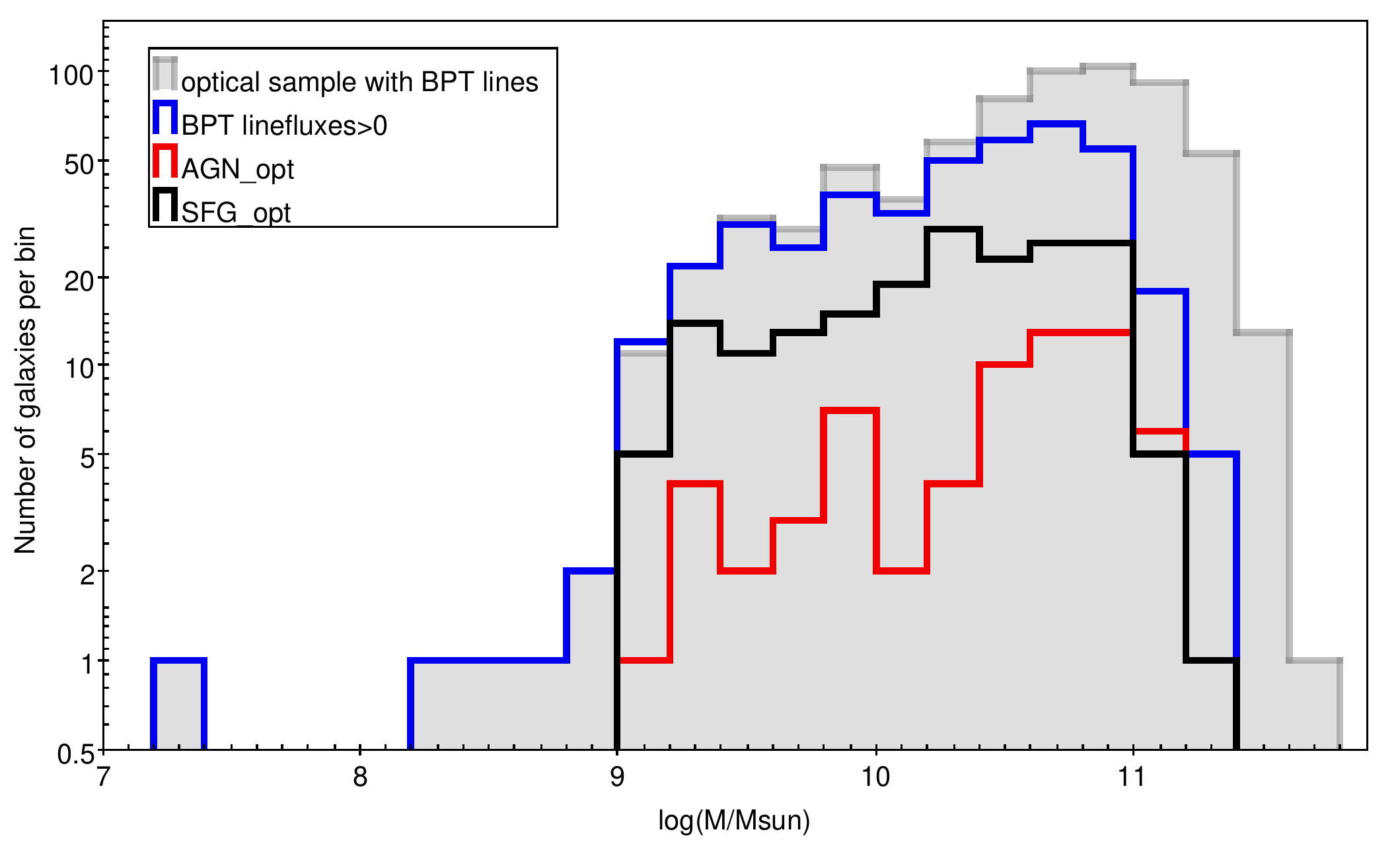}
	\includegraphics[width=\columnwidth]{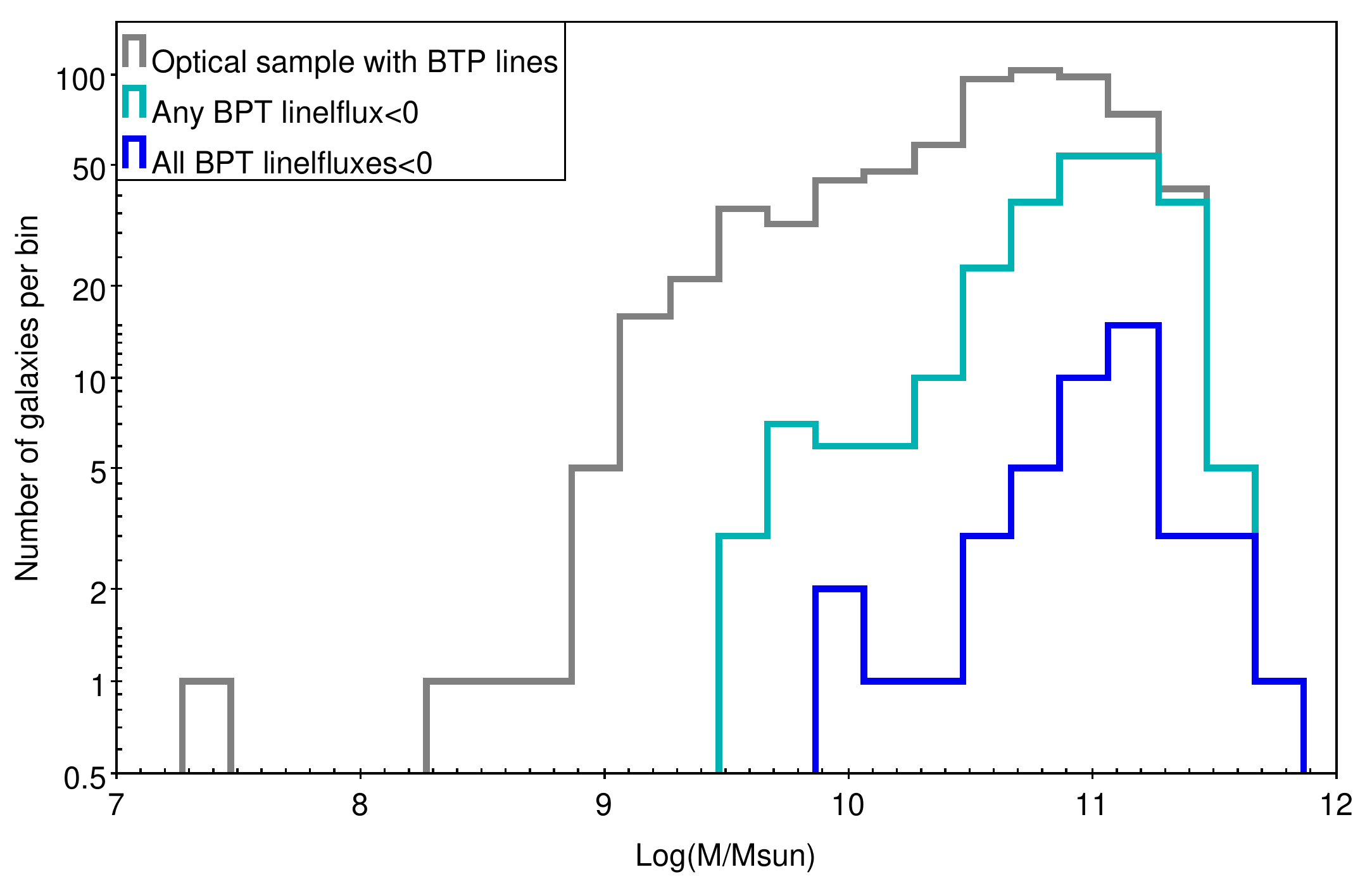} 
    \caption{
      The distribution of masses  (number of galaxies per mass bin, in units of log($M/M_{\odot}$))
     for the different categories of G23 galaxies with
     936\,MHz emission, all with measured BPT line fluxes. 
     Top panel: the full radio/optical sample (grey), those with all 4 BPT lines in emission (blue), 
     SFG$_{opt}$ with BPT lines in emission (black), and AGN$_{opt}$ with BPT lines in emission (red).
     Bottom panel: the full radio/optical sample (grey), those with at least one BPT line in absorption (cyan), 
     and those with all 4 BPT lines in absorption (blue). The requirement of a 936\,MHz detection constrains the sample
     to relatively high masses, and the mass distributions of the optically classified AGN and SFGs with radio detections
     are similar. 
     }
    \label{fig:fig960massDistBPT}
\end{figure}

Next, the BPT diagnostic was applied to the radio detected 936/G23  sample.
Figure~\ref{fig:fig960massDistBPT} shows the mass distributions for the 936/G23 galaxies and their five sub-categories. 
Comparison of the mass distributions in Figure~\ref{fig:fig960massDistBPT} with those in Figure~\ref{fig:figmassDistBPT} shows
that masses for galaxies with 936\,MHz emission are higher compared to galaxies without radio emission:
AGN$_{opt}$ with 936\,MHz emission  has mean mass higher by factor 2.5 than the whole AGN$_{opt}$ set; and
SFG$_{opt}$ with 936\,MHz emission has mean mass higher by factor 1.3 than the whole SFG$_{opt}$ set. 
Similar results to that for 936/G23 are obtained when we apply the BPT diagnostic to 936/G23/WISE  sample.

\begin{figure}
		\includegraphics[width=\columnwidth]{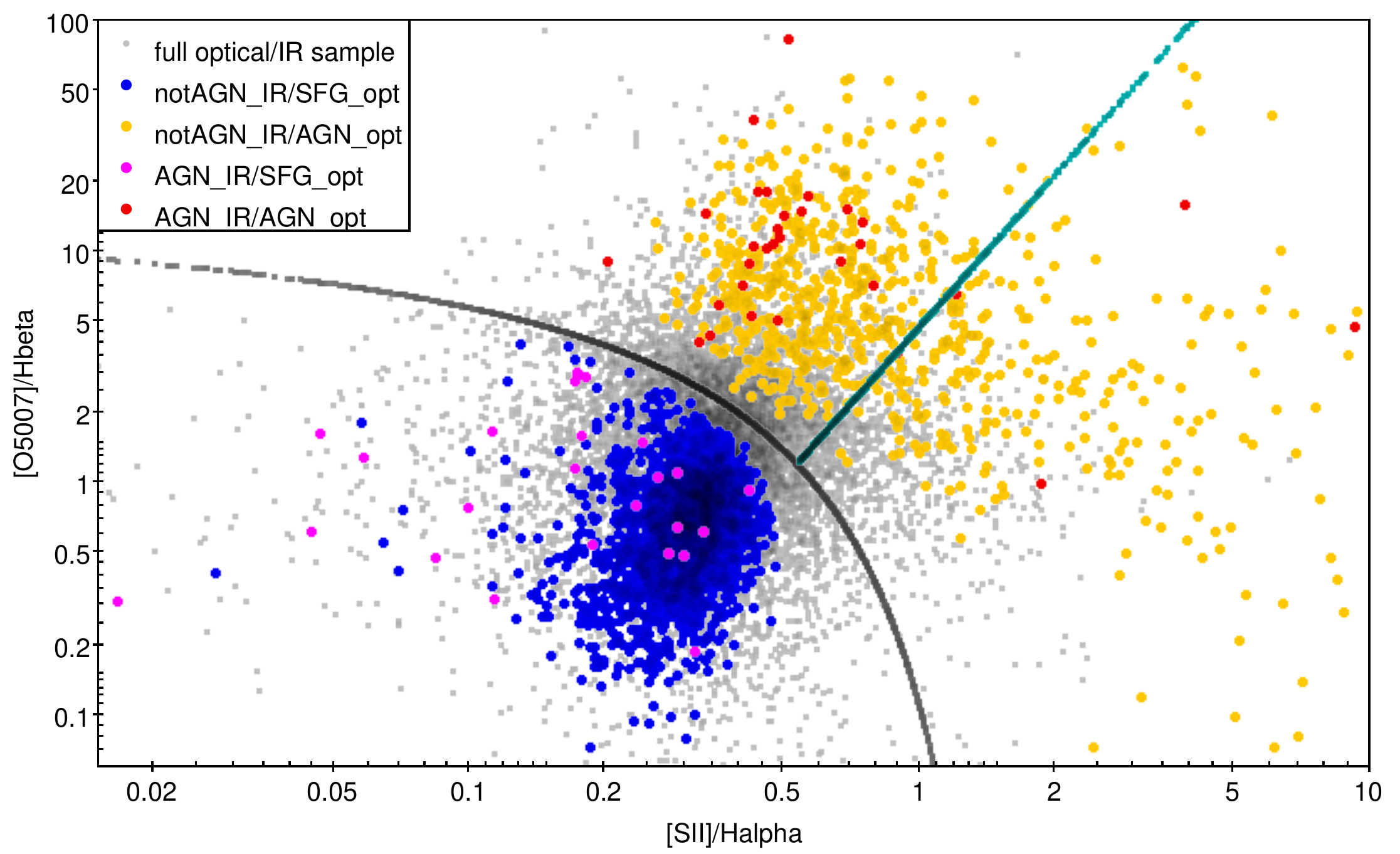} 
		\includegraphics[width=\columnwidth]{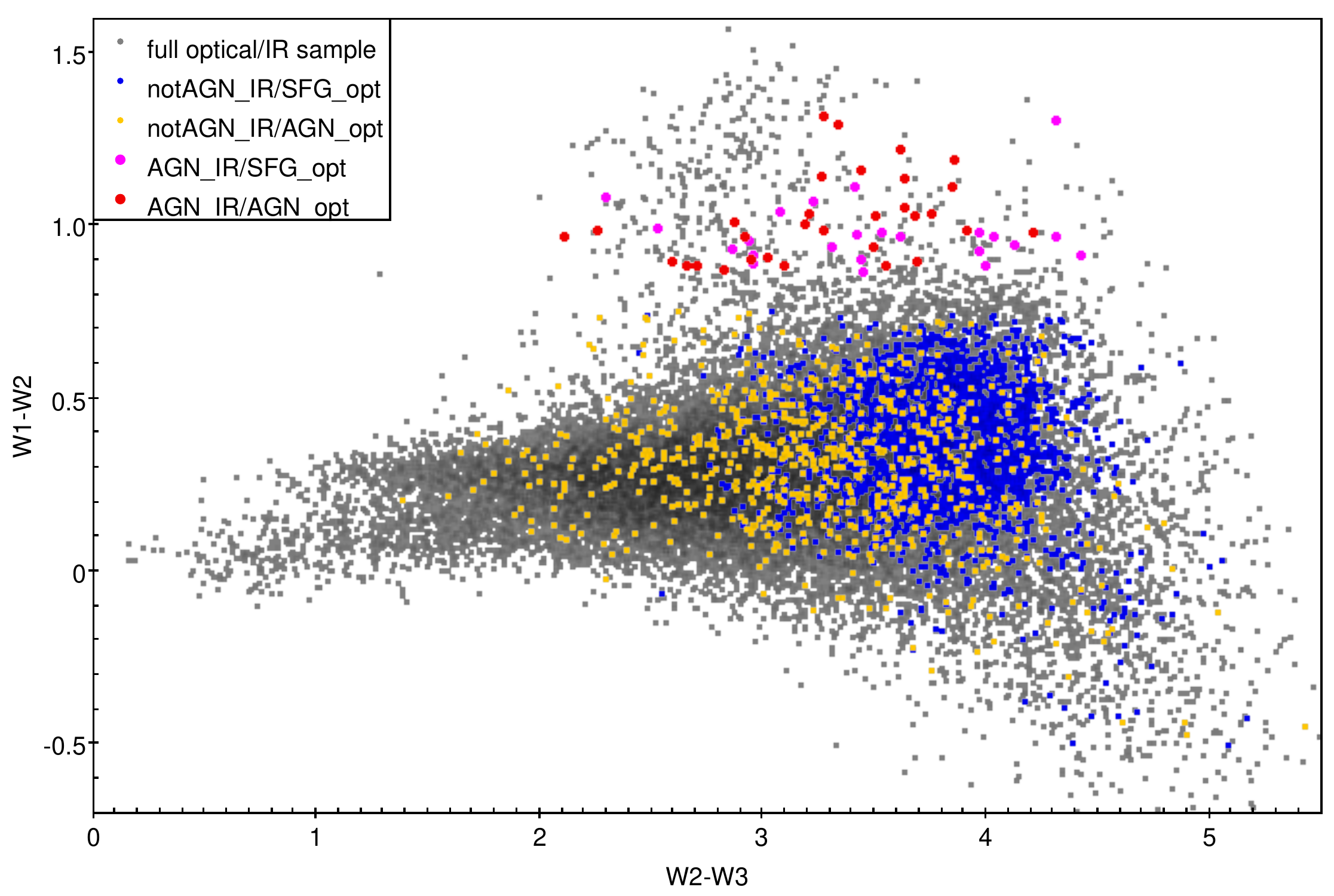}
		\includegraphics[width=\columnwidth]{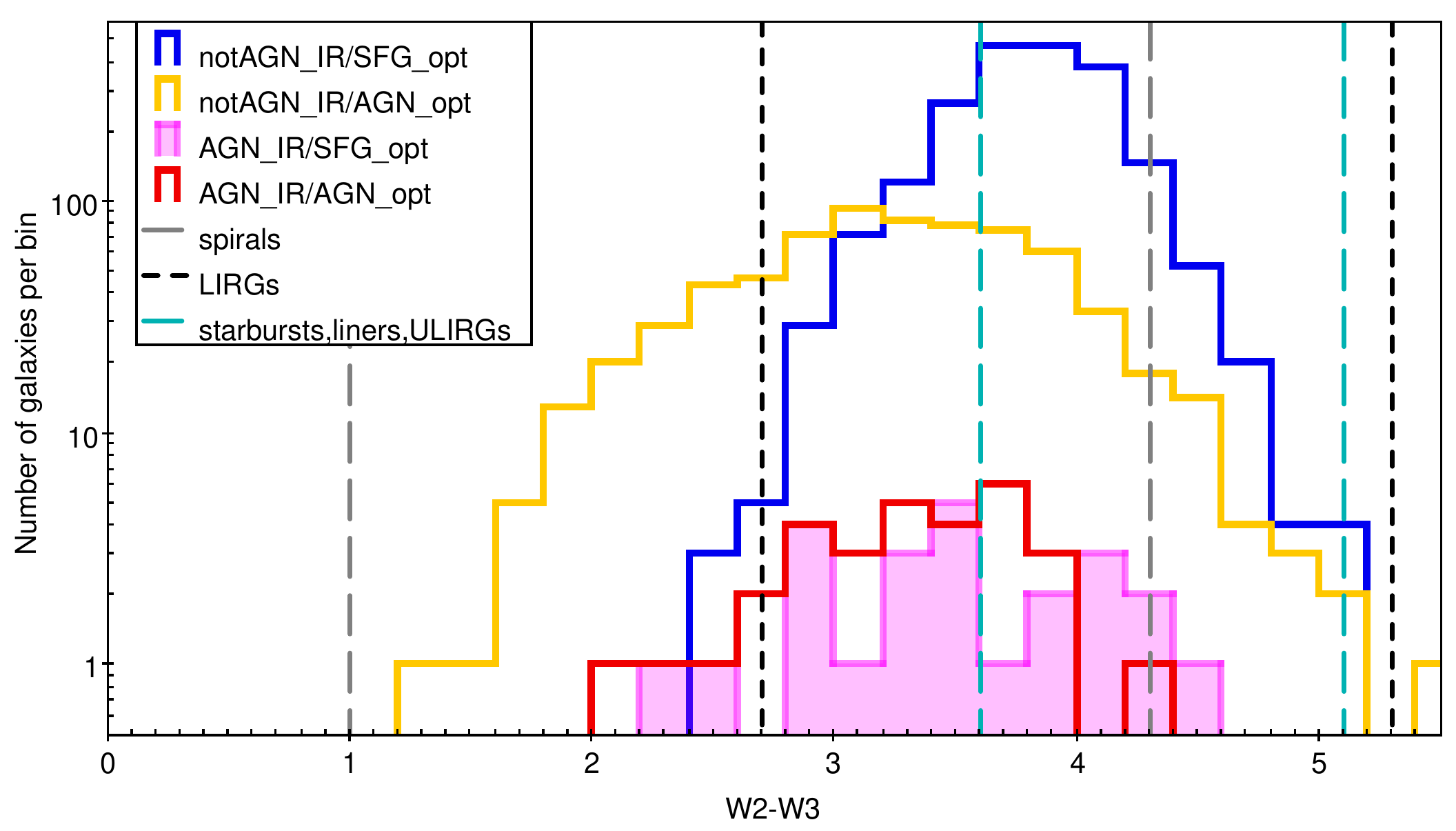}
    \caption{
    G23/WISE galaxies, classified as AGN or notAGN/SFG by the BPT and W1-W2 diagnostics.
    The full optical/IR set of galaxies is shown in grey, 
    notAGN$_{IR}$ and SFG$_{opt}$ galaxies are marked in blue,
    galaxies classified as notAGN$_{IR}$ and AGN$_{opt}$ are marked in yellow, 
    galaxies classified as AGN$_{IR}$ and SFG$_{opt}$ are marked in magenta, 
   and those classified as AGN$_{IR}$ and AGN$_{opt}$ are marked in red.  
    The top panel shows the BPT diagram, with 
    the BPT line \protect\citep{2001KewleySBmodel} separating SFG$_{opt}$ and AGN$_{opt}$ in black; and with
    the line that marks the division \protect\citep{2010Sharp}  between AGN-photo-ionized gas (upper-left of the cyan line)
    and shock-ionized gas (lower-right of the cyan line).
     The middle panel shows W1-W2 colour vs W2-W3 colour for sources with W3snr$\ge$3.
         The division between AGN$_{IR}$ and notAGN$_{IR}$ is at W1-W2=0.8.
     The bottom panel shows the distribution of W2-W3 colours for the different subsets (histograms) and the
     colour ranges (pairs of vertical dashed lines) for spirals, for LIRGs and for starbursts/liners/ULIRGs from Fig. 12 of
     \protect\citet{2010Wright}.
    }
    \label{fig:figSharpLine}
\end{figure}

\subsubsection{Comparison of optical-classified and IR-classified samples}

The G23/WISE galaxies were separated into four categories: notAGN$_{IR}$/SFG$_{opt}$; notAGN$_{IR}$/AGN$_{opt}$;  AGN$_{IR}$/SFG$_{opt}$;
and   AGN$_{IR}$/AGN$_{opt}$.
These are shown on a BPT diagram in the top panel of Figure~\ref{fig:figSharpLine}.
The AGN$_{IR}$ and  notAGN$_{IR}$  galaxies are scattered on both sides of the line which forms the BPT criterion
demonstrating that there is no clear correlation seen between the BPT and W1-W2 diagnostics.

The middle panel of Figure~\ref{fig:figSharpLine} shows the
various classes of galaxies placed in the WISE colour-colour diagram. 
\citet{2010Wright} and \citet{2011Jarrett} showed that 
the notAGN$_{IR}$ types include: ellipticals with W2-W3 of $\simeq0.5-1.5$; spirals with W2-W3 of $\simeq1.0-4.3$;
 luminous IR galaxies (LIRGs) with W2-W3 of $\simeq2.7-5.3$; and
starbursts, liners and  ultraluminous  IR galaxies (ULIRGs) with W2-W3 of $\simeq3.6-5.1$.
The colour limits for these groups are shown in the bottom panel of  Figure~\ref{fig:figSharpLine} overlaid
on the distribution of W2-W3 colors.
The notAGN$_{IR}$/SFG$_{opt}$ have a W2-W3 colour range consistent with that for LIRGs or a mixture of LIRGS 
and starbursts/liners/ULIRGs.
The notAGN$_{IR}$/AGN$_{opt}$ have a wide W2-W3 colour range, consistent with a mixture of spirals 
and one or both of the other groups of galaxies (LIRGs and starbursts/liners/ULIRGs). 

Table~\ref{tab:TBLw1w2vsBPT} shows that there is little quantitative correlation between the W1-W2 criterion and BPT criterion.
Generally, the BPT classes of W1-W2 classified sources show poor correlation, e.g.
for the G23/WISE sample, AGN$_{IR}$ are 6.4\% classified as SFG$_{opt}$ and 8.2\% classified as AGN$_{opt}$.
The W1-W2 classes of both SFG$_{opt}$ and AGN$_{opt}$ show most to be classified as notAGN$_{IR}$. 

\begin{table*}
	\centering
	\caption{Numbers of AGN/notAGN/SFG using W1-W2 and BPT criteria, or of Emission-line Galaxies.}
	\label{tab:TBLw1w2vsBPT}
	\begin{tabular}{lccccr}
		\hline
		 Class$^{a}$: & notAGN$_{IR}$ & AGN$_{IR}$ & SFG$_{opt}$ & AGN$_{opt}$ & Emission-line \\	
		 & No./\% & No./\% & No./\% & No./\% &  No./\% \\
			Sample  & & & & & \\
		\hline
		G23/WISE  &  29983/95.3\% & 374/1.2\% & 2202/7.0\% & 765/2.4\% & 13149/41.8\% \\
	       subgroups & & & & &\\
		notAGN$_{IR}$  &  29983/100\% & 0/0\% & 2041/6.8\% & 695/2.3\% & 12121/40.4\% \\
	        AGN$_{IR}$  &  0/0\% & 374/100\% & 24/6.4\% & 31/8.2\% & 108/28.9\% \\
		SFG$_{opt}$  &  2014/92.7\% & 24/1.1\% & 2202/100\% & 0/0\% & 2202/100\% \\
	        AGN$_{opt}$  &  695/90.8\% & 31/4.1\% & 0/0\% & 765/100\% & 765/100\% \\
%	           & & & & \\
		\hline
		936/G23/WISE  &  813/91.5\% & 51/5.7\% & 140/15.7\% & 56/6.3\% & 352/39.6\% \\ 
	       subgroups & & & & &\\
		notAGN$_{IR}$  &  813/100\% & 0/0\% & 130/16.0\% & 38/4.7\% & 308/37.9\% \\
	        AGN$_{IR}$  &  0/0\% & 51/100\% & 1/2.0\% & 13/25.4\% & 26/51.0\% \\
		SFG$_{opt}$  &  130/92.9\% & 1/0.7\% & 140/100\% & 0/0\% & 140/100\% \\
	        AGN$_{opt}$  &  38/67.8\% & 13/23.2\% & 0/0\% & 56/100\%  & 56/100\% \\		
		\hline
		\hline
		 Class$^{b}$: & AGN$_{IR}$/photo  & notAGN$_{IR}$/photo &   AGN$_{IR}$/shock  & notAGN$_{IR}$/shock & \\	
		 & No. & No. & No. & No. &  \\
%			Sample  & & & & & \\
		\hline	
		G23/WISE AGN$_{opt}$  &  26 & 446 & 5 & 249 &  \\ 
		\hline
	\end{tabular}
			\begin{tablenotes}\footnotesize
	\item[*] a.The samples are listed in the first column; the numbers and percentages within each sample are listed across the rows
	in columns two to six.
	\item[*] a.The breakdown of AGN$_{opt}$ types into photo-ionized and shock-ionized categories is given in
	the lowest section of the Table.
\end{tablenotes}	
\end{table*}

\begin{figure*}
\includegraphics[width=\columnwidth]{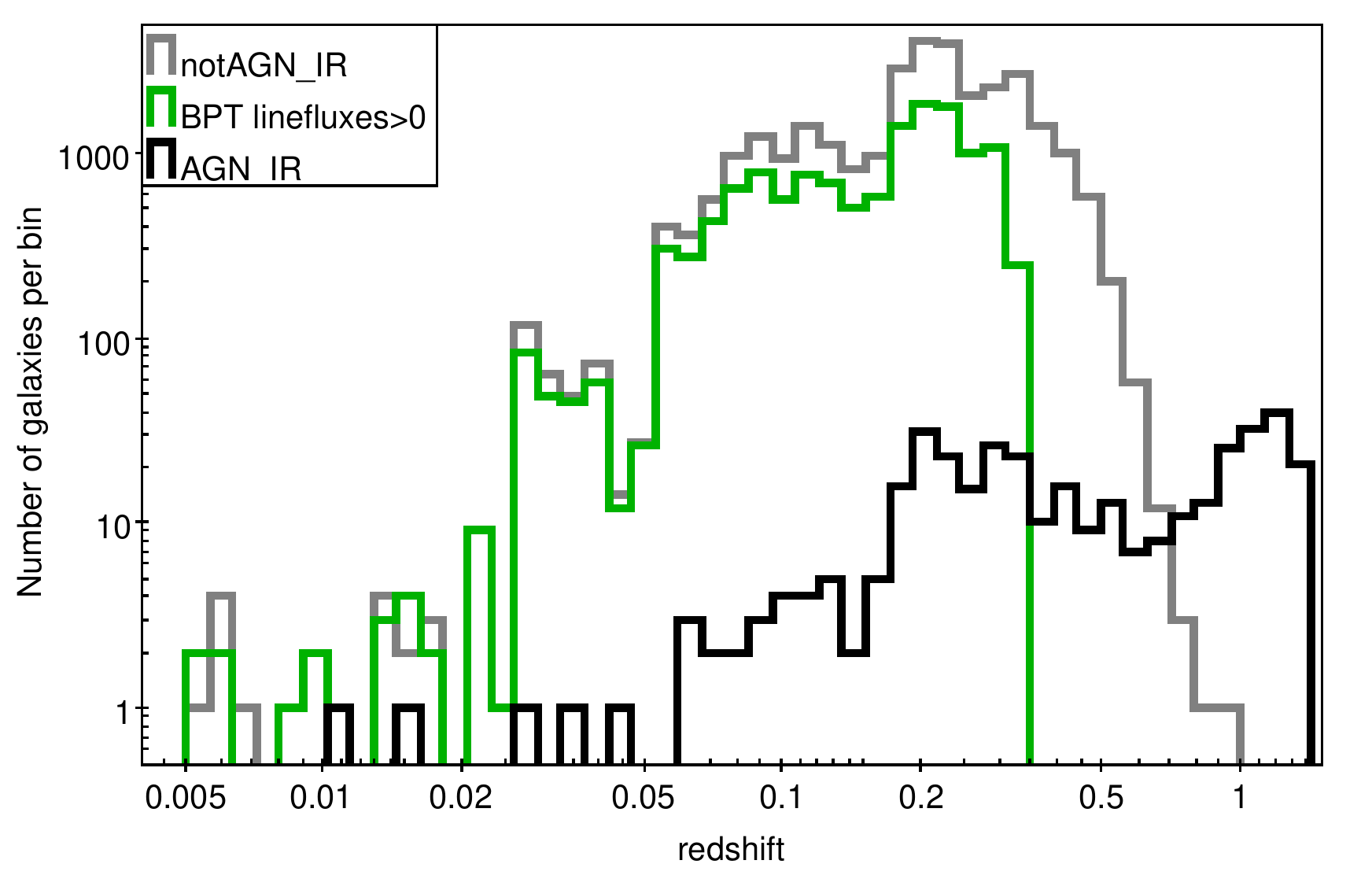} 
\includegraphics[width=\columnwidth]{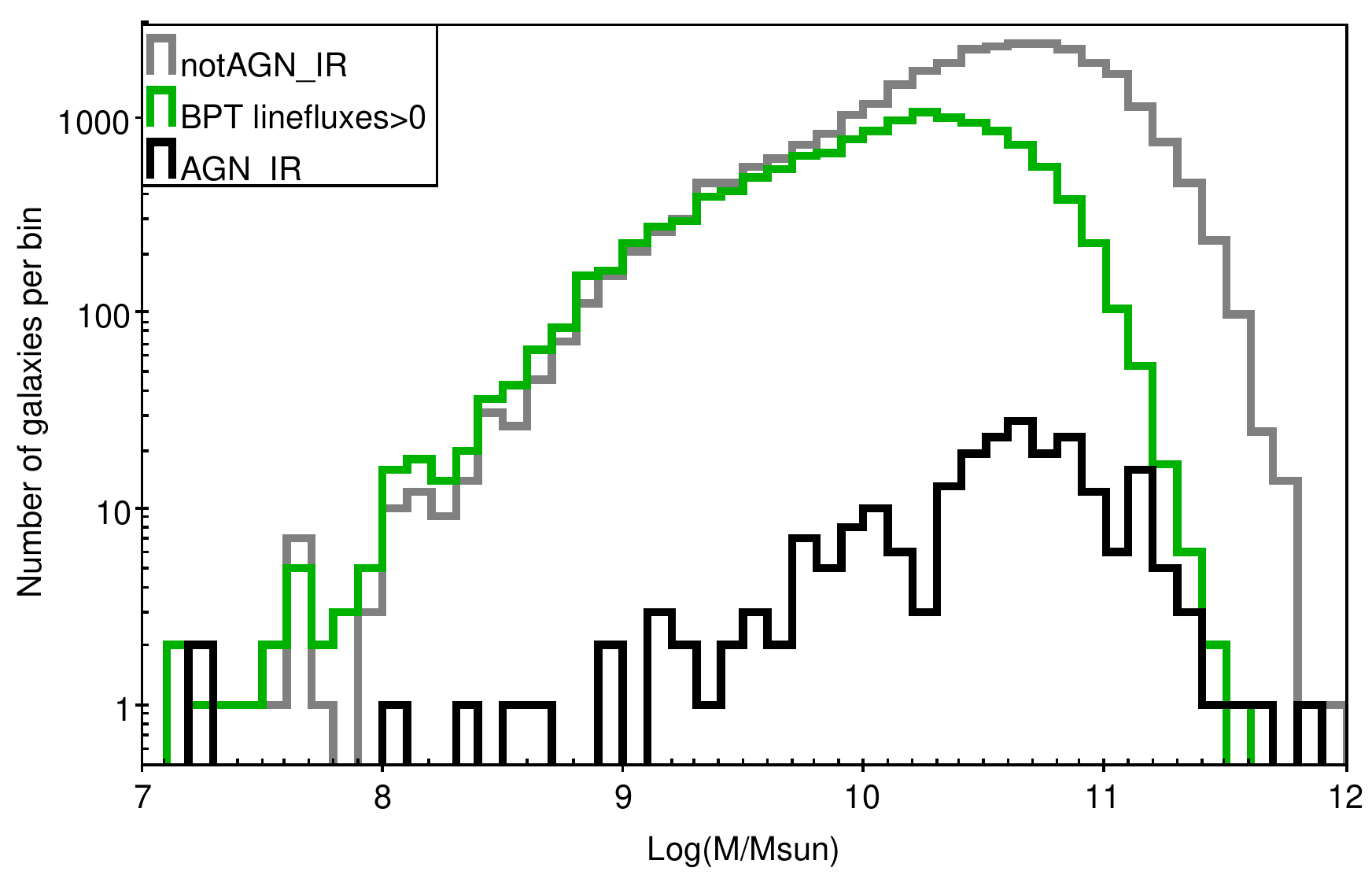} 
\includegraphics[width=\columnwidth]{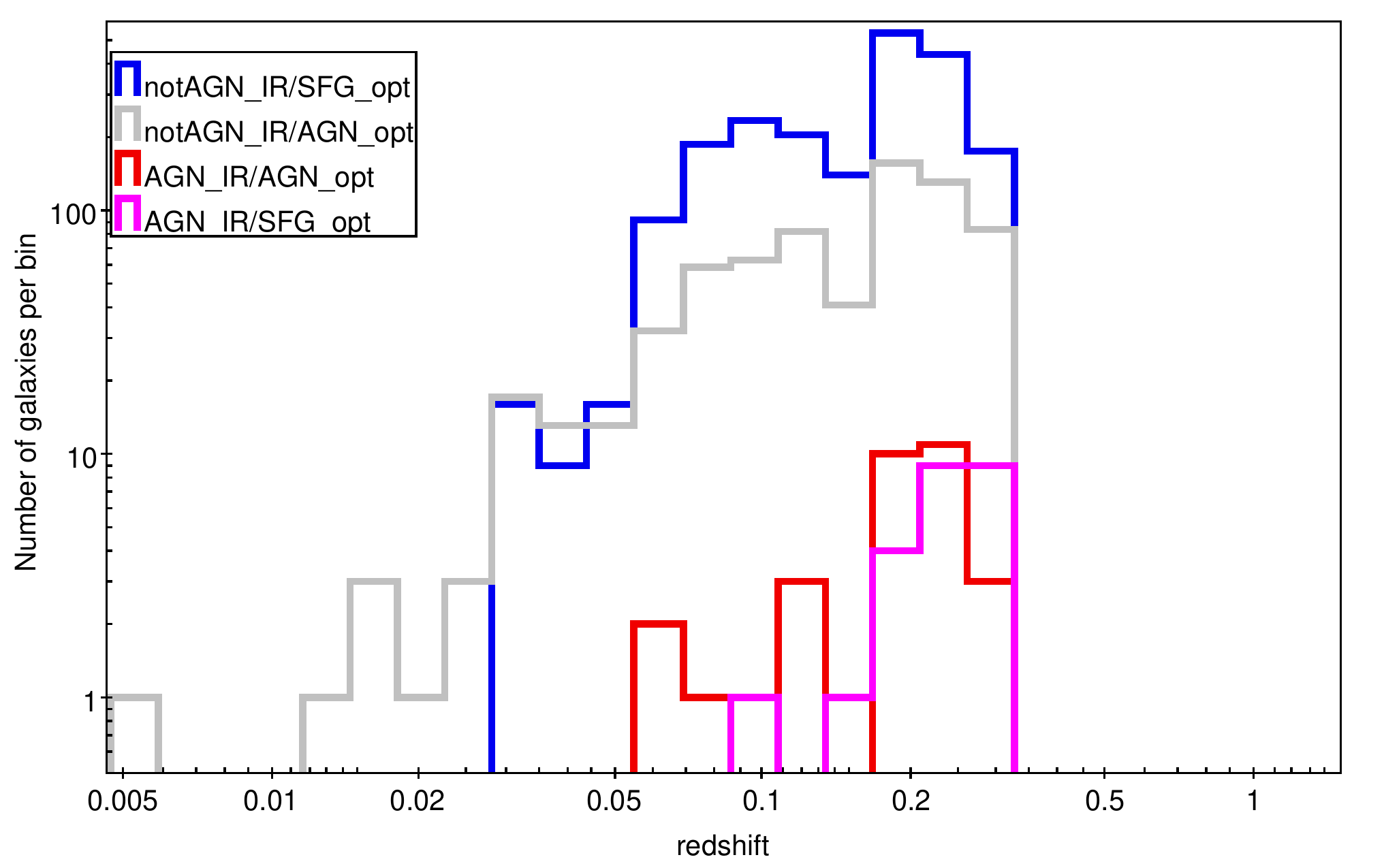} 
\includegraphics[width=\columnwidth]{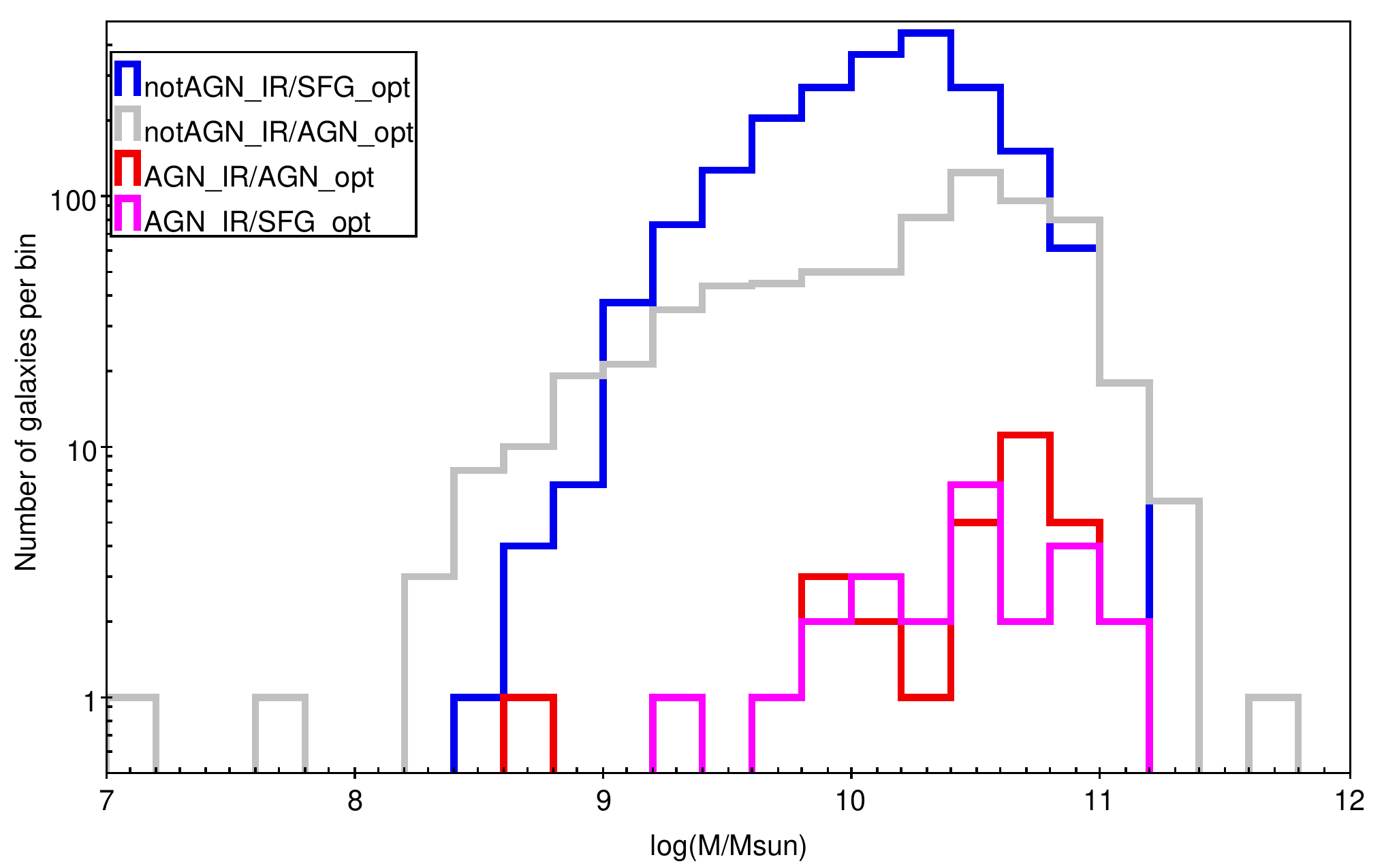} 
\includegraphics[width=\columnwidth]{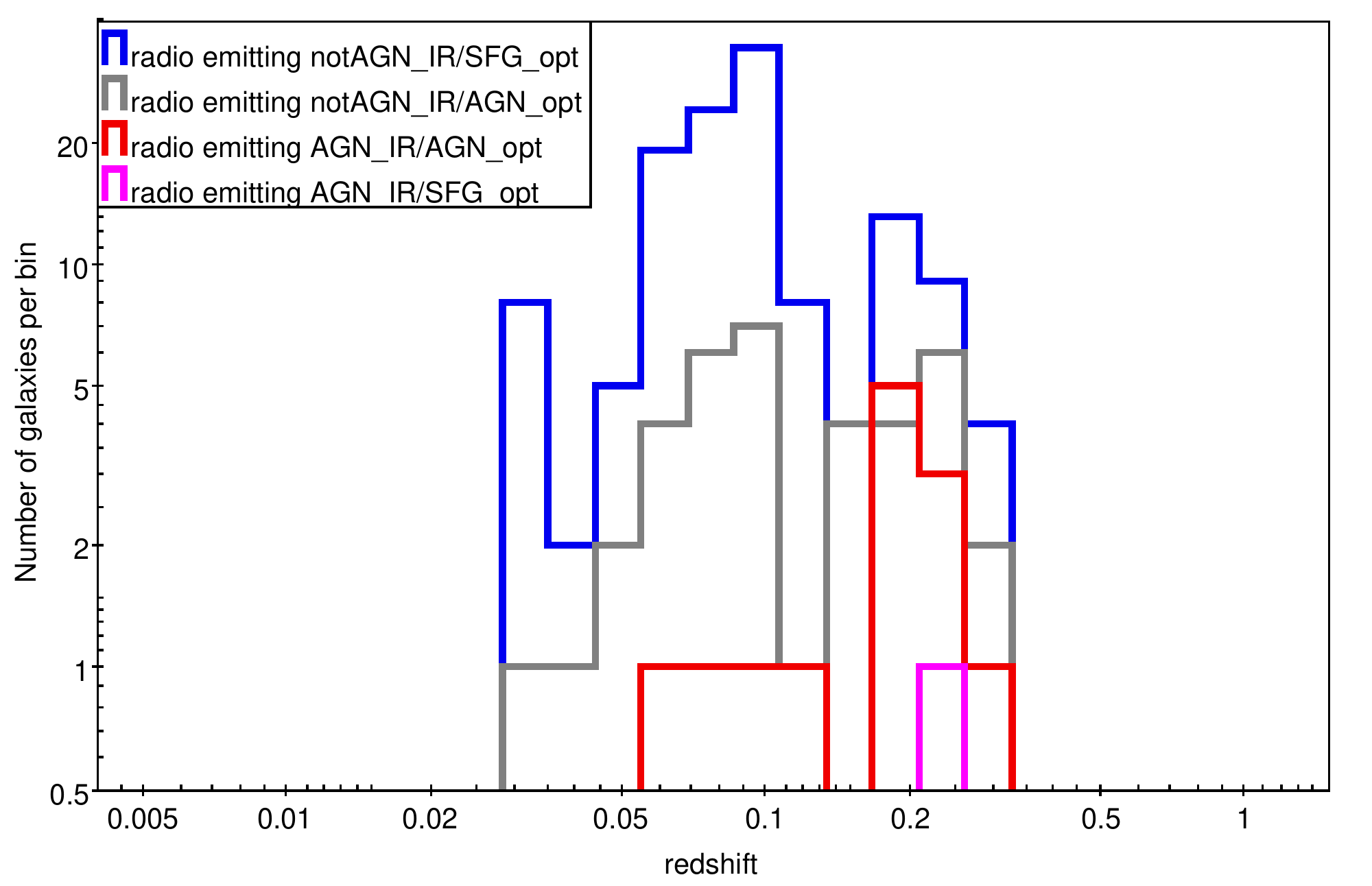}
\includegraphics[width=\columnwidth]{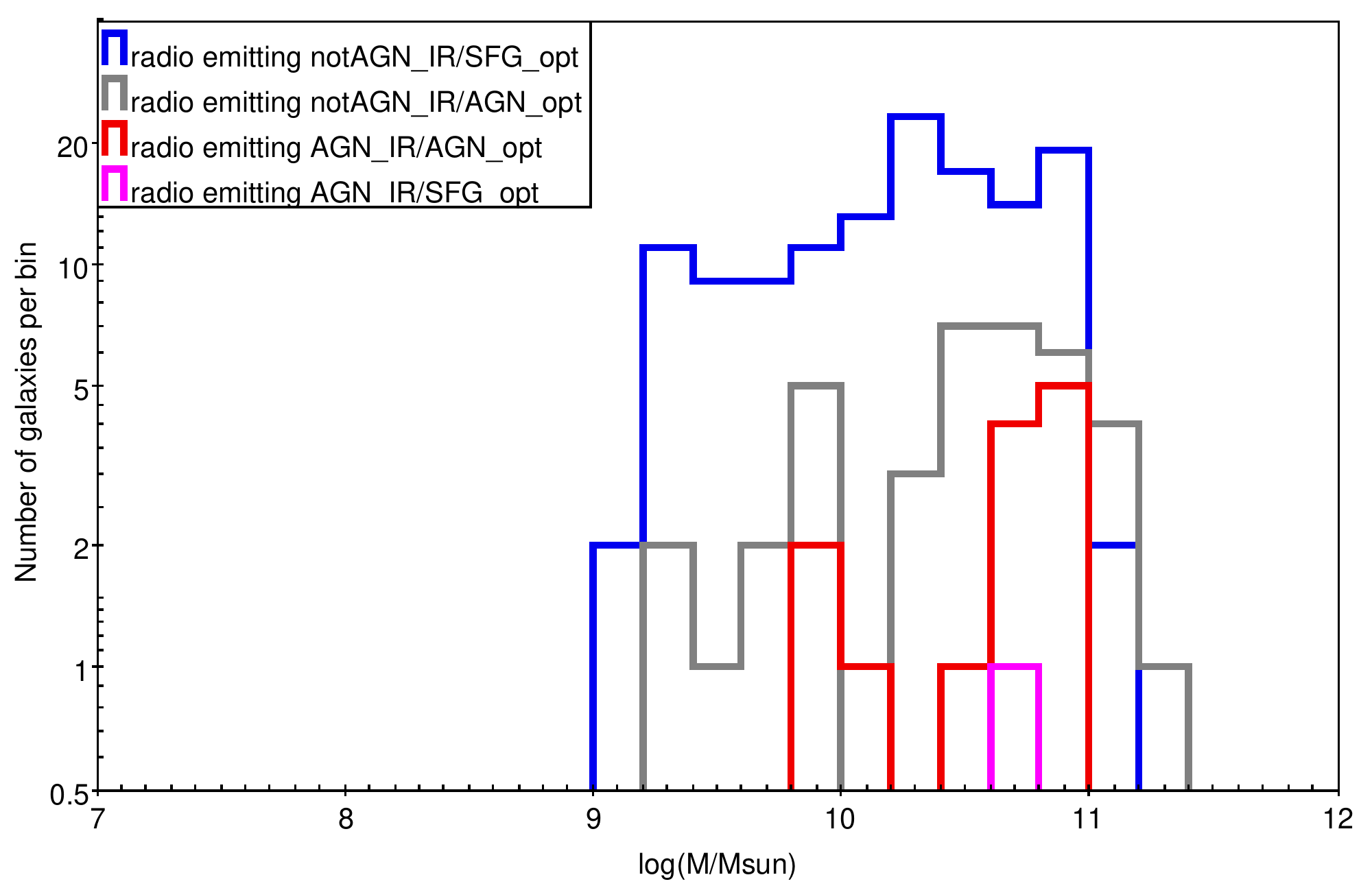} 
    \caption{
    Distributions of redshift (number of galaxies per redshift bin, left panels) and distributions of mass 
  (number of galaxies per mass bin, units of log($M/M_{\odot}$), right panels) for  the various subsets of G23/WISE galaxies. 
    Subsets shown in the top two panels are notAGN$_{IR}$ galaxies (grey), 
    AGN$_{IR}$ galaxies (black), and G23/WISE with all BPT lines in emission (green).
     Subsets shown in the middle and bottom panels are notAGN$_{IR}$/SFG$_{opt}$ (blue), notAGN$_{IR}$/AGN$_{opt}$ (grey), AGN$_{IR}$/AGN$_{opt}$ (red) and
    AGN$_{IR}$/SFG$_{opt}$ (magenta).
       }
    \label{fig:figG23WISEz}
\end{figure*}

The redshift distributions for the various categories of G23/WISE galaxies are shown in the left panels of Figure~\ref{fig:figG23WISEz}.
The Kolmogorov-Smirnov test was applied and showed that the 
notAGN$_{IR}$/AGN$_{opt}$ subset has a different redshift distribution (lower redshifts) than the AGN$_{IR}$/SFG$_{opt}$ subset at >99.9\% confidence.
The notAGN$_{IR}$/AGN$_{opt}$ have lower redshifts 
than notAGN$_{IR}$/SFG$_{opt}$ at the 99\% confidence level, and
the AGN$_{IR}$/AGN$_{opt}$  have  lower redshifts than AGN$_{IR}$/SFG$_{opt}$ at only the 97.5\% confidence level.
The mass distributions for the various categories of G23/WISE galaxies are shown in the right panels of Figure~\ref{fig:figG23WISEz}.

\begin{figure}
	\includegraphics[width=\columnwidth]{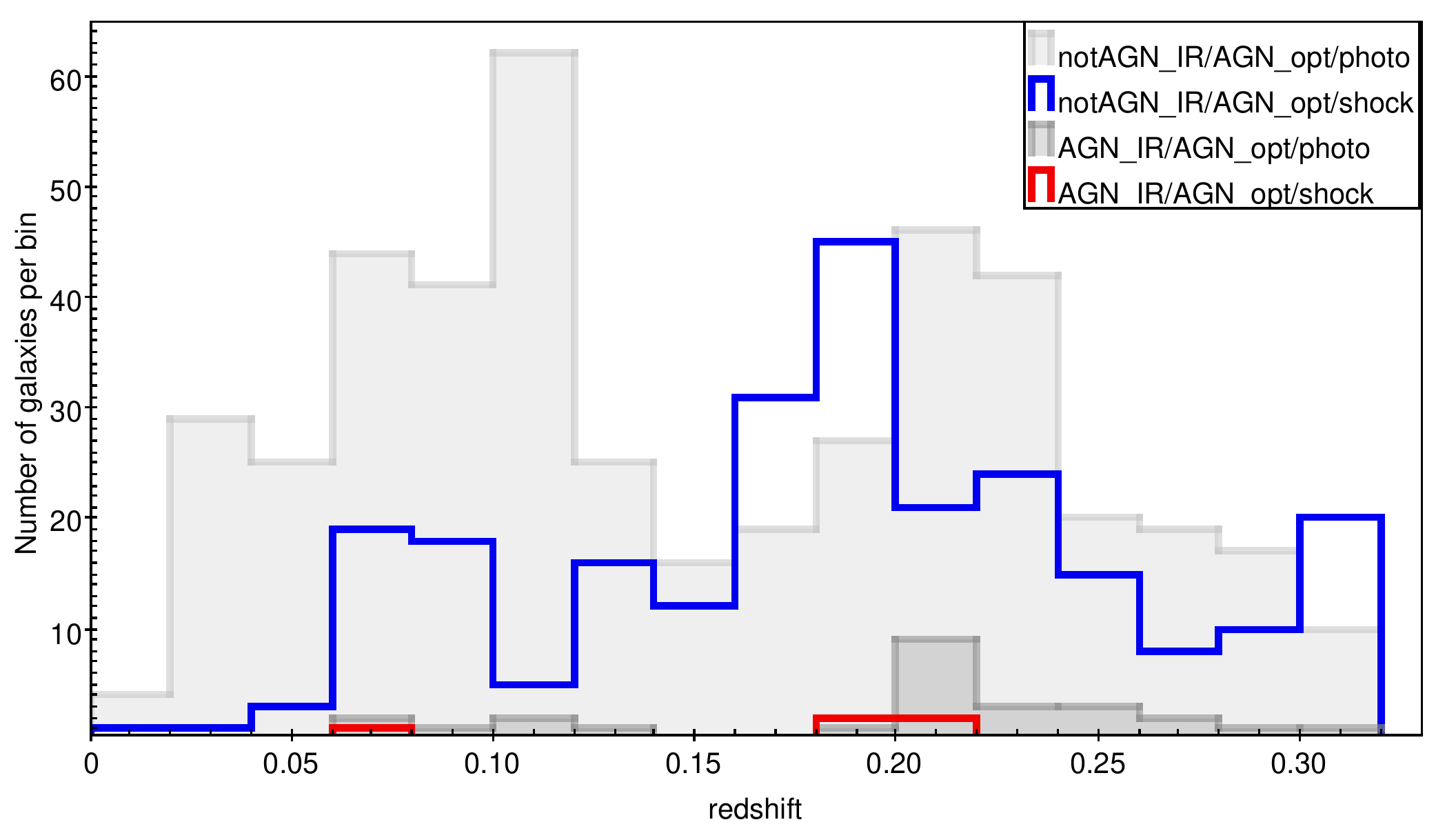}
	\includegraphics[width=\columnwidth]{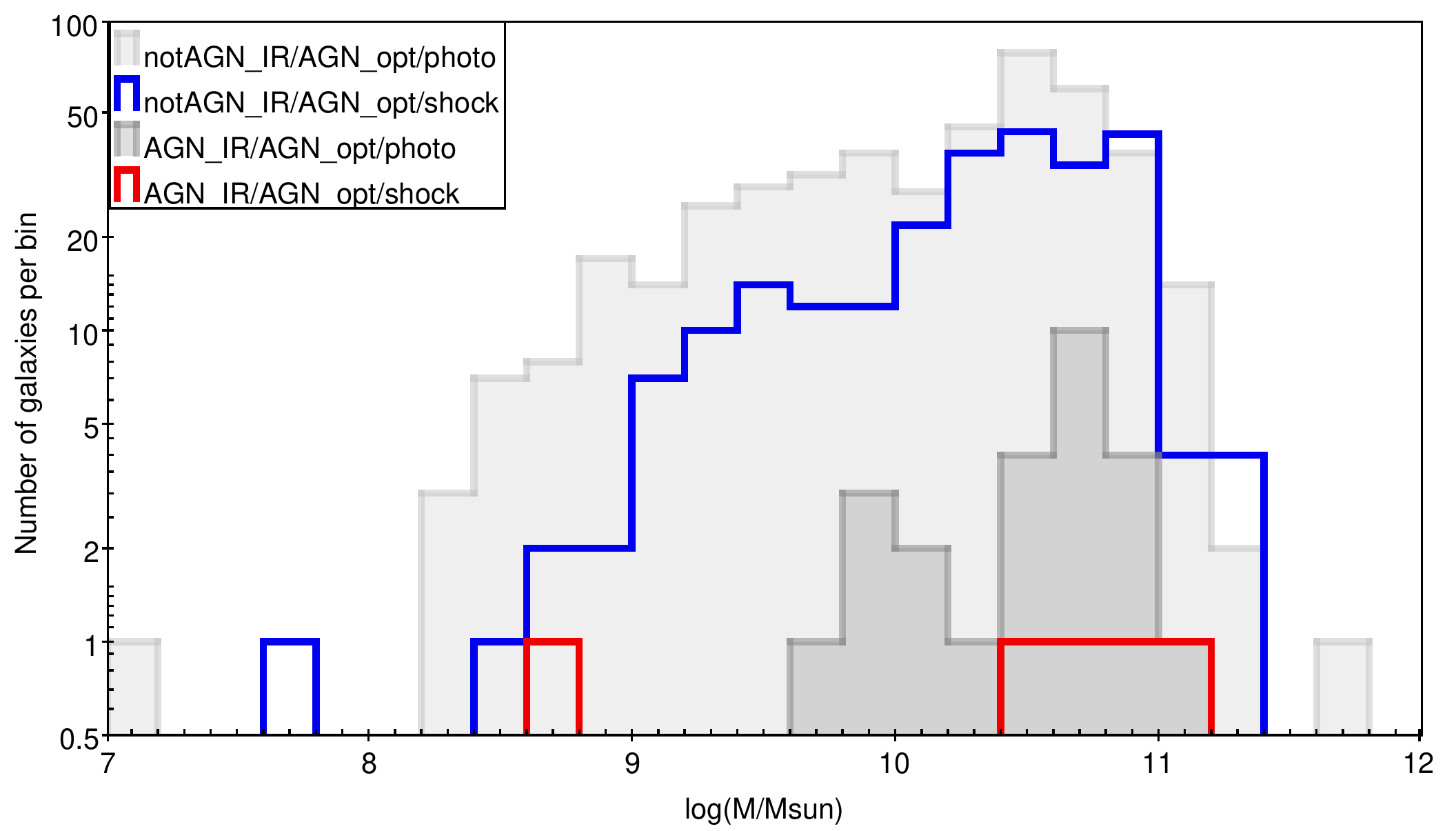}
    \caption{
    Distributions of redshift and mass for  
 G23/WISE AGN$_{opt}$ galaxies split into a subset dominated by photo-ionization (``/photo'' in the label) and a subset dominated by 
   shock ionization  (``/shock'' in the label). 
   The sets shown are  notAGN$_{IR}$/AGN$_{opt}$/photo (light grey),  notAGN$_{IR}$/AGN$_{opt}$/shock (blue), 
   AGN$_{IR}$/AGN$_{opt}$/photo (dark grey) and AGN$_{IR}$/AGN$_{opt}$/shock (red).
       }
    \label{fig:figshockM}
\end{figure}

We further split the G23/WISE AGN$_{opt}$ galaxies into subsets based on the dominant mechanism of ionization \citep{2010Sharp}: 
gas dominated by photo-ionization is left of the line  (cyan line in the top panel of Figure~\ref{fig:figSharpLine}) 
and shock-ionized gas is right of the line. 
Essentially the same line in the BPT diagram \citep{2006MNRAS.372..961K} separates Seyferts (left of the line) from LINERs (right of the line). 
64\% of the AGN$_{opt}$ galaxies are photo-ionized/Seyferts  and 36\% are shock-ionized/LINERs. 

The redshift and mass distributions of the photo-ionized and shock-ionized subsets of AGN$_{opt}$ galaxies 
are shown in Fig.~\ref{fig:figshockM}.
The AGN$_{IR}$ subsets are small in number with 26 in the photo-ionized subset and 5 in the shock-ionized subset. 
Thus the redshift distributions cannot be distinguished.
The AGN$_{IR}$/photo subset has a higher mean mass than the AGN$_{IR}$/shock subset,  by a factor of $\sim1.5$.

Comparison of the notAGN$_{IR}$/photo subset with the notAGN$_{IR}$/shock subset, shows
that the shock-ionized fraction increases with redshift:
from near 0 at $z<0.01$ to $\sim$50\% for $z>0.15$.
The notAGN$_{IR}$/shock subset has a narrower mass distribution than the photo-ionized subset and higher mean mass,
by a factor of $\sim1.6$.
For the notAGN$_{IR}$, the shock-ionized subsets are shifted to higher redshift and to higher mass compared to the photo-ionized subsets. 
Because of the correlation of mass with redshift  (see Fig.~\ref{fig:figmassVSz}), which is a selection 
effect of the galaxy survey, these are not independent effects. 

\subsubsection{Comparison of optically-classified and IR-classified samples for the radio sources}

Table~\ref{tab:TBLw1w2BPTstats} lists basic properties of the our different samples of G23 galaxies, with additional selections based on
detection by WISE, by 936 MHz emission or by WISE W4 ($22~\mu m$ band). 
The goal here is to determine how the radio emitters differ from their parent samples.
The SFG radio emitters (936/G23/SFG$_{opt}$) have lower mean redshift but similar mean mass, than the parent galaxy sample (G23/SFG$_{opt}$).
The AGN radio emitters (936/G23/AGN$_{opt}$) have lower mean redshift but higher mean mass, by factor 2.5, than the parent galaxy sample (G23/AGN$_{opt}$).
The lower redshift can be understood as a selection effect, but the higher mass should be intrinsic.

The WISE selected radio and non-radio emitting samples, 936/G23/WISE and G23/WISE, are compared.
SFG$_{opt}$ and AGN$_{opt}$ have lower redshift for the radio emitters, but the
mean masses of  936/G23/WISE SFG$_{opt}$ and AGN$_{opt}$ are essentially the same as for G23/WISE.
With WISE selection, there are additional AGN$_{IR}$ and notAGN$_{IR}$ categories. 
The radio emitters have lower redshift  for AGN$_{IR}$ and notAGN$_{IR}$ than for the G23/WISE sample.
The mean masses of radio emitting AGN$_{IR}$ and the G23/WISE/AGN$_{IR}$ are the same, but the
radio emitting notAGN$_{IR}$ have higher mean mass than G23/WISE/notAGN$_{IR}$, by factor 2.1.
The W1 magnitudes, related to the mass of the galaxy, of the radio emitters (936/G23/WISE) are brighter by 
$\sim$1 magnitude for all 4 subsets than for G23/WISE.

The radio (936 MHz) emitting subsets of AGN$_{IR}$,  notAGN$_{IR}$, SFG$_{opt}$ and AGN$_{opt}$ sets
are next compared.
 Table~\ref{tab:TBLw1w2BPTstats} lists some properties of these subsets, with the 936 label.
 First we consider the 936/G23/WISE sets. 
The highest redshift set is the AGN$_{IR}$ subset.
The notAGN$_{IR}$ has highest mean mass (by a factor of $\sim2-3$ over the other subsets). 
AGN$_{opt}$ and AGN$_{IR}$ are of similar mean masses and SFG$_{opt}$ are of lowest mean mass. 
The mean 936\,MHz radio flux density of AGN$_{IR}$ is much higher than any of the other subsets 
(by a factor of $\sim9$ above the overall mean for 936/G23/WISE galaxies). 
notAGN$_{IR}$  have the second highest radio flux density, higher than either AGN$_{opt}$ or SFG$_{opt}$, by a factor of $\sim2.5$. 

396 of the 936/G23/WISE sources have W4 luminosity with SN$\ge$3 
(labelled 936/G23/WISE w4snr$\ge$3). The subsets' sample sizes are small except for notAGN$_{IR}$.
The AGN$_{IR}$ are the most luminous subset at 936 MHz and at $22~\mu m$.
The sub-categories, AGN$_{IR}$/AGN$_{opt}$, AGN$_{IR}$/SFG$_{opt}$ and AGN$_{IR}$/lineflux$<0$,
 are consistent with the same luminosity in the $22~\mu m$ band, and the same flux density at 936\,MHz.

 Table~\ref{tab:TBLw1w2vsBPT} shows the breakdown of G23/WISE  and 936/G23/WISE samples into
 IR AGN categories, BPT categories and the emission line galaxies category.
The G23/WISE sample has a small fraction (1.2\%) classified as AGN$_{IR}$ compared to notAGN$_{IR}$.
It has several percent classified as SFG$_{opt}$ or AGN$_{opt}$.
For the radio emitting sample, 936/G23/WISE, 
the fraction of AGN$_{IR}$ is 5.7\%, larger by factor $\sim5$ .
 The fraction of SFG$_{opt}$ or AGN$_{opt}$ for 936/G23/WISE  is larger by a factor of $\sim$2-2.5 compared to G23/WISE.
 The rest of  Table~\ref{tab:TBLw1w2vsBPT} is discussed in Section~\ref{sec:disc} below.

\section{Discussion}\label{sec:disc}

\subsection{Radio Source Distribution}

We compared  flux density distributions of the single, double, and triple sources (Fig.~\ref{fig:figSglDblTpl}), and found that the peak for doubles and triples
occured at $\sim$20 mJy compared to $\sim$1 mJy for the singles.
All of the doubles and triples (except one) are AGN$_{IR}$.
The single notAGN$_{IR}$ triple does not have a G23 optical galaxy counterpart or redshift.
In contrast, the singles are dominated by the lower radio luminosity SFG (813 notAGN$_{IR}$ vs 51 AGN$_{IR}$).

\begin{figure}
	\includegraphics[width=\columnwidth]{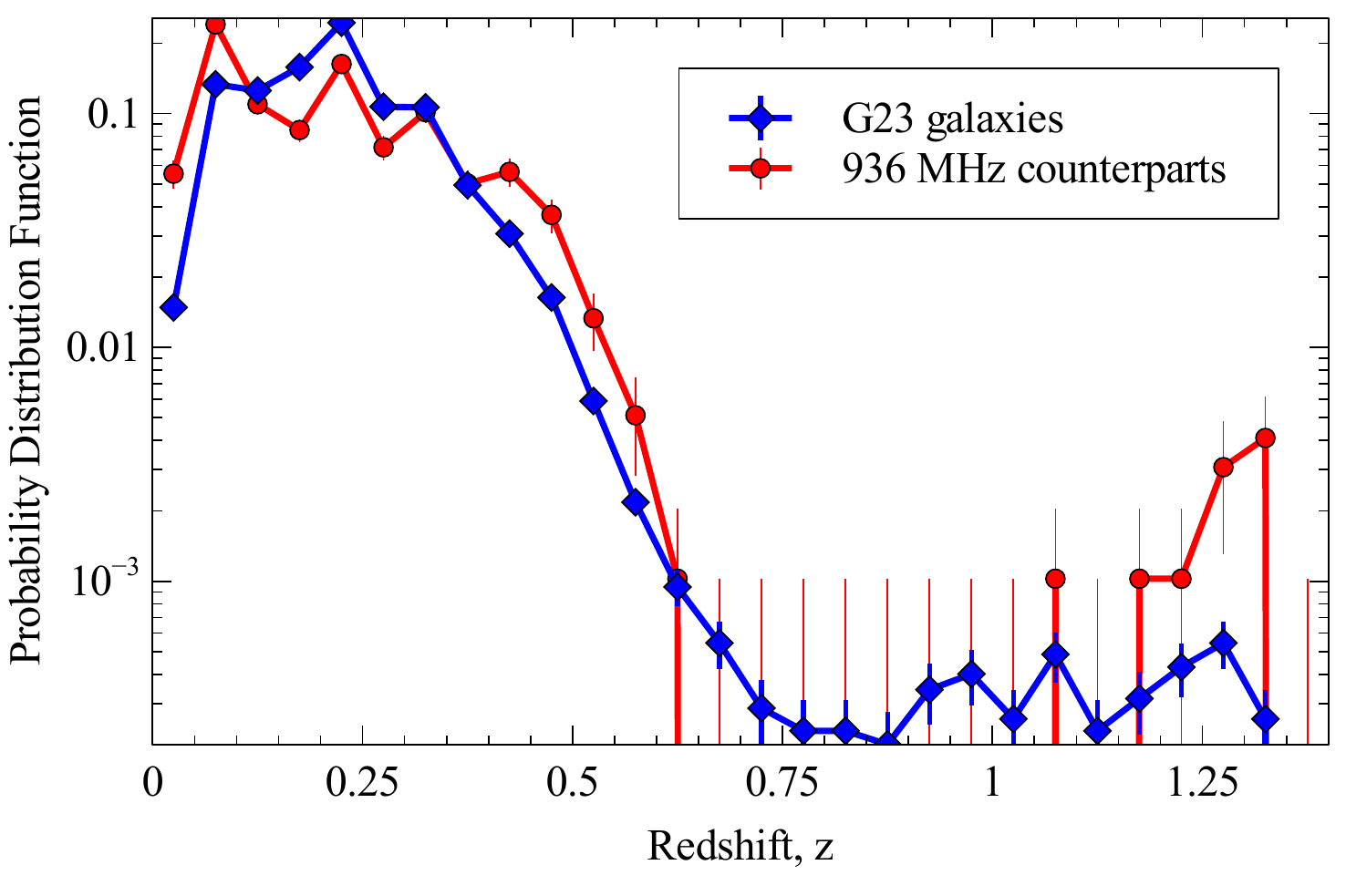}
    \caption{
    Redshift distributions ($P_i$ for redshift bin $i$) for G23 galaxies and for 936\,MHz counterparts are shown by the blue and red curves. 
    These are normalized by $\Sigma_{i} P_i=1$ .
There is an excess of  radio counterparts for some redshift ranges and a deficit for other ranges.
 The $P_i$  for 1320\,MHz counterparts is almost identical with that for 936\,MHz counterparts, except that it has slightly larger error bars.
    }
    \label{fig:pdfratio}
\end{figure} 

The redshift distributions of the G23 galaxies and those with 936 MHz emission 
 are shown in Figure~\ref{fig:pdfratio}. 
The radio emitters have similar redshifts to G23 galaxies,  but they have 
real ($> 3\sigma$)  excesses at redshifts of $\simeq$0.0-0.10 and 0.35-0.60 and a deficit at redshifts of 0.10-0.30.
There is a weak ($2\sigma$) excess for $z>$1.2. 
This may be a result from having multiple types of radio sources (SFG and AGN), each with a different spatial distribution. 
We find that the median redshifts for doubles and triples (which are all AGN) 
are 0.39 and 0.37 whilst the median redshift for single sources (which includes AGN and SFG) is 0.21.

\begin{figure}
	\includegraphics[width=\columnwidth]{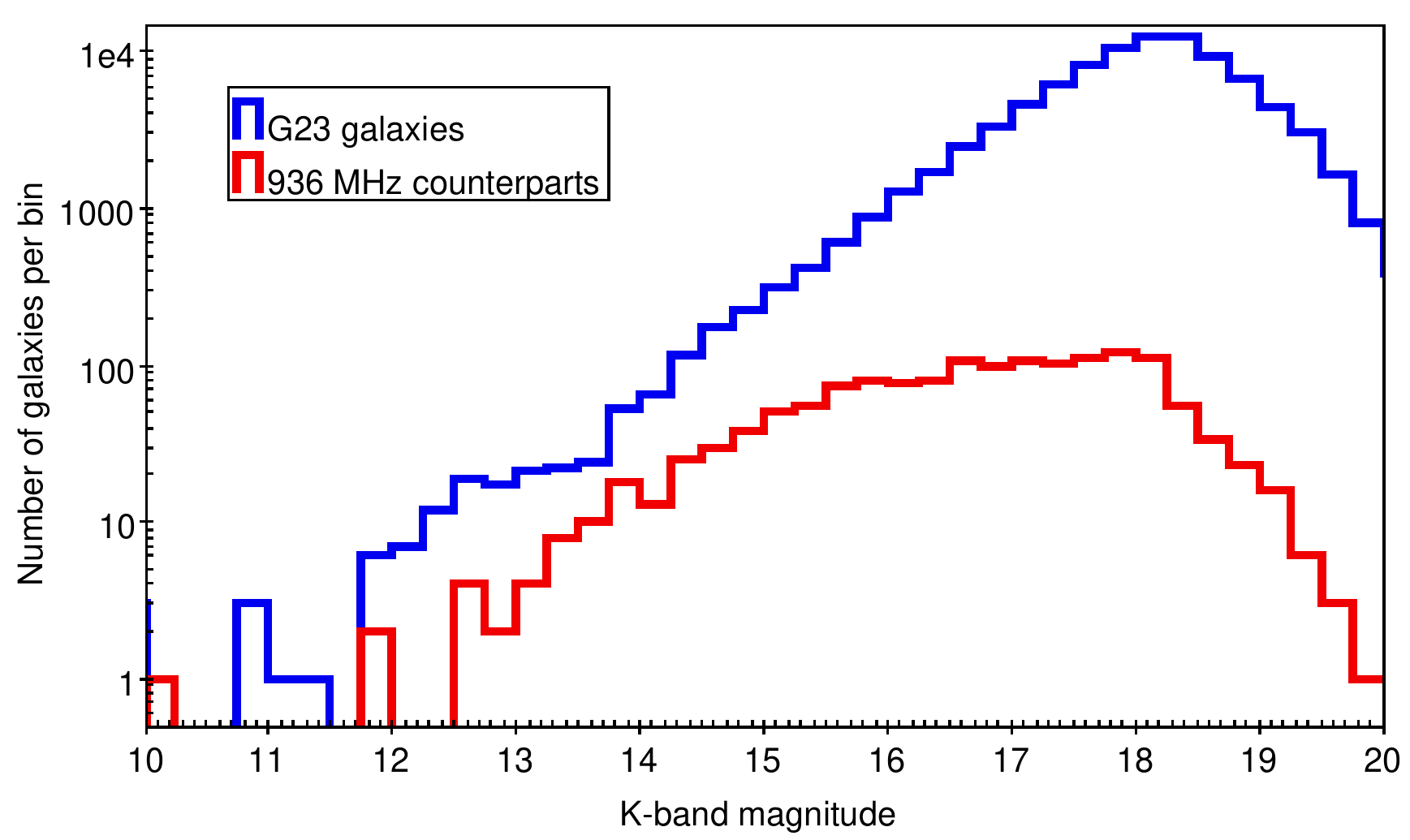} 
          \includegraphics[width=\columnwidth]{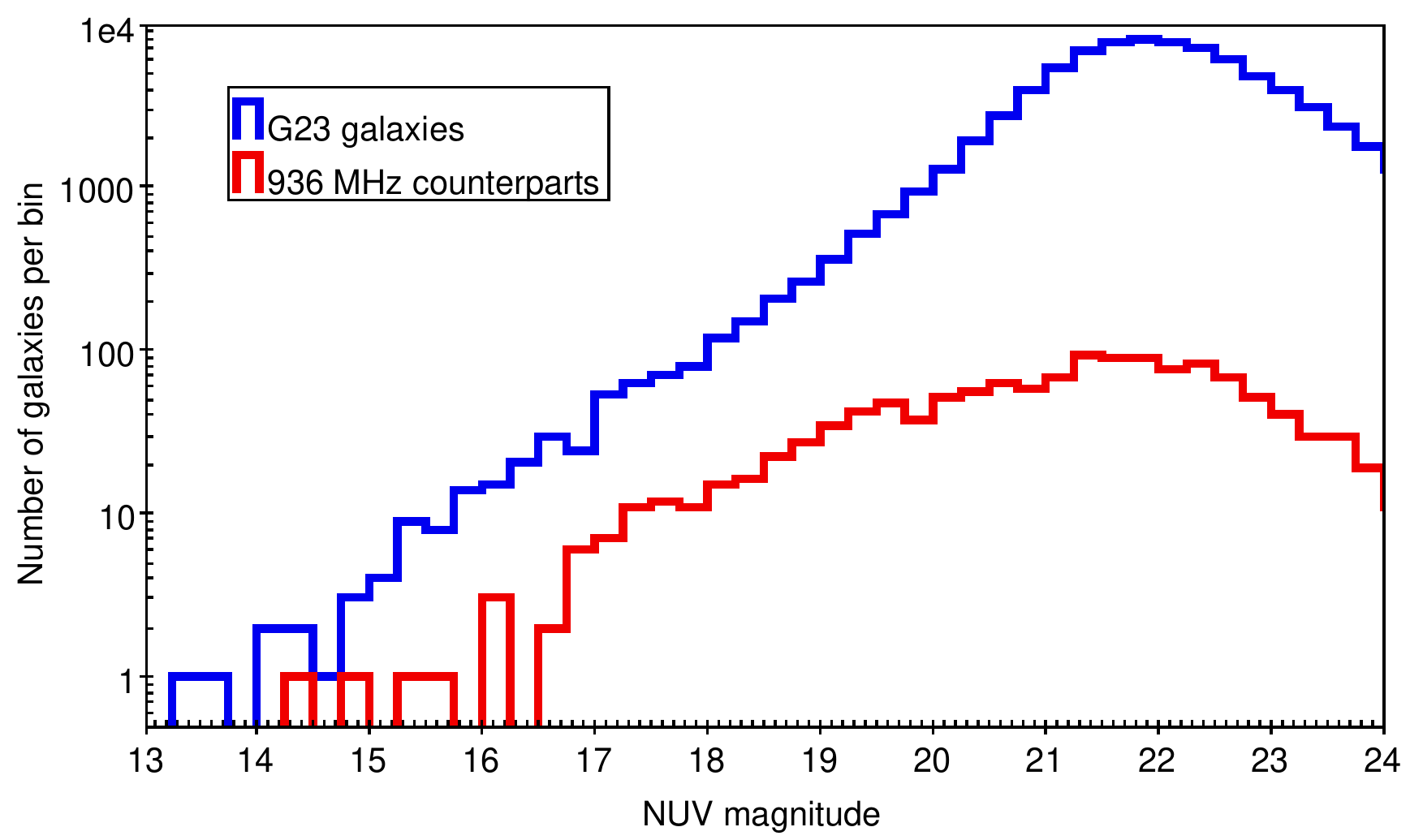} 
    \caption{
    Top panel: K-band magnitude distributions for G23 galaxies with photometry (blue histogram) and for 936\,MHz counterparts (red histogram). 
   Bottom panel: NUV magnitude distributions for G23 galaxies with photometry (blue histogram) and for 936\,MHz counterparts (red histogram). 
    }
    \label{fig:KNUV}
\end{figure} 

Figures~\ref{fig:figmassVSz} and \ref{fig:figmassHBabsdist} illustrated that galaxies that are radio sources are more massive than average. 
The mean mass is higher by a factor of $\simeq2.3$ and the peak mass of the distribution is shifted higher by a factor of $\simeq3.2$.

We found  K band and NUV magnitudes for G23 galaxies and radio counterparts by obtaining  
photometry from the G23 input catalogue. 
First we selected the set of G23 galaxies with photometry for the same sky area as the 936 MHz map.
We also selected the subset with photometry and spectra, to check for selection effects of requiring spectra.
We then cross-matched the 936 MHz sources to the G23 set and G23 subset with spectra
to obtain photometry for the radio counterparts.
Table~\ref{tab:KNUVstats} shows the means and standard deviations (SD) for the K band and NUV magnitudes of the galaxies and the radio counterparts.
The mean K band flux for the radio-emitting galaxies is higher by a factor of $\simeq3$, consistent with the mass excess.
The mean NUV band flux for the radio emitters is higher by a factor of $\simeq2$.
The main effect of the additional requirement of having spectra (sp in Table~\ref{tab:KNUVstats}) is
the removal of the faint end of the magnitude distribution, hence reduction in the mean magnitudes by
a small amount. 

Figure~\ref{fig:KNUV} shows the distributions of K-band magnitudes and NUV magnitudes for the G23 galaxies with photometry
and for the radio emitting subset. The radio emitters have much flatter distribution to low (brighter) magnitudes for both K and NUV. 
Radio-emission is detected much more frequently at high mass ($>10^{11}M_{\odot}$) (Fig.~\ref{fig:figmassHBabsdist}), 
at bright K magnitude ($<15.5$) and at bright 
NUV magnitude ($<19$), by factors which range from $\sim 5-10$.
Mass and K magnitude are expected to behave similarly because the mass is dominated by K and M stars (main-sequence and giant),
and the optical light is dominated by K and M giants. 
The excess UV magnitude can come from active star formation or from the accretion disk associated with an AGN.

\begin{table}
	\caption{K and NUV magnitudes of G23 galaxies and 936 MHz counterparts}
	\label{tab:KNUVstats}
	\begin{tabular}{lccccccc} 
		\hline
		Sample & band & sample   size$^{a}$           & mean (SD)        \\
		       &    &      & (magnitude)      \\		
		\hline
		G23ph$^{b}$  % &   &   &   \\ 
		 & K & 92838 & 18.00(0.95)\\ 
		  & NUV & 81981 & 21.99(1.17)\\ 
		G23ph/936$^{c}$ %  &   &   &  &  \\ 
		 & K & 1467 & 16.74(1.33) \\
		   & NUV & 1296 & 21.18(1.65)\\ 
		\hline
		G23ph/sp$^{d}$ %  &   &   &   \\ 
		 & K & 35236 & 17.42(0.88)\\ 
		  & NUV & 32523 & 21.64(1.23)\\ 
		G23ph/sp/936$^{c}$ % &   &   &  &  \\ 
		 & K & 930 & 16.33(1.12) \\
		   & NUV & 856 & 20.94(1.62)\\ 
		  \hline
	\end{tabular}
				\begin{tablenotes}\footnotesize
	\item[*] a. Galaxies with measured K band magnitudes ($<21$) and with measured NUV
	magnitudes ($<25$) were selected for K and NUV samples, respectively.
	\item[*] b. G23ph is the set of G23 galaxies with photometry.
	\item[*] c. 936 indicates the sample is cross-matched with the 936 MHz source catalogue using radius of $7^{\prime\prime}$. 
	\item[*] d. G23ph/sp is the set of G23 galaxies with photometry and spectra.
\end{tablenotes}	
\end{table}

The fraction of galaxies that host a radio-loud AGN is known to rise strongly with stellar mass \citep{2005MNRAS.362...25B},
from nearly zero below stellar mass of $10^{10}~M_{\odot}$ to $>30\%$ for mass $>5\times10^{11}~M_{\odot}$.
For massive early-type galaxies, a strong correlation of radio power with K-band absolute magnitude has been demonstrated
\citep[e.g.,][]{1989MNRAS.240..591S,1989ApJ...347..127F,2011ApJ...731L..41B}.
For our full optical sample of galaxies  (Figure~\ref{fig:figmassHBabsdist}), we find that there are no radio emitters below $2\times10^8 M_{\odot}$ (with one exception). 
For the absorption line galaxies, which should be  mostly massive early type galaxies, there are no radio emitters below $6\times10^9 M_{\odot}$.
This is consistent with the results of \citet{2011ApJ...731L..41B} and \citet{2005MNRAS.362...25B}, despite the difference in galaxy samples.

\subsection{AGN-SFG discrimination}

In Section \ref{AGNSFG}, we used two tests to distinguish SFG from AGN: W1-W2 and BPT. 
Both tests have weaknesses. 
Two main types of AGN (for a review, see \citealt{2012Best}) are High Excitation Radio Galaxies (HERGs) and  Low Excitation Radio Galaxies (LERGs).
The W1-W2 AGN diagnostic accepts some galaxies that are not AGN and misses AGN with weak mid-IR emission, such as LERGs.
The BPT test can miss AGN with optically obscured nuclei and
cannot be applied to the large fraction of galaxies that only present absorption lines.

\citet{2012ApJ...753...30S} showed that W1-W2 has high reliability using empirical AGN spectra and the galaxy spectral
templates of \citet{2010Assef}.
However, that study did not consider all galaxy types.
Some AGN$_{IR}$ are not true AGN: \citet{2014ApJS..212...18B} used observed spectra of nearby galaxies to show 
that observed spectral energy distributions of low metallicity 
blue compact dwarfs and extremely dusty lumininous IR galaxies (LIRGS) give W1-W2 colours like AGN.
LIRGs are found predominantly in older, massive galaxies with strong stellar absorption lines, thus would be found among
 the $\sim$13000 optical galaxies with absorption lines and not classified by the BPT criterion.

The WISE colors of LERGs and HERGs were investigated by \citet{2014MNRAS.438.1149G}. 
They show that LERGs are found in the WISE color-colour diagram with W1-W2$<$0.8 and W2-W3 in a broad range of $\sim$2-6.
The LERGs stand out most clearly from the other classes of AGN in the WISE $22~\mu m$ vs radio 151 MHz luminosity diagram
(their Fig.8). 
We converted the boundaries of the regions containing HERGs and LERGs to 936 MHz luminosity 
using a spectral index of $-0.7$ and plotted those boundaries on Figure~\ref{fig:figL22vL936}.
The HERG and LERG regions are distinct from the region occupied by SFGs.
 Most galaxies in our sample of 
radio emitting galaxies are located in a region consistent with SFGs, and few are consistent with HERGs or LERGs.
  
The $22~\mu m$ vs radio luminosity diagram was made for the subcategories of
notAGN$_{IR}$/SFG$_{opt}$, notAGN$_{IR}$/AGN$_{opt}$, AGN$_{IR}$/AGN$_{opt}$ and AGN$_{IR}$/ SFG$_{opt}$. 
We find notAGN$_{IR}$/AGN$_{opt}$ and  notAGN$_{IR}$/ SFG$_{opt}$ are not distinguishable in $22~\mu m$ vs 
radio luminosity.
The  AGN$_{IR}$/AGN$_{opt}$ and  AGN$_{IR}$/SFG$_{opt}$ are at higher $22~\mu m$ and radio luminosity 
than the two notAGN$_{IR}$ sets. However all 4 sets lie in the region of the blue points in the top panel of 
Figure~\ref{fig:figL22vL936}\footnote{The requirement of optical classification (redshift $<$0.32) removes all of the high luminosity AGN in
 both  $22~\mu m$ and 926 MHz bands. Fig.~\ref{fig:figL22vL936} does not include the optical (redshift) selection, so includes 
 those high luminosity AGN.}, and are in the region of SFGs rather than the regions of HERGs or LERGs.
 
The optical and IR classifications as AGN do not agree for a significant number of galaxies (see Table 5).
24 of the SFG$_{opt}$ are AGN$_{IR}$ (vs 2014 notAGN$_{IR}$) and AGN$_{opt}$ are mostly notAGN$_{IR}$ (695) with few AGN$_{IR}$ (31).
What causes the difference between BPT and W1-W2 classifications?

The W1-W2 classification is based on the slope of the 3.4 to 4.6 $\mu m$ IR spectrum.
This is on the long-wavelength Rayleigh-Jeans side of the stellar continuum peak for all stars, whether they are hot, massive and young, or old and cool. 
The presence of a strong AGN implies a significant contribution to the IR continuum by dust which is reradiating the AGN continuum.
This results in a mid IR spectrum flatter than that seen in typical starbursts.

An AGN can have high extinction from the surrounding interstellar medium of the galaxy preventing the optical BPT lines from the AGN from being seen, 
while emission from star formation in the regions of the galaxy outside the
high-extinction region may be detected. In this scenario the result is that such a galaxy will be
classified as SFG$_{opt}$, despite the presence of an AGN.
Thus the SFG$_{opt}$/ AGN$_{IR}$ galaxies can be explained as high-extinction AGN,
accounting for one of the two conflicting classification pairs.

The extinction of galaxies is too small in the 3.4 to 4.6 $\mu m$ range to affect the spectral slope, so the W1-W2 diagnostic can distinguish strong AGN from SFG.
 Faint AGN where the stellar continuum has a significant contribution ($\gtrsim$50\%) in the 3.4 to 4.6 $\mu m$ range 
have W1-W2<0.8 (Figure~2 of \citealt{2012ApJ...753...30S}), which results in faint AGN being classified as  notAGN$_{IR}$/SFG$_{opt}$ galaxies.

\citet{2010Wright} (their Figure~10) shows the WISE colour-colour 
diagram with different classes of galaxies in partially overlapping regions. 
The WISE colour-colour diagram for our sample (middle panel of Figure~\ref{fig:figSharpLine}) shows that 
 notAGN$_{IR}$/AGN$_{opt}$ galaxies are at
smaller W2-W3 than notAGN$_{IR}$/SFG$_{opt}$ galaxies.
Both of our sets overlap the spiral, LIRG and starburst regions. 
From the lower 2 panels of Fig.~\ref{fig:figSharpLine}, we find that notAGN$_{IR}$ galaxies with optical emission 
dominated by star formation (SFG$_{opt}$, blue histogram) are most consistent with W2-W3
colors of LIRGs. 
The notAGN$_{IR}$/AGN$_{opt}$ (yellow histogram) have W2-W3 colors consistent with mixture of
spirals and LIRGs and possibly starbursts.  
The two AGN$_{IR}$ categories (SFG$_{opt}$ and AGN$_{opt}$) have few objects but spread across
the LIRG, Seyfert, starburst, QSO and ULIRG regions of  Figure~10 of \citet{2010Wright}.

The notAGN$_{IR}$/AGN$_{opt}$ galaxies have optical lines showing AGN-like ionization, 
but their mid IR is powered mainly by stellar emission. 
The W1-W2$<$0.8  notAGN$_{IR}$ criterion corresponds to the AGN contributing $<$62\%
of the mid IR emission \citep{2012ApJ...753...30S} for redshift z<0.32 (the upper limit of galaxies classified by BPT).
The left panel of Fig.2 of \citet{2012ApJ...753...30S} shows that almost the same limit ($<$58\%) applies for E(B-V)=1
and the doesn't change much for higher extinction if z<0.32. 
Thus the AGN can contribute up to 60\% to the mid IR emission, yet a galaxy will be classified as notAGN$_{IR}$.
We refer to this case as a weak AGN.

The  notAGN$_{IR}$/AGN$_{opt}$ could be weak AGN where the stellar mid IR dominates over the AGN mid IR,
but the AGN optical emission would have to dominate over the stellar optical emission to give the AGN$_{opt}$ classification.
Using the \citet{2006MNRAS.372..961K} criterion, we find 64\% of AGN$_{opt}$ are Seyferts and  36\% are LINERs.
Because the vast majority of both sets are notAGN$_{IR}$, this implies both Seyferts and LINERs, with few exceptions, 
are not classified as AGN using IR colors.
 
Some fraction of AGN should be in the phase where the AGN has shut off but the AGN photo-ionization remains. 
Such AGN would be seen as AGN$_{opt}$/notAGN$_{IR}$/photo, whereas active AGN would be seen as
AGN$_{opt}$/AGN$_{IR}$/photo. 
Table~\ref{tab:TBLw1w2vsBPT} gives
the observed number of seen as these two types (446 and 26, respectively).
The ratio of these two types of $\sim17$ in the current sample would imply the timescale for AGN photo-ionization 
to linger is greater than the AGN active phase by a factor of $\sim17$.  
The estimated timescale for lingering photo-ionization in AGN is $10^4$ to $10^5$yr \citep{2015MNRAS.451.2517S},
and the AGN active timescale is estimated at  $10^5$yr.
Thus the expected ratio of AGN$_{opt}$/notAGN$_{IR}$/photo to AGN$_{opt}$/AGN$_{IR}$/photo should be 
$\sim0.1$ to 1, rather than $\sim17$. 
So, either: a) only $\sim$3 to 30 of the 446 AGN$_{opt}$/notAGN$_{IR}$/photo can be explained by lingering photo-ionization; or  
b) the AGN on and off timescales are inaccurate by 1 to 2 orders of magnitude.
In the first case, there must be another explanation for most of the AGN$_{opt}$/notAGN$_{IR}$/photo.

In summary, there is the explanation of extinction for the apparent conflicting classification AGN$_{IR}$/SFG$_{opt}$,
but no single convincing explanation for notAGN$_{IR}$/AGN$_{opt}$.

\section{Summary and Conclusion}\label{sec:summary}

We have processed ASKAP commissioning observations of the GAMA G23 field in two frequency bands centred at
936\,MHz and 1320\,MHz, and have demonstrated that ASKAP produces excellent image quality, allowing identification of radio sources
down to $\sim$1 mJy with positions accurate to $\sim5^{\prime\prime}$.
We extracted the radio components and identified multiple component radio sources to produce 936\,MHz and 1320\,MHz radio source catalogues.

The radio sources were position matched to GAMA G23 galaxies, to WISE IR sources and to G23 galaxies with WISE IR counterparts. 
The masses for the G23 galaxies were based on masses of photometry-matched galaxies from the GAMA equatorial fields.
 A summary of our analysis of the multi-wavelength data set is as follows. 
\begin{itemize}
\item
Of the $\sim$5800 radio sources, $\sim$1000 have counterparts amongst the G23 galaxies, $\sim$3000 have WISE IR counterparts
and $\sim$900 have both G23 galaxy and WISE counterparts.
Some fraction of the radio sources without WISE counterparts will be AGN too faint to be detected by WISE \citep[IR-faint radio sources; e.g.][]{2014MNRAS.439..545C}.
 The G23 galaxy radio sources tend to be those at the  
high-mass end  ($\gtrsim 10^{10}$M$_{\odot}$) of the galaxy distribution (top panel of Figure~\ref{fig:figmassVSz}).
\item
Galaxies that are detected in radio are observed to be more massive, and brighter at K-band and UV, than those without radio sources.
Radio emission from AGN is expected to be brighter for more massive galaxies, which are brighter in K band. 
Similarly, radio emission from star formation and UV emission are stronger for galaxies with active star formation. 
\item
The majority of the ASKAP radio sources show the well-known correlation \citep{2009Rieke} for SFG between the radio 
and mid-IR luminosities (Figure~\ref{fig:figL22vL936}).
A few tens of sources have excess radio emission, and are likely to host AGN, although not all are classified as AGN by their IR colour.
\item
The G23 galaxies were classified as AGN$_{IR}$ or notAGN$_{IR}$, based on the W1-W2 indicator.
$\sim$380 galaxies are classified as  AGN$_{IR}$, compared to $\sim$30,000 classified as notAGN$_{IR}$. 
AGN$_{IR}$ have higher redshifts than notAGN$_{IR}$ (factor $\sim3$) but similar masses.
AGN$_{IR}$ have much higher $22~\mu m$ luminosities (factor $\sim30$).
Only $\sim$50 of these AGN$_{IR}$ are radio sources.
\item
Using the BPT diagnostic, galaxies were classified as AGN$_{opt}$ or SFG$_{opt}$. 
$\sim$770 galaxies are classified as  AGN$_{opt}$ and $\sim$2200 classified as SFG$_{opt}$.
AGN$_{opt}$ have similar redshifts to SFG$_{opt}$ but a broader mass distribution.
\item
We find disagreement between these two AGN indicators. 
24 of  55 AGN$_{IR}$ are classified as  SFG$_{opt}$ and $\sim$700 of 2800 notAGN$_{IR}$ are classified as AGN$_{opt}$
(Table~\ref{tab:TBLw1w2vsBPT}).
The AGN$_{IR}$/SFG$_{opt}$ galaxies can be explained in terms of obscured AGN$_{IR}$.
The reasons why  AGN$_{opt}$ are classified as notAGN$_{IR}$ are not clearly understood yet. 
\item
The radio sources were separated into AGN$_{IR}$, notAGN$_{IR}$, AGN$_{opt}$ and SFG$_{opt}$ subsets.
These radio subsets are smaller by $\sim$1 order of magnitude than the galaxy subsets.
Each of the 4 radio subsets has indistinguishable redshift distribution from that of the corresponding parent galaxy subset
(Figure~\ref{fig:figG23WISEz}, left panels).
We find that the level of disagreement between optical and IR AGN diagnostics is similar for the radio emitting subsets as for the corresponding parent galaxy subset (Table~\ref{tab:TBLw1w2vsBPT}).
\end{itemize}

The results presented  in this paper are a first step towards combining large deep radio survey data with large well-studied samples of galaxies in the nearby Universe with extensive multi-wavelength data. 
The EMU survey will be almost an order of magnitude more sensitive than the observations presented here, and this, combined with even deeper  optical and IR observations, 
will provide even greater overlap between the radio, optical, and IR samples of galaxies, enabling radio observations to become another tool that will be used routinely to characterise and understand  galaxies in the local Universe.

\section*{Acknowledgements}
The Australian SKA Pathfinder is part of the Australia Telescope National Facility which is managed by CSIRO. 
Operation of ASKAP is funded by the Australian Government with support from the National Collaborative Research Infrastructure Strategy. 
ASKAP uses the resources of the Pawsey Supercomputing Centre. 
Establishment of ASKAP, the Murchison Radio-astronomy Observatory and the Pawsey Supercomputing Centre are initiatives of the Australian Government, with support from the Government of Western Australia and the Science and Industry Endowment Fund. 
We acknowledge the Wajarri Yamatji people as the traditional owners of the Observatory site.
GAMA is a joint European-Australasian project based around a spectroscopic campaign using the Anglo-Australian Telescope. The GAMA input catalogue is based on data taken from the Sloan Digital Sky Survey and the UKIRT Infrared Deep Sky Survey. Complementary imaging of the GAMA regions is being obtained by a number of independent survey programmes including GALEX MIS, VST KiDS, VISTA VIKING, WISE, Herschel-ATLAS, GMRT and ASKAP providing UV to radio coverage. 
GAMA is funded by the STFC (UK), the ARC (Australia), the AAO, and the participating institutions. The GAMA website is http://www.gama-survey.org/ .
This publication makes use of data products from the Wide-field Infrared Survey Explorer, which is a joint project of the University of California, Los Angeles, and the Jet Propulsion Laboratory/California Institute of Technology, funded by the National Aeronautics and Space Administration.
DL is supported by a grant from the Natural Sciences and Engineering Research Council of Canada.
The referee is acknowledged for a number of suggestions that led to improvements in this paper.

\bibliographystyle{mnras}
\bibliography{ASKAPG23_2019} % if your bibtex file is called example.bib

%%%%%%%%%%%%%%%%% APPENDICES %%%%%%%%%%%%%%%%%%%%%
%\appendix
% Don't change these lines
%\bsp	% typesetting comment
\label{lastpage}
\end{document}